\newif\ifNotes\Notesfalse
\newif\ifAnon\Anonfalse
\newif\ifDraft\Draftfalse
\newif\ifArxiv\Arxivtrue
\newif\ifCamera\Camerafalse

\ifCamera
\Notesfalse
\Anonfalse
\Draftfalse
\Arxivfalse
\fi

\PassOptionsToPackage{hyphens}{url}
\PassOptionsToPackage{dvipsnames,table}{xcolor}

\documentclass[letterpaper,twocolumn,10pt]{article}
\usepackage{usenix-2020-09}

\usepackage{epsfig,endnotes}
\ifCamera
\else
\fi
\usepackage{color}
\usepackage{gensymb}
\usepackage[sort,numbers]{natbib}
\usepackage{combelow}
\usepackage{booktabs}
\usepackage{xspace}
\usepackage{textcomp}
\usepackage{hyperref}
\usepackage[flushleft]{threeparttable}
\usepackage{pifont}
\usepackage{courier}
\usepackage{array}
\usepackage{listings}
\usepackage[geometry]{ifsym}

\usepackage[inline]{enumitem}
\usepackage{caption}
\usepackage{subcaption}
\usepackage{xspace}
\usepackage{graphicx}
\usepackage{soul}
\usepackage[cachedir=minted-cache, frozencache=true]{minted}
\usepackage{MnSymbol}
\usepackage{multirow}
\usepackage{tcolorbox}
\usepackage[htt]{hyphenat}
\usepackage[compact]{titlesec}

\newtcolorbox{takeaway}[1][]
{
    top=0pt,
    left=0pt,
    bottom=0pt,
    right=0pt,
    boxsep=2pt,
    arc=0pt,
    before=\vspace{1pt}\par\noindent,
    #1,
}

\setminted[c]{breaklines,xleftmargin=2em,linenos,tabsize=4,obeytabs,frame=single,framesep=5pt}
\setminted[js]{breaklines,xleftmargin=2em,linenos,tabsize=4,obeytabs,frame=single,framesep=5pt}
\setminted[gas]{breaklines,xleftmargin=2em,linenos,tabsize=4,obeytabs,frame=single,framesep=5pt}

\setitemize{itemsep=1ex,topsep=1ex,parsep=0pt,partopsep=0pt}

\let\origthelstnumber\thelstnumber
\makeatletter
\newcommand*\Suppressnumber{%
	\lst@AddToHook{OnNewLine}{%
		\let\thelstnumber\relax%
		\advance\c@lstnumber-\@ne\relax%
	}%
}

\newcommand*\Reactivatenumber[1]{%
	\setcounter{lstnumber}{\numexpr#1-1\relax}
	\lst@AddToHook{OnNewLine}{%
		\let\thelstnumber\origthelstnumber%
		\refstepcounter{lstnumber}
	}%
}

\makeatother

\ifDraft
\usepackage{draftwatermark}
\definecolor{watermarkcolor}{rgb}{0.8,0.8,1}
\SetWatermarkText{DRAFT}
\SetWatermarkColor{watermarkcolor}
\fi

\definecolor{linkcolor}{rgb}{0.65,0,0}
\definecolor{citecolor}{rgb}{0,0.4,0}
\definecolor{urlcolor}{rgb}{0,0,0.65}
\definecolor{TolDarkGreen}{HTML}{117733}
\hypersetup{hyperindex=true,pdfpagemode=UseNone,pdfstartview=FitH,colorlinks=true, linkcolor=linkcolor, urlcolor=urlcolor, citecolor=citecolor}

\usepackage{color}
\usepackage[english]{babel}
\usepackage{blindtext}
\usepackage{tikz}

\newcommand{\swallow}[1]{}
\ifNotes
\newcommand{\colorcomment}[2]{\leavevmode\unskip\space{\color{#1}#2}\xspace}
\else
\newcommand{\colorcomment}[2]{\leavevmode\unskip\relax}
\fi

\ifArxiv

\else

\fi

\definecolor{darkviolet}{HTML}{9400D3}
\definecolor{PineGreen}{HTML}{01796F}
\definecolor{neonpink}{HTML}{FF10F0}

\definecolor{lightgray}{rgb}{0.95, 0.95, 0.95}
\definecolor{darkgray}{rgb}{0.4, 0.4, 0.4}
\definecolor{editorGray}{rgb}{0.95, 0.95, 0.95}
\definecolor{editorOcher}{rgb}{1, 0.5, 0}
\definecolor{editorGreen}{rgb}{0, 0.5, 0}
\definecolor{orange}{rgb}{1,0.45,0.13}		
\definecolor{olive}{rgb}{0.17,0.59,0.20}
\definecolor{brown}{rgb}{0.69,0.31,0.31}
\definecolor{purple}{rgb}{0.38,0.18,0.81}
\definecolor{lightblue}{rgb}{0.1,0.57,0.7}
\definecolor{lightred}{rgb}{1,0.4,0.5}

\lstdefinelanguage{JavaScript}{
	morekeywords={typeof, new, true, false, catch, function, return, null, catch, switch, var, if, in, while, do, else, case, break, const},
	morecomment=[s]{/*}{*/},
	morecomment=[l]//,
	morestring=[b]",
	morestring=[b]',
	morestring=[b]/,
	stringstyle=\color{olive}
}

\lstdefinelanguage{HTML5}{
	language=html,
	sensitive=true,	
	alsoletter={<>=-},	
	morecomment=[s]{<!-}{-->},
	tag=[s],
	otherkeywords={
		<!DOCTYPE,
		</html, <html, <head, <title, </title, <style, </style, <link, </head, <meta, />,
		</body, <body,
		</div, <div, </div>, 
		</p, <p, </p>,
		</script>, <script>,
		<canvas, /canvas>, <svg, <rect, <animateTransform, </rect>, </svg>, <video, <source, <iframe, ></iframe>, </video>, <image, </image>, <header, </header, <article, </article
	},
	ndkeywords={
		=,
		charset=, src=, id=, width=, height=, style=, type=, rel=, href=,
		fill=, attributeName=, begin=, dur=, from=, to=, poster=, controls=, x=, y=, repeatCount=, xlink:href=,
		margin:, padding:, background-image:, border:, top:, left:, position:, width:, height:, margin-top:, margin-bottom:, font-size:, line-height:,
		transform:, -moz-transform:, -webkit-transform:,
		animation:, -webkit-animation:,
		transition:,  transition-duration:, transition-property:, transition-timing-function:,
	}
}

\lstdefinestyle{web} {
	basicstyle={\footnotesize\ttfamily},   
	frame=single,
	xleftmargin={0.75cm},
	numbers=left,
	stepnumber=1,
	firstnumber=1,
	numberfirstline=true,	
	identifierstyle=\color{black},
	keywordstyle=\color{blue}\bfseries,
	ndkeywordstyle=\color{editorGreen}\bfseries,
	stringstyle=\color{editorOcher}\ttfamily,
	commentstyle=\color{brown}\ttfamily,
	language=HTML5,
	alsolanguage=JavaScript,
	alsodigit={.:;},	
	tabsize=2,
	showtabs=false,
	showspaces=false,
	showstringspaces=false,
	extendedchars=true,
	breaklines=true,
	literate=%
	{Ö}{{\"O}}1
	{Ä}{{\"A}}1
	{Ü}{{\"U}}1
	{ß}{{\ss}}1
	{ü}{{\"u}}1
	{ä}{{\"a}}1
	{ö}{{\"o}}1
}

\newcommand{\parhead}[1]{\vspace{0pt plus 1pt}\par\noindent\textbf{#1}\hspace{.75em plus .5em minus .5em}}

\usepackage[nameinlink,capitalize]{cleveref}
\crefname{figure}{Figure}{Figures}
\crefname{equation}{Equation}{Equations}
\crefformat{equation}{#2{}Equation~#1{}#3}
\crefrangeformat{line}{Lines~#3#1#4--#5#2#6}

\DeclareMathAlphabet\mathbfcal{OMS}{cmsy}{b}{n}

\newcommand{\js}{JavaScript\xspace}

\ifAnon
\newcommand{\authorlist}[1]{\relax}
\else
\newcommand{\authorlist}[1]{\author{#1}}
\fi

\usepackage{lipsum}

\begin{document}
\title{Hot Pixels: Frequency, Power, and Temperature Attacks on GPUs and Arm SoCs}
\ifAnon
\author{Paper \#596}
\else
\author{
	{\rm Hritvik Taneja}\\
	Georgia Tech\\
	\rm htaneja3@gatech.edu
	\and 
	{\rm Jason Kim}\\
	Georgia Tech\\
	\rm nosajmik@gatech.edu
	\and
	{\rm Jie Jeff Xu}\\
	Georgia Tech\\
	\rm jxu680@gatech.edu
	\and
	{\rm Stephan van Schaik}\\
	University of Michigan\\
	\rm stephvs@umich.edu
	\and
	{\rm Daniel Genkin}\\
	Georgia Tech\\
	\rm genkin@gatech.edu
	\and
	{\rm Yuval Yarom\thanks{Work partially done while affiliated with the University of Adelaide}}\\
	 Ruhr University Bochum\\
	\rm yuval.yarom@rub.de
}
\fi
\maketitle

\begin{abstract}
The drive to create thinner, lighter, and more energy efficient devices has resulted in modern SoCs being forced to balance a delicate tradeoff between power consumption, heat dissipation, and execution speed (i.e., frequency). While beneficial, these DVFS mechanisms have also resulted in software-visible hybrid side-channels, which use software to probe analog properties of computing devices. Such hybrid attacks are an emerging threat that can bypass countermeasures for traditional microarchitectural  side-channel attacks. 

Given the rise in popularity of both Arm SoCs and GPUs, in this paper we investigate the susceptibility of these devices to information leakage via power, temperature and frequency, as measured via internal sensors. We demonstrate that the sensor data observed correlates with both instructions executed and data processed, allowing us to mount software-visible hybrid side-channel attacks on these devices.

To demonstrate the real-world impact of this issue, 
we present JavaScript-based pixel stealing and history sniffing attacks on Chrome and Safari, with all side channel countermeasures enabled. Finally, we also show website fingerprinting attacks, without any elevated privileges.
\end{abstract}

\section{Introduction}
Since their discovery about 70 years ago~\cite{tempest}, side channel attacks have traditionally been divided into two main categories:
either physical side channels (e.g., power consumption and electromagnetic radiation)~\cite{agrawal2002side, gandolfi2001electromagnetic, quisquater2001electromagnetic, van1985electromagnetic, kocher1999differential, mangard2008power} measured by external equipment, or microarchitectural attacks (e.g., cache contention and transient execution)~\cite{kocher1996timing, osvik2006cache, last_level_cache_practical, flush+reload, spectre, meltdown, foreshadow} mounted via resident software.

Recently, however, side channel research has uncovered an intermediate category, where attackers measure analog leakage using software-accessible mechanisms, instead of external measurement devices. Indeed, recent works have demonstrated using Intel's RAPL interface~\cite{Lipp2020Platypus} to perform software-only power analysis, exploiting Dynamic Voltage Frequency Scaling (DVFS) to break constant-time code~\cite{hertzbleed, liu2022frequency} and even mounting electromagnetic attacks via audio interfaces~\cite{ear}. These software-based analog attacks pose a paradigm shift in side channel research, as they allow attackers to bypass microarchitectural-attack countermeasures previously considered sufficient to mitigate software-based side channels.  

Another change brought about in the recent evolution of computing hardware is the departure from \texttt{x86}-based architectures as the sole source of high performance computing. Indeed, the past few years have seen the introduction of highly-performant Arm-based hardware, as well as a steady growth in the capabilities and integration of GPUs. Aiming to create thinner, lighter, and more energy efficient devices, modern CPUs and GPUs are forced to balance a delicate three-way tradeoff between power consumption, heat dissipation and execution speed (frequency). While exceptions do exist~\cite{dipta2022df}, the side channel implications of the DVFS mechanism were primarily studied  on (properly cooled and powered) Intel platforms~\cite{Lipp2020Platypus, hertzbleed, liu2022frequency}, despite the
increased reliance on DVFS in GPUs and high-performance Arm SoCs.

Thus, in this paper we study the following main questions:	

\smallskip
\emph{
	Are software-based physical side channels present on GPUs and high-end Arm SoCs?  What would it take to create such attacks and what information can be extracted using it?
}

\subsection{Our Contribution}
In this paper, we show that SoCs exhibit instruction- and data-dependent behaviors as they struggle to balance the three-way tradeoff between frequency, power, and temperature.  Moreover, we demonstrate how this behavior is present across high-end Arm SoCs as well as  GPUs,  resulting in side channel leakage via two properties when the third becomes an operational constraint. Remarkably, the ever-changing behavior of these SoCs is also visible via internal measurement sensors, allowing us to distinguish between executed instructions, and even different operands of the same instruction. Next, as access to internal  frequency, power, and temperature sensors remains open to unprivileged users in most platforms, we can exploit these for mounting website fingerprinting attacks using native code running on the target device.
Finally, we show that the frequency throttling behavior of the SoC  is both data-dependent and observable via timing channels, even from \js code running inside browsers. Capitalizing on this observation, we design pixel stealing and history sniffing attacks against recent versions of Chrome and Safari, with all side channel countermeasures enabled. 

\parhead{Observing Instruction and Data-Dependent CPU Behavior.} 
In \cref{sec:arm_cpu} we investigate the frequency, power, and thermal behavior of  Arm SoCs as they execute different workloads. We find that passively cooled devices (e.g., phones and M1-based MacBook Air laptops) are often thermally constrained, and thus adjust their frequency and power while aiming not to exceed a certain temperature. In contrast, we find that actively cooled devices (e.g., M1-based MacBook Pro and Mac Mini) are usually frequency constrained, aiming to complete the workload as fast as possible, and thus leak information via power and temperature. For both categories of devices we show that it is possible to distinguish between different instructions being executed, as well as between different operands to the same instruction, using data from only internal measurement sensors. Finally, we model CPU leakage in the Hamming distance (HD) model, showing a relationship between operand HD and the CPU power, frequency, and temperature.

\parhead{Observing the Behavior of Integrated and Discrete GPUs.} 
Moving away from CPUs to discrete and integrated GPUs (dGPUs and iGPUs), in \cref{sec:gpu} we
show that GPU devices likewise continuously adjust their power, frequency and temperature aiming to meet operational restrictions. We show that these adjustments are also visible via software-only measurements, allowing us to distinguish between different instructions and different operands. Finally, we also analyze leakage from GPU devices using the Hamming Distance and Hamming Weight models. 

\parhead{Browser-Based Pixel Stealing Attacks.} Having modeled the leakage of both CPU and GPU devices, in \cref{sec:attacks-intensive} we demonstrate \js-based pixel stealing attacks from cross origin \texttt{iframe}s. At a high level, we design an SVG filter stack, which generates high computational load, causing throttling in a way which depends on the pixel's color. Observing the throttling behavior via timing, we recover images from cross origin \texttt{iframe}s and extract the user's browsing history in recent versions of Chrome and Safari, with all side channel countermeasures enabled. 
In particular, our pixel-stealing attacks demonstrate that constant-cycle code is insufficient to prevent leakage from SVG filters, when the underlying hardware exhibits color-dependent DVFS behavior.

\parhead{Fingerprinting Websites Through Internal Frequency and Power Sensors.} As our final contribution, in \cref{sec:attacks-light} we demonstrate that DVFS-based attacks are still possible even without inducing high computational load. More specifically, we show that websites cause frequency and power bursts on Apple iGPUs at different times and intensity. These bursts form a unique  signature, allowing us to fingerprint websites with high accuracy across different devices. With Apple providing access to GPU frequency and power via unprivileged code, we show how these can be exploited to mount software-only physical side channel attacks on Apple devices without inducing high workloads. 

\parhead{Summary of Contributions.} We contribute the following:
\begin{itemize}[nosep, leftmargin=*]
	\item We demonstrate instruction- and data-dependent behavior of Arm CPUs, iGPUs and dGPUs, leaking information via frequency, power and, temperature (\cref{sec:arm_cpu,sec:gpu}).
	\item We present browser-based pixel stealing and history sniffing attacks against recent versions of Chrome and Safari, with all side channel countermeasures enabled (\cref{sec:attacks-intensive}).
	\item We show website fingerprinting attacks on Apple devices using internal  power and frequency measurements from unprivileged software (\cref{sec:attacks-light}).
\end{itemize}

\subsection{Responsible Disclosure}
We initially disclosed our findings to the product security teams of Apple, Nvidia, AMD, Qualcomm, and Intel and the Chrome team at Google on March 2023. All vendors have acknowledged the issues described in this paper. We have further discussed mitigation to our work with the Chrome security team (see \cref{sec:limit+mit}).

\section{Background}

\subsection{Hybrid and GPU Side Channel Attacks}
We begin by surveying prior side channel techniques. 

\parhead{Physical Side Channels.}
Physical side channels leak information via physical properties of the device, such as its power consumption~\cite{kocher1999differential, mangard2008power}, electromagnetic emanations (EM)~\cite{agrawal2002side, gandolfi2001electromagnetic, quisquater2001electromagnetic}, and acoustics~\cite{genkin2014rsa}. 
Traditionally, physical side channel research has focused on small computing devices such as FPGAs and embedded microcontrollers.
However, more recent works also consider laptops~\cite{genkin2015get, genkin2015stealing}, phones~\cite{genkin2016ecdsa, belgarric2016side}, and GPUs~\cite{zhan2022graphics, liang2022clairvoyance}. 

\parhead{Hybrid Attacks.} 
Hybrid attacks aim to use software to measure physical properties.
These include using built-in power measurements for power analysis~\cite{Lipp2020Platypus} and reading CPU temperature~\cite{thermalbleed}.
EM and power analysis attacks can even be conducted using internal audio interfaces~\cite{ear} and Rowhammer-induced bit flips~\cite{cohen2022hammerscope}. Moreover, software-based fault attacks have also been demonstrated~\cite{plundervolt, kenjar2020v0ltpwn, clkscrew} via the CPU's DVFS mechanisms.

\parhead{DVFS-based Attacks.} Recently, \cite{hertzbleed, liu2022frequency, dipta2022df} showed that power limits on Intel CPUs can result in data-dependent frequency throttling, demonstrating key extraction attacks against constant-time SIKE~\cite{hertzbleed} and AES-NI~\cite{liu2022frequency} implementations, as well as fingerprinting attacks~\cite{dipta2022df}. Finally, in a concurrent independent work, \citet{2h2b} demonstrated the extension of \cite{hertzbleed} to additional cryptographic targets, as well as showing how CPU throttling can be affected by data processed during GPU computations, resulting in pixel stealing attacks in the Chrome browser.

In this paper, we show that  frequency throttling is part of the three-way trade-off between power consumption, execution speed, and heat dissipation. Moving away from x86 devices, we show that GPUs and ARM SoCs also exhibit browser-observable data dependent behavior, especially in the case that one of the three variables becomes a operational constraint.

\parhead{GPU Side-Channel Attacks.}
Recent work shows that manipulating the power curve of a GPU can induce targeted misclassifications in machine learning models~\cite{lightning}.
Similarly, monitoring the GPU performance counter or memory traces allows for identification of browsing activity, detecting keystroke timing, and inferring neural network structure~\cite{naghibijouybari2018rendered}.
Finally, side channel attacks on GPUs can facilitate a Rowhammer-based sandbox escape on Firefox's Android application~\cite{frigo2018grand}.

Website fingerprinting attacks exploiting GPUs have also been demonstrated.
Here, websites can be identified by memory allocation patterns~\cite{naghibijouybari2018rendered}, contention with other GPU workloads~\cite{wu2022rendering}, power consumption on integrated GPUs~\cite{zhang2021red}, and EM radiation from discrete GPUs~\cite{zhan2022graphics}. 

\subsection{Pixel Stealing and History Sniffing Attacks}
The rendered image of a webpage may contain private information that should be isolated from scripts running on the page.
Examples include embeddings of cross-domain content through the use of \texttt{iframe} elements, and the rendering of hyperlinks, which indicates whether they have been visited.
Over the years, many attacks that expose such private information have been devised.
As several of them exploit a feature called SVG filters, we first describe this feature and then proceed to describe the related attacks.

\parhead{SVG Filters.}
SVG filters specify image transformations, such as blurring or re-coloring, that are applied to the rendered contents of a web page before the page is displayed~\cite{W3C19}.
Basic filters, supplied by the browser, can be parametrized and combined to achieve customized effects.

\parhead{Pixel Stealing.}
Because filters have the unique ability to compute over arbitrary pixels, \citet{barth2011} postulated that they may leak pixel colors. Indeed, practical pixel-stealing attacks surfaced shortly after data-dependent branches were discovered in SVG filters ~\cite{kotcher2013, pixel_perfect}.
In an attempt to remedy this, browsers eliminated data-dependent branches in their SVG filter implementations~\cite{chrome_mitigation_1, firefox_mitigation_1, safari_mitigation_1}.
However, \citet{FPU_leaky} showed that branchless filter code is still vulnerable to microarchitectural side channels, designing a filter that caused computation on white pixels to take longer than black pixels on Intel CPUs.
Three years later, follow-up work found that all major browsers were still vulnerable to the same side channel~\cite{Kohlbrenner2017OnTE}. This in turn has prompted browser vendors to further harden SVG filter implementations, attempting to eliminate data-dependent microarchitectural behavior~\cite{chrome_mitigation_2, firefox_mitigation_2, safari_mitigation_2}.

\parhead{History Sniffing.}
History sniffing attacks attempt to recover the identity of websites a user has visited, potentially revealing web surfing habits of users and exposing private information.
To mount the attack, the attacker's page typically includes a link to a website.
The attacker then observes the browser's rendering behavior to identify previously visited websites. 

The first published attack exploited caching mechanisms by measuring the time to render the target website or to resolve its domain~\cite{felten2000timing}.
Later attacks exploited properties of rendering the CSS \texttt{visited} selector~\cite{clover39css, mozilla_visited_image, mozilla_visited} to identify whether the link has been visited.
When browsers restricted the CSS properties of the \texttt{visited} selector to always report `not visited' to \js in the webpage, history sniffing attacks shifted to
exploiting SVG filters and similar transformations~\cite{pixel_perfect,kotcher2013, smith2018browser, huang2020adaptive} and deceptive user interaction~\cite{summer, kikuchi2016interactive}.

\subsection{Dynamic Voltage Frequency Scaling}\label{sec:background-dvfs}
Dynamic Voltage Frequency Scaling (DVFS) is a power management technique that aims to
manage the system's energy consumption based on available resources (e.g., power and temperature) and workload demands. More specifically, the system constantly adjusts the SoC's frequency and voltage based on its current workload, while trying to maintain its power limits and thermal budget and frequency limitations. These voltage-frequency pairs are known as Performance States (P-states) or Operating Performance Points (OPPs), and the range of attainable P-states together represents the system's DVFS curve.
While the notations may vary across vendors, in this paper, we denote the highest P-state as the P-state providing maximum performance.

\parhead{Power Management Hardware.}
While older hardware required the operating system to manage P-states directly through dedicated registers, modern CPUs and GPUs typically feature a separate microcontroller or co-processor to regulate the voltage and frequency.
However, the OS can still provide the DVFS policy and limit the available P-states to save energy or to keep the system within thermal limits, or select the desired frequency the CPU or GPU cores should run at.
Furthermore, depending on the hardware, the main operating system can or must provide the power management data, including the DVFS states, when initializing the GPU.

\parhead{Sources of Throttling.} 
We distinguish between two causes of frequency throttling in DVFS mechanisms:
\begin{itemize}[nosep,leftmargin=*]
	\item \parhead{Power-Induced Throttling.} Adjusts the operating frequency to limit power consumption.
	\item \parhead{Thermal-Induced Throttling.} Adjusts the operating frequency to avoid overheating.
\end{itemize}
While the underlying causes for throttling vary, both types reduce the operating frequency. This reduction can be observed through dedicated hardware interfaces, as well as by measuring the time it takes to complete computations.

\section{Threat Model}
In this paper we focus on hardware made by Apple, AMD, Nvidia, Google, and Qualcomm. 

\parhead{Accessing Internal Sensors.}
Our experiments in \cref{sec:arm_cpu,sec:gpu} require reading internal sensors for frequency, power, and temperature.
While frequency readings are available to unprivileged users on all platforms, access to power and temperature readings is vendor-specific.
For example, Apple allows unprivileged access to both, whereas Google makes temperature readings privileged while allowing unprivileged access to power. Finally, we note that our attacks presented in \cref{sec:attacks-intensive,sec:attacks-light} do not require any elevated privileges.

\parhead{Browser Versions.}
For our attacks, we assume that the system has been updated to the latest browser versions at the time of writing: Chrome 108 and Safari 16.2.
We also assume the browsers are in their default configurations, with all side channel countermeasures enabled. 

\section{Frequency Side Channels on Arm CPUs}
\label{sec:arm_cpu}
In this section, we establish the presence of power, frequency and thermal side channels on Arm CPUs manufactured by Apple, Qualcomm and Google which are visible using internal sensors. Firstly, we show that executing different instructions results in distinguishable CPU frequency, temperature and power distributions. We then extend our results to show that these behaviors are also data-dependent.

Next, we investigate the source of our observations showing how some CPUs leak via power and frequency while attempting to satisfy thermal constraints, while others present variable power and thermals while running at a fixed frequency. Finally, we build and test a fine-grained leakage model for data-dependent frequency throttling on the Apple M1. 

\parhead{Device Classifications.}
Throughout the paper, we classify devices into three categories depending on their cooling and power budget as below.
\begin{itemize}[nosep,leftmargin=*]
	\item \parhead{Frequency Constrained.} Does not throttle, but leaks information through variations in power and temperature.
	\item \parhead{Power Constrained.} Power-induced throttling leaks information through frequency and temperature.
	\item \parhead{Thermally Constrained.} Thermal-induced throttling leaks information through frequency and power.
\end{itemize}

\parhead{Experimental Setup.}
\cref{table:cputested} lists our testing devices and the frequency scaling capabilities of their CPUs.
Both MacBooks run macOS Ventura, while the other devices run Android 13.
Our experiments require measuring CPU frequency, power consumption, and CPU core temperature over time.
We can access this data on macOS by querying \texttt{IOReport}, an internal library used by the IOKit framework.
We modify \texttt{SocPowerBuddy}~\cite{SocPowerBuddy} to periodically query these values  and write  to a file along with timestamps.

For both Android phones, we obtain the same measurements from  the power supply, thermal zone, and CPU frequency modules of \texttt{sysfs}.
As mobile phones have the poorest cooling budget due to lack of space for a fan or heatsink, we place only the Pixel 6 Pro and OnePlus 10 Pro on a cooling pad to prevent excessive thermal throttling from masking instruction- or data-dependent behavior.

\begin{table}[htb]
	\scriptsize
  \setlength{\tabcolsep}{4pt}
	\begin{tabular}{lllrrr}
		\toprule
		\textbf{Device} 		&\textbf{CPU} & \textbf{Architecture}	 & \textbf{\#C} & \textbf{Freq. (MHz)}& \textbf{\#P} \\
		\midrule
		MacBook Air &M1  & Firestorm (P) & 4 	 & 600 - 3204 & 15  \\
		\rowcolor{blue!5}
		MacBook Air	&M2  & Avalanche (P) & 4 	 & 660 - 3504 & 17  \\
		Pixel 6 Pro	&Tensor& Cortex-X1 (P)  & 2	 & 500 - 2802 & 17\\
		\rowcolor{blue!5}
		OnePlus 10 Pro &{\scriptsize Snapdragon 8 Gen 1} &  Cortex-X2 (P)\dagger & 1	 & 806 - 2995 & 21 \\
		\bottomrule
	\end{tabular}
	\vspace{-1em}
	\caption{Test devices and  CPU information. \#C is the number of performance (P) cores, and \#P is the number of P-states. $\dagger$: This CPU uses an Arm Cortex-X2-based Kryo Prime core.}
	\label{table:cputested}
\end{table}

\subsection{Instruction-Dependent Behavior}
\label{sec:cpu_diff_instr}
To investigate whether different instructions exhibit different long-term system behavior,
we follow the methodology of \cite{x86-instructions} which surveyed the most commonly used instructions in application binaries, partitioning them into different buckets.
We then selected one Arm instruction from each data-processing bucket,
testing stores (\texttt{str}), AES instructions (\texttt{aese, aesmc}), rotate right (\texttt{ror}), bitwise and (\texttt{and}), and both integer and floating-point addition (\texttt{add}, \texttt{fadd}) and multiplication (\texttt{mul}, \texttt{fmul}).
We run each instruction in a loop on all available P-cores on each test device.
We start with an idle device at room temperature, and sample the power consumption, frequency, and temperature every 10\,ms.

\begin{figure}[h]
	\centering
	\includegraphics[width=\linewidth]{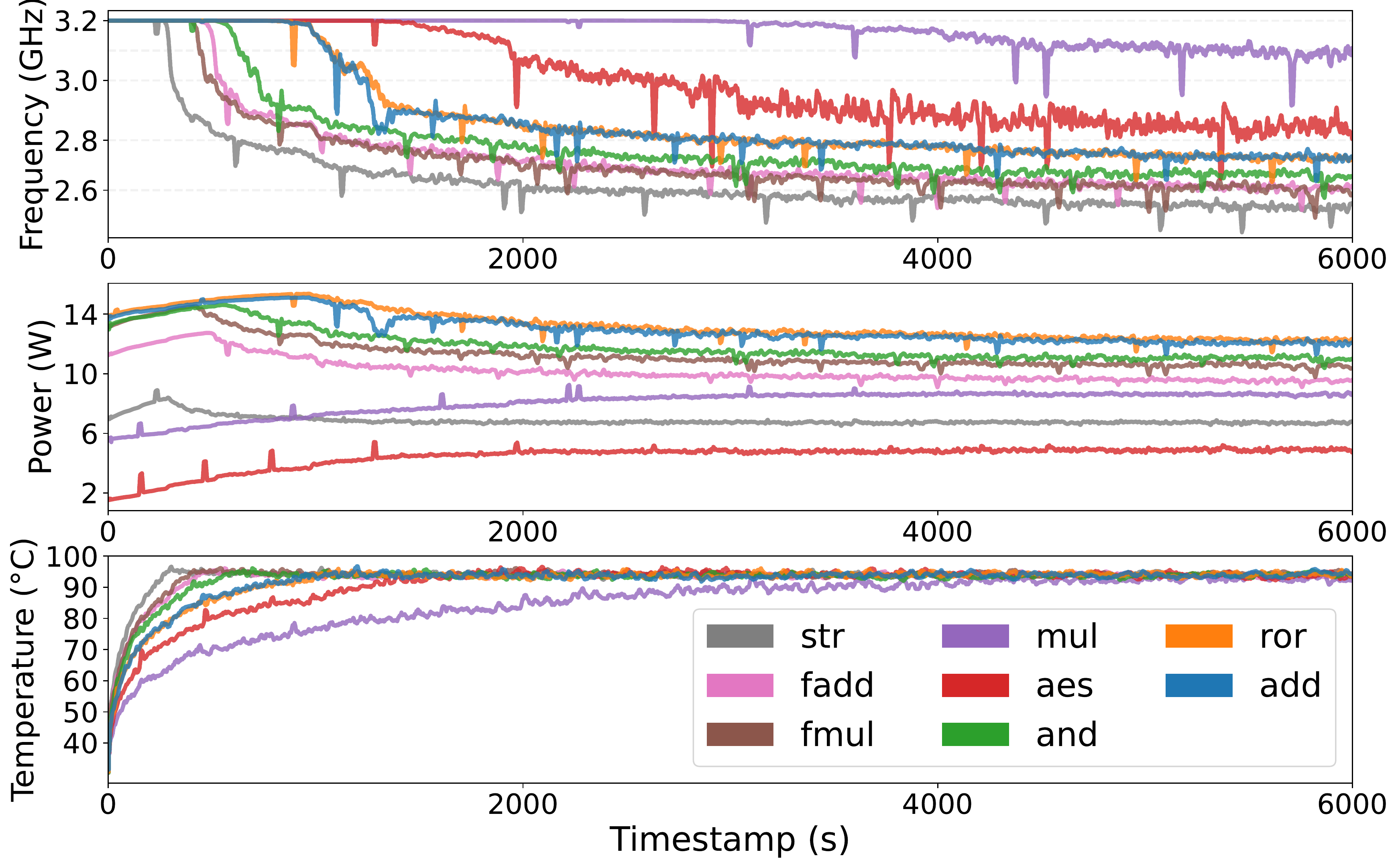}	
	\vspace{-2em}
	\caption{Traces of frequency (top), power (middle), and temperature (bottom) from running our selected workloads on a MacBook Air with Apple M1 CPU. The \texttt{aes} curve represents the \texttt{aese} and \texttt{aesmc} instructions.}
	\label{fig:cpu_diff_instr}	
\end{figure}

\parhead{Distinguishing Instructions on Apple Silicon.}
\cref{fig:cpu_diff_instr} presents the frequency (top), power consumption (middle), and temperature (bottom) while running the workloads for 6000 seconds on an M1 CPU.
We observe the time to throttle (starting from an idle state) varies greatly between instructions, starting at about 300 seconds for stores and going up to 3000 seconds for integer multiplication. Once the M1 reaches steady state, most instructions except the pair of  (\texttt{ror}, \texttt{add}) can be distinguished by their frequency and power.

On the temperature graph, we see the M1 gradually achieves thermal equilibrium, but the time to equilibrium is directly proportional to the time to throttling: that is, the instructions that throttle quickly are the fastest to converge. Furthermore, we observe a counterintuitive result where the instructions that throttle quickly or more severely do not always consume the most power. While the exact cause is uncertain, we hypothesize that this occurs due to the varied power density across different parts of the M1 chip, which subsequently affects heat dissipation capability. Finally, we observe similar results on the Apple M2.

\parhead{Source of Throttling.}
Prior work has reported identical steady-state power consumption but different frequencies for instruction-dependent throttling on x86 CPUs, concluding that the behavior is caused by power limits~\cite{hertzbleed,liu2022frequency}.
In contrast, our experiments show differences in both power and frequency on the Apple M1, 
with the CPU  maintaining a fixed temperature of around 93\,\degree C at steady state.
As the Macbook Air only uses passive cooling, we conjecture that the CPU is thermally constrained, and that it adjusts the power and the frequency to avoid exceeding its temperature limit.
 
\parhead{Confirming Thermally-Constrained Behavior.}
 We now confirm that our M1 MacBook Air indeed exhibits different power and frequency behaviors  between instructions due to thermal limitations. To that aim, we run our workloads on devices that use the same M1 SoC, but have different cooling capacities. 
 More specifically, we rerun the \texttt{add} and \texttt{fadd} workloads for 2000 seconds, on a passively cooled MacBook Air with and without a laptop cooling pad, a MacBook Pro with fans, and a Mac Mini with even higher-capacity fans.
 
 \begin{table}[htb]
 	\scriptsize
 	\begin{tabular}{@{}lrr|rr|rr|rr}
 		& \multicolumn{2}{c|}{MacBook Air} & \multicolumn{2}{c|}{Air+Pad} & \multicolumn{2}{c|}{MacBook Pro} & \multicolumn{2}{c}{Mac Mini} \\
 		& \multicolumn{1}{c}{fadd} & \multicolumn{1}{c|}{add} 	& \multicolumn{1}{c}{fadd} & \multicolumn{1}{c|}{add} & 
 		\multicolumn{1}{c}{fadd} & \multicolumn{1}{c|}{add}   & \multicolumn{1}{c}{fadd} & \multicolumn{1}{c}{add} \\
 		\cline{2-9} 
 		\multicolumn{1}{@{}l|}{Temp. (\degree C)} & 91.7 & {90.6} & 84.3 & {77.8} & 
		70.3 & {66.5} & 46.9 &  44.3 \\
 		\multicolumn{1}{@{}l|}{Freq. (GHz)} & 2.8 & {3.0} & 3.1 & {3.2} & 3.2 & {3.2} & 3.2 &  3.2 \\
 		\multicolumn{1}{@{}l|}{Power (W)} & 10.9 & {14.0} & 12.4 & {14.8} & 12.3 & {14.6} & 11.8 & 13.7 \\                     
 	\end{tabular}
 	\vspace{-1em}
 	\caption{Average temperature,  frequency, and power consumption of the 
 		\texttt{add} and \texttt{fadd} workloads on M1-based devices.}
 	\label{table:M1-cooling}
 \end{table}
 
\parhead{Observing the Effect of Cooling.}
In the first two columns of \cref{table:M1-cooling}, we observe a notable effect of the external cooling pad on the passively cooled MacBook Air.
Without the cooling pad, the CPU operates at around 91\,\degree C for both workloads, presumably due to thermal constraints.
This allows us to distinguish between instructions using power and frequency. With the cooling pad, the MacBook Air throttles much less, operating at a frequency of  3.1--3.2 GHz. Thus, with the cooling pad, the MacBook Air is more frequency constrained than thermal constrained, allowing us to use temperature and power consumption to distinguish instructions.
 
We observe similar effects on our Mac Mini and MacBook Pro devices (\cref{table:M1-cooling}, right columns). Here, both devices maintain the highest P-state (3.2\,GHz) for both workloads,  resulting in no frequency difference between them. However, both devices show instruction-dependent temperature and power consumption. We conjecture that the onboard fans of these devices cool them sufficiently to become frequency-constrained.  
 
\parhead{Temperature and Power Consumption Correlation.}
Inspecting the right three setups of \cref{table:M1-cooling}, we note that the fan-cooled M1 can maintain the highest frequency, albeit with  stark differences in temperature going from the coolest Mac Mini to hottest MacBook Air with cooling pad. We also notice a similar upward trend in power consumption despite the use of identical SoCs  in the three devices. We conjecture that this can be attributed to heat-induced increase in power consumption of CMOS circuits~\cite{hong2010integrated, butts2000static, kim2003leakage}. 

\parhead{The Effects of Instruction Level Parallelism.}
Finally, we note a counterintuitive observation where 
in \cref{fig:cpu_diff_instr} (middle) and \cref{table:M1-cooling}, the \texttt{add} workload consistently consumes more power than the \texttt{fadd}.
We conjecture that, while individual ALUs are less complex than FPUs, the \texttt{add} workload can draw more power due to the presence of more ALU ports~\cite{dougallj} which allows for greater instruction-level parallelism.

\smallskip
 \begin{takeaway}
	\textbf{Takeaway:} Passively cooled CPUs are usually thermally constrained, leaking information via power and frequency. Actively cooled CPUs are usually frequency constrained, leaking information via temperature and power.  
\end{takeaway}

\subsection{Observing Instructions on Arm Cortex}
\begin{figure}[h]
	\centering
	\includegraphics[width = 0.48\textwidth]{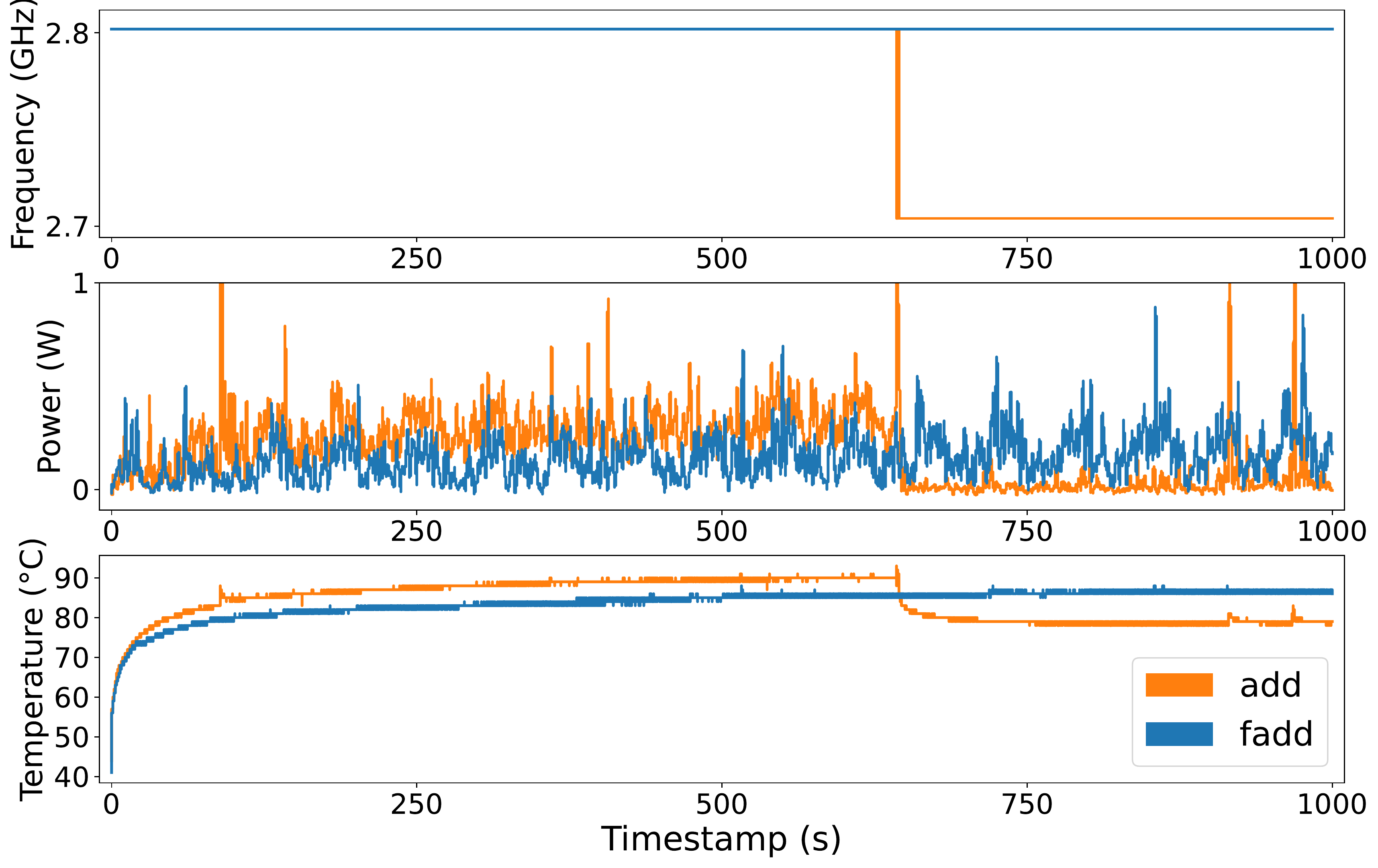}	
	\vspace{-2em}
	\caption{Frequency (top), power (middle), and temperature (bottom) from \texttt{add} and \texttt{fadd} workloads on Pixel 6 Pro.}
	\label{fig:cpu_diff_instr_x1}	
\end{figure}

Moving away from Apple devices, 
\cref{fig:cpu_diff_instr_x1} presents our results on the Cortex-X1 cores of our Google Pixel 6 Pro. On this test target, we measure the \texttt{add} and \texttt{fadd} workloads from our experiments on the Apple M1.

Similarly to Apple CPUs, we observe the \texttt{add} workload consuming more power than the \texttt{fadd} workload, albeit by a much smaller margin.
Both workloads start running at the highest P-state of 2.8\,GHz.
However, after 643 seconds, the \texttt{add} workload starts to throttle, dropping to 2.7\,GHz, whereas the \texttt{fadd} workload does not throttle for the entire duration of this experiment (1000 seconds).
Because the initial difference in power consumption between \texttt{add} and \texttt{fadd} is small, we observe the power consumption of \texttt{add} dropping slightly below \texttt{fadd} once throttling occurs.

We observe similar behavior in the Cortex-X2-based core of the OnePlus 10 Pro's Snapdragon SoC.
Notably, in addition to clearly discernible steady-state power consumption on the Pixel 6 Pro, we find another indicator of throttling due to thermal limits in the temperature graph at \cref{fig:cpu_diff_instr_x1} (bottom), with the CPU throttling aggressively when it reaches 90\degree\,C.

\parhead{Intractability of Throttling on Efficiency-Focused Cores.}
We further experimented with executing the workload on medium-performance cores and on E-cores of the Arm Cortex devices.
We observe that on either type of cores the workloads fail to generate enough heat, even when running at their highest P-states.
That is, the core temperature  rarely exceeded 50\,\degree C on the Pixel 6 Pro and OnePlus 10 Pro.

MacOS does not provide a direct method for pinning processes to cores.
Instead, it provides a quality-of-service mechanism where tasks can be declared as interactive or background.
While setting a task as background will reliably cause it to execute on the E-cores, we empirically observed that this
limits the E-cores' frequency to 1 GHz on the M1. With E-cores having a peak power consumption of about 1.4\,W~\cite{ecore-power}, we did not observe any instruction-dependent frequency, power, or temperature changes on our Mac devices even after one hour of execution. 

To conclude, we conjecture that the cooling and power budgets required for good performance from the P-cores on our devices can sustain prolonged workloads on any of the lower-performance cores without any frequency throttling.

\subsection{Data-Dependent Leakage}
\label{sec:cpu-data-dependent}
So far, we observed that the frequency, temperature, and power consumption traces can be used to distinguish instructions executing on a target CPU. We now demonstrate that these traces also have sufficient fidelity to distinguish between different operands of the same instruction.

\begin{table}[htb]
\captionsetup{type=figure}
\centering
\footnotesize
\setlength{\tabcolsep}{7pt}
\begin{tabular}{l|l}
	\multicolumn{1}{c}{\textbf{Experiment 1}} & \multicolumn{1}{c}{\textbf{Experiment 2}} \\
	\hline
	{\tt uint64\_t val = 0;} & {\tt uint64\_t val = 0;} \\
  {\tt while (1) \{val = val + 0;\}} & {\tt while (1) \{val = val + 1;\}}\\
\end{tabular}
\vspace{-1em}
\caption{Workloads for testing for data-dependent leakage.}
\label{fig:cpu-data}
\end{table}

\parhead{Experimental Setup.}
Following the methodology of \cref{sec:cpu_diff_instr}, we run the 
\texttt{add} instruction with different operands on our M1-based MacBook Air and Pro.
More specifically, we use two workloads (\cref{fig:cpu-data}) which repeatedly add a constant to a variable on all P-cores. 
One of the workloads (Experiment~1) adds the constant 0, whereas the other (Experiment~2) adds 1.
We expect that adding 1 will cause bit flips, and because the power consumed by CMOS circuits and the heat produced correlate with the number of bit flips~\cite{chandrakasan1995minimizing}, we expect to see distinguishable frequency and power consumption. 

\begin{figure}[h]
		\centering
		\includegraphics[width = 0.49\textwidth]{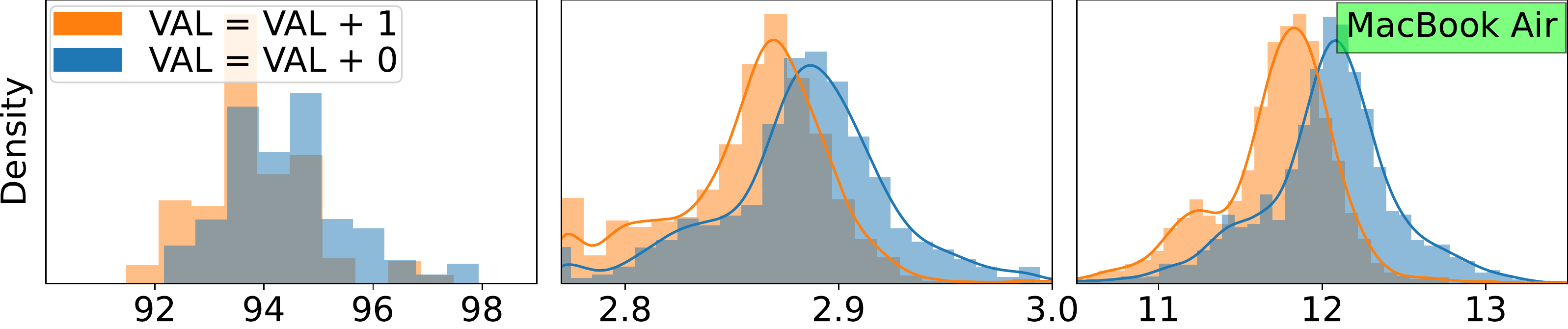}
		\includegraphics[width = 0.49\textwidth]{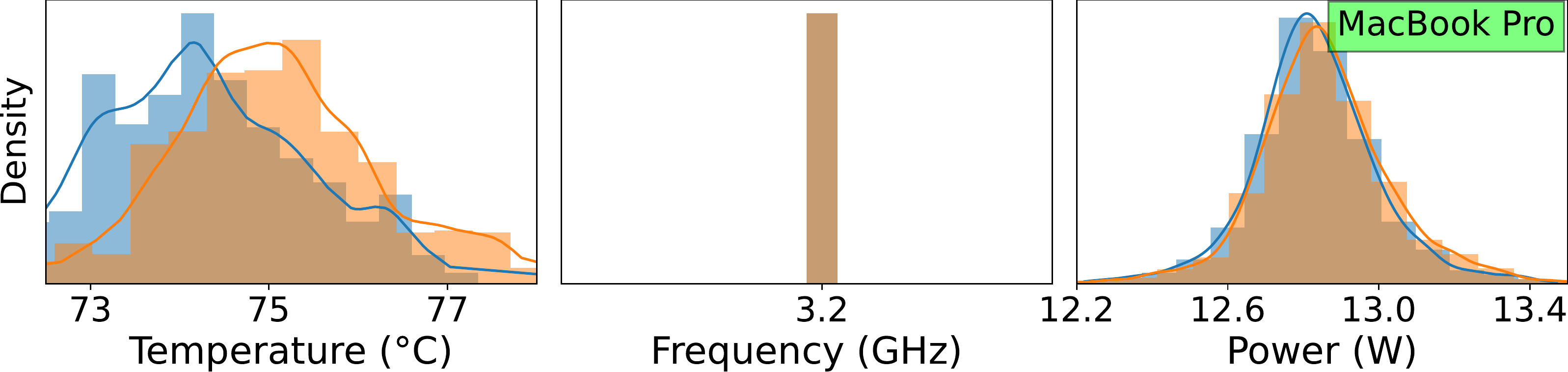}
		\vspace{-1.5em}
		\caption{Histograms of temperature (left), frequency (middle), and power (right) for \texttt{add} with different data running on the M1 MacBook Air (Top) and M1 MacBook Pro (Bottom).}
		\label{fig:cpu_diff_data}	
\end{figure}

\parhead{Results on Apple Silicon.}
\cref{fig:cpu_diff_data} shows a histogram of steady state frequency, power, and temperature on the M1 MacBook Air and MacBook Pro. Since the MacBook Air is a temperature constrained device, we do not observe significant temperature differences between the two experiments, see \cref{fig:cpu_diff_data} (top left). However, at steady state, we observe that the mean 
power consumption and frequency in Experiment~2 are 2.85\,GHz and 11.7\,W, lower than the 2.88\,GHz and 12\,W in Experiment~1. We observe similar results on an M2-based MacBook Air, demonstrating data-dependent leakage.

Repeating the experiment on an M1-based MacBook Pro with an internal cooling fan in \cref{fig:cpu_diff_data} (bottom), we observe little differences in frequency. However, at the steady state, the mean temperature and power consumption for Experiment~2 are 75.2\,\degree C and 12.86\,W, higher than the 74.3\,\degree C and 12.83\,W in Experiment~1. This aligns with our conclusions from \cref{sec:cpu_diff_instr}, where actively cooled devices are typically frequency constrained, leaking through power and temperature.

\begin{figure}[h]
	\includegraphics[height = 0.1\textwidth]{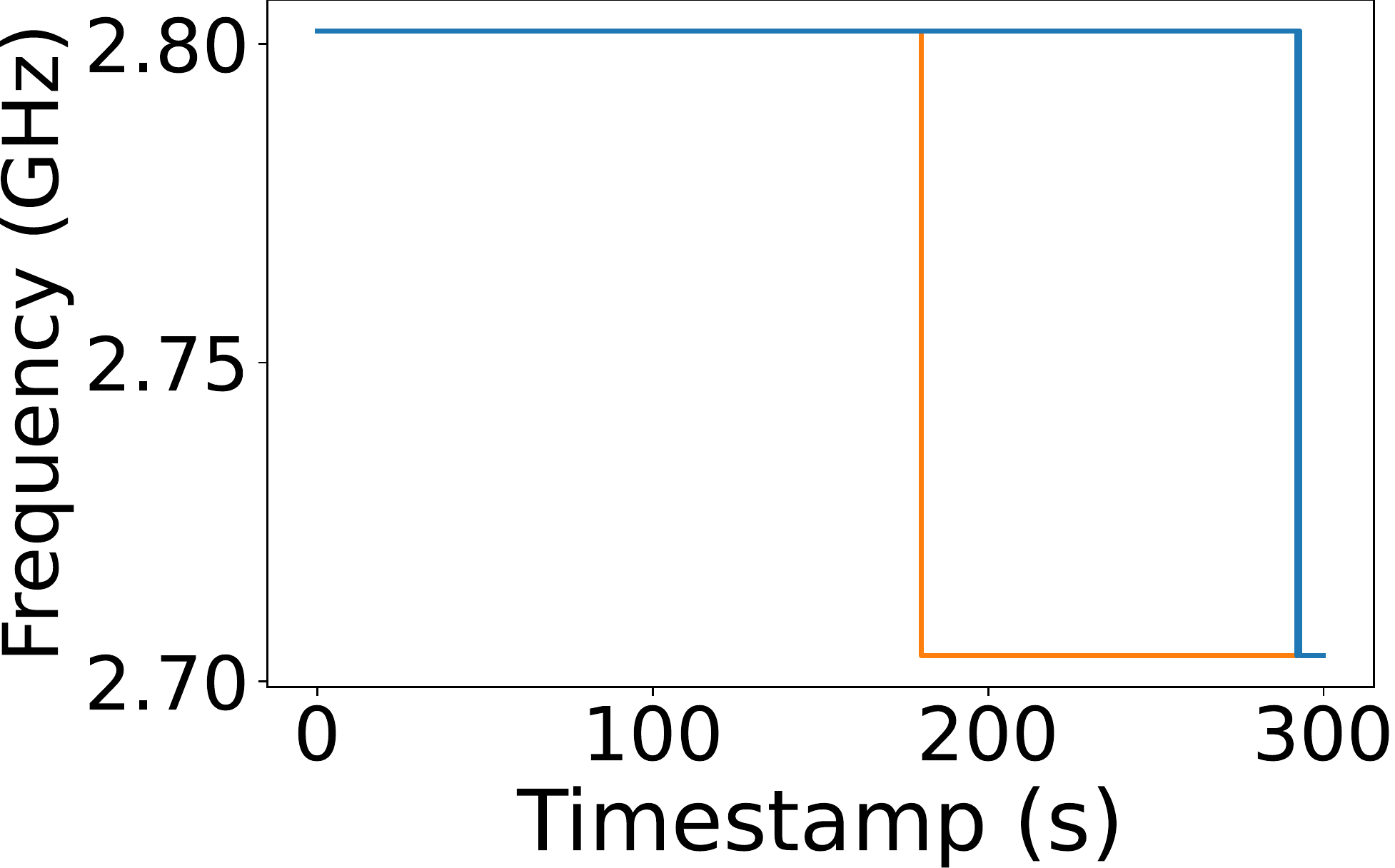}
	\includegraphics[height = 0.1\textwidth]{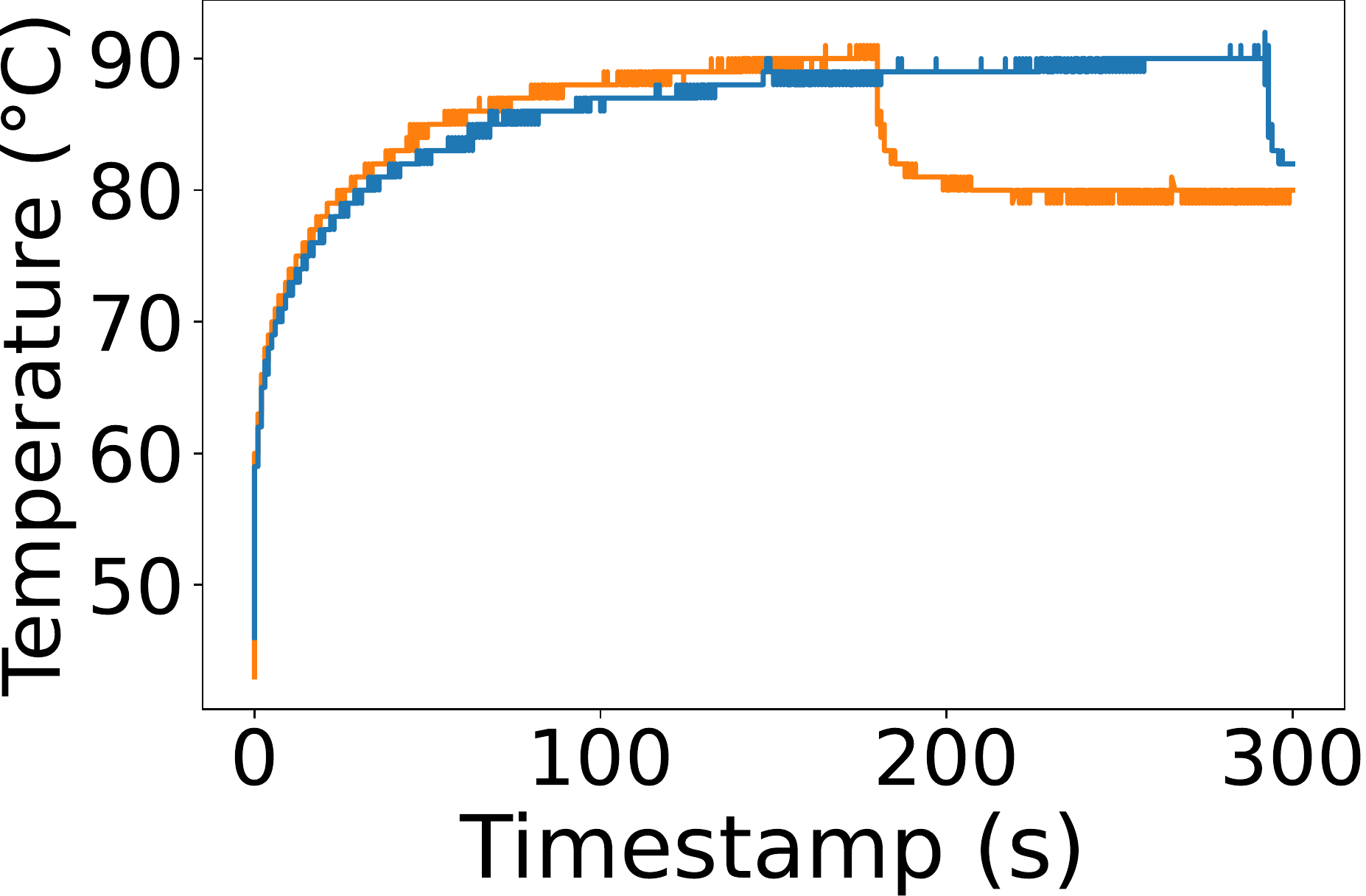}
	\includegraphics[height = 0.1\textwidth]{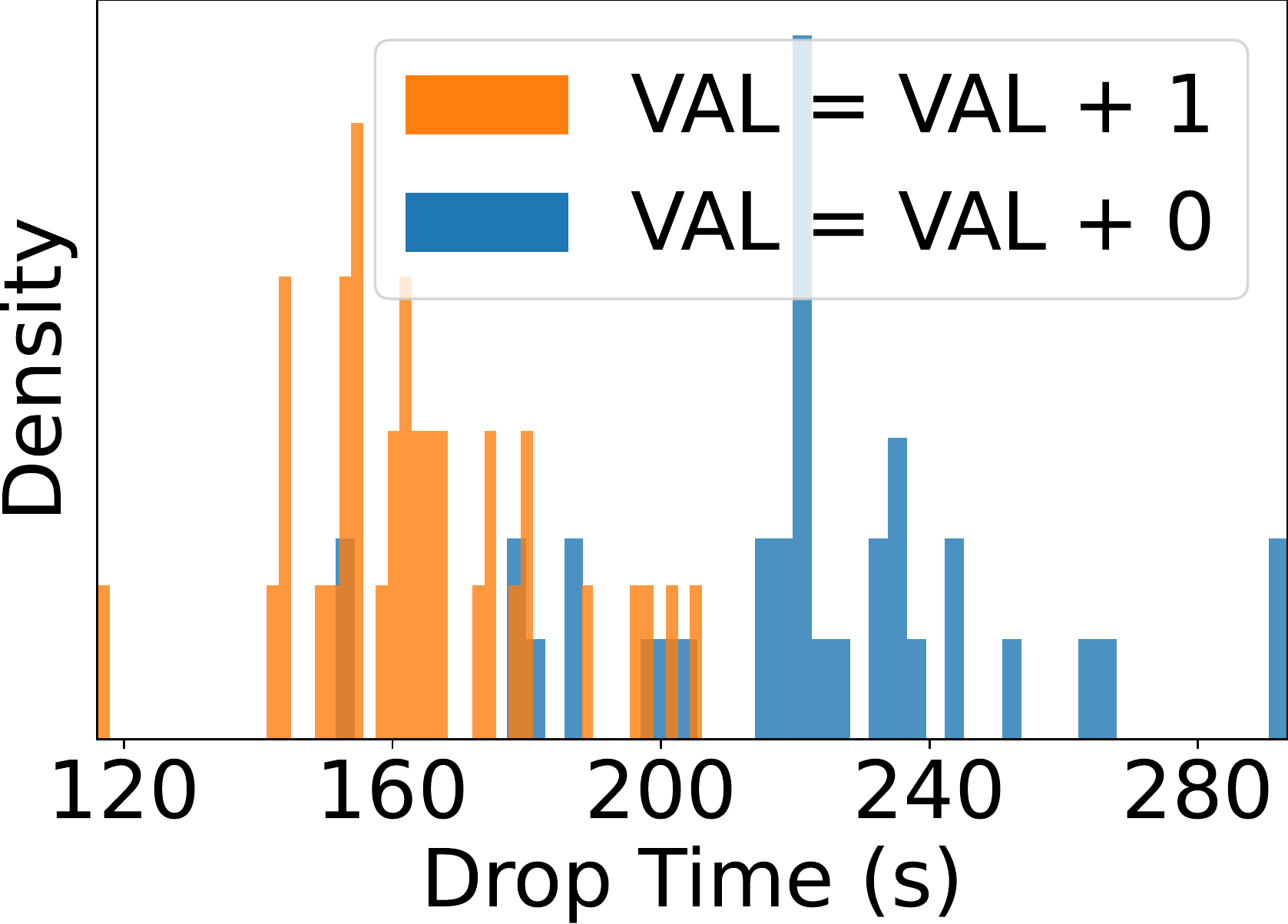}
	\vspace{-0.5em}
	\caption{(Left) Frequency, (Center) temperature, and (Right) distribution of throttling start time on a Pixel 6 Pro.}
	\label{fig:cpu_diff_data_x1}	
\end{figure}

\parhead{Results on Arm Cortex.}
We observe different behavior on the Cortex-X1 of the Pixel 6 Pro, where both experiments throttle to a steady-state frequency of 2.7\,GHz instead of stabilizing at different P-states.
However, we observe notable differences in temperature and time to throttling, which we show in \cref{fig:cpu_diff_data_x1}.
On \cref{fig:cpu_diff_data_x1} (Left), Experiment~2 induces throttling after 180 seconds, while Experiment~1 needs 290 seconds.
We find the cause on \cref{fig:cpu_diff_data_x1} (Center), where Experiment~2 reaches the CPU's thermal limit at 90\,\degree C first.

To confirm the consistency of these results, we measure the time to reach steady state for 70 runs and show a histogram on \cref{fig:cpu_diff_data_x1} (Right).
We observe that throttling appears consistently earlier in Experiment~2 than in Experiment~1, revealing the operand of the add instruction.
Lastly, we observe similar behavior on the Cortex-X2-based CPU of the OnePlus 10 Pro.
\smallskip

\begin{takeaway}
	\textbf{Takeaway:} Data dependent leakage can be observed on Arm CPUs via the distribution of temperature, power, and frequency measurements.
\end{takeaway}

\subsection{Modeling Hamming Distance Leakage}
\label{subsec:hd_cpu_diff_data}
We have established that the frequency, power and temperature measurements of Arm CPUs correlate with the data they process.
In this section, we model the correlation using the Hamming distance (HD) of instruction operands.

\parhead{Modeling the Bit Shifter.}
As a case study, we focus on three shift instructions of the 
Armv8 ISA: \texttt{ror}, \texttt{lsl}, and \texttt
{lsr}.
The first cyclically shifts the value in the input register to the right by the number of bits specified by a second (shift-by).
The latter two perform non-cyclic shifts to the left and the right, respectively. As the M1 CPU contains six ALU ports supporting these instructions~\cite{dougallj}, we always execute workloads that repeatedly perform six identical instructions. This maximizes the signal-to-noise ratio from the frequency, temperature and power consumption traces.  

 \begin{figure}[htb]
 	\centering
	\includegraphics[width=0.8\linewidth]{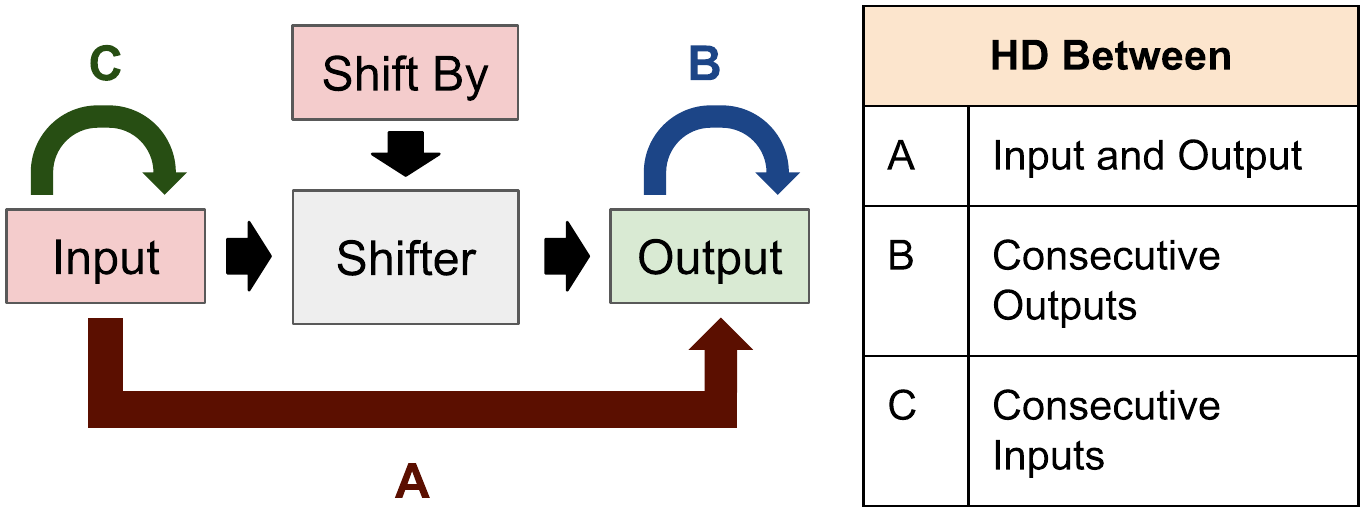}
	\vspace{-0.5em}
 	\caption{Our model of data flow in the bit shifter of an Apple M1 CPU. A, B, and C denote our models of HD.}
 	\label{fig:shifter-model}	
 \end{figure}

We consider three potential components of the HD model, illustrated in \cref{fig:shifter-model}.
The first (A) measures the HD between the input and the output of the shift.
The second (B) measures the HD between successive outputs, and the third (C) measures the HD between successive inputs.
In the evaluation we combine the models, first showing that Component~A shows no correlation between the HD and the measurements.
We then combine it with Components~B and~C to show that these two components show a clear correlation.
Finally, we combine all three components, showing that the effect is additive.

\parhead{Testing Component~A.}
To isolate Component~A, we use the workload in \cref{fig:cpu-ror1}, varying \textsc{shift} between~0 and~16. 
We note that all the inputs are identical, and that for a given shift, all outputs are identical.
Hence, Components~B and~C are always 0.
Component~A, however, depends on the shift, and its value is  $4 \cdot \textsc{shift}$.

\begin{listing}[htb]
\begin{minted}{c}
x8 = 0xFFFF0000FFFF0000
x9 = SHIFT
.Lloop0:
    ror     x19, x8, x9
    ror     x20, x8, x9
    ror     x21, x8, x9
    ror     x22, x8, x9
    ror     x23, x8, x9
    ror     x24, x8, x9
b   .Lloop0
\end{minted}
	\vspace{-1.49em}
	\caption{\texttt{ror} workload for isolating the effects of HD between inputs and outputs.}
	\label{fig:cpu-ror1}
\end{listing}

\begin{figure}[htb]
	\centering 
	\includegraphics[width = 0.49\textwidth]{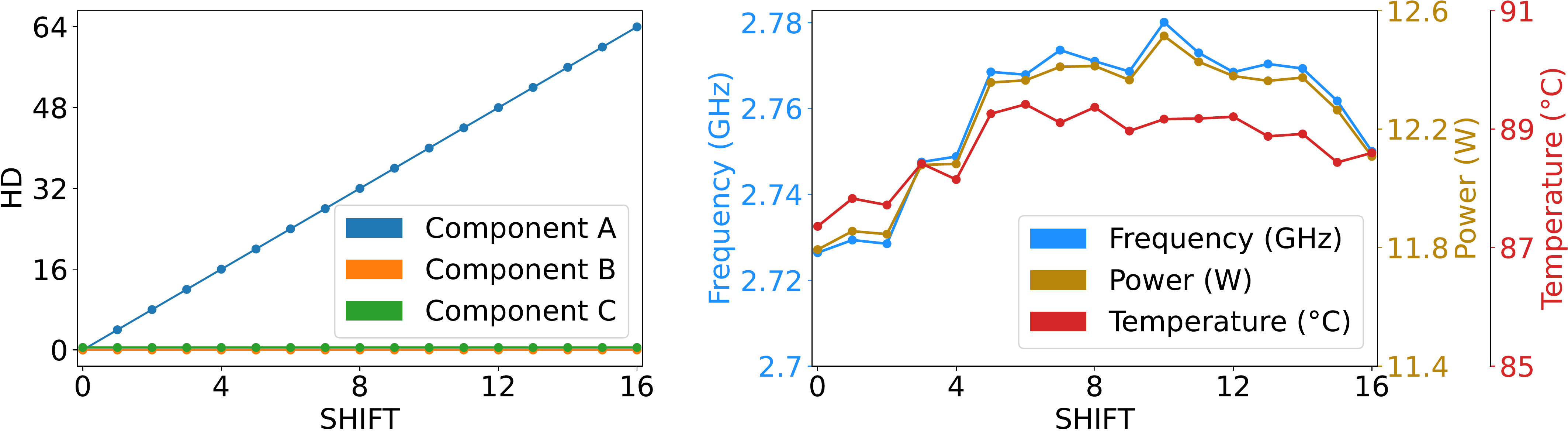}
	\vspace{-1.5em}
	\caption{HD model components vs.\ average measurements for the workload in \cref{fig:cpu-ror1}.}
	\label{fig:m1_ror1}	
\end{figure}

\cref{fig:m1_ror1} contrasts the HD model (left) with the average frequency, power consumption, and temperature as a function of the shift (right).
While there are some variations in the measurements, the correlation between those and Component~A is not clearly evident.

\parhead{Testing Component~B.}
We now try to observe the effect of  Component~B, HD between consecutive outputs. 
We note however, that  keeping identical consecutive inputs, and a fixed HD between inputs and outputs, severely restricts the choices of outputs.
Hence, instead of isolating Component~B, we vary both Components~A and~B and rely on the absence of observed correlation of the measurements with Component~A.

 \begin{listing}[htb]
\begin{minted}{c}
x8  = 0x00000000FFFFFFFF
x11 = 0xFFFFFFFF00000000

x12 = SHIFT
.Lloop0:
    lsl     x9,  x8, x12  // x9  = x8 << x12
    lsl     x10, x8, x12  // x10 = x8 << x12
    lsl     x19, x8, x12  // x19 = x8 << x12
    lsl     x20, x8, x12  // x20 = x8 << x12
    lsl     x21, x8, x12  // x21 = x8 << x12
    lsl     x22, x8, x12  // x22 = x8 << x12
    
    lsr     x23, x11, x12 // x23 = x11 >> x12
    lsr     x24, x11, x12 // x24 = x11 >> x12
    lsr     x25, x11, x12 // x25 = x11 >> x12
    lsr     x26, x11, x12 // x26 = x11 >> x12
    lsr     x27, x11, x12 // x27 = x11 >> x12
    lsr     x28, x11, x12 // x28 = x11 >> x12
b   .Lloop0
\end{minted}
 	\vspace{-0.99em}
 	\caption{Workload for testing the effect of Component~B on the measurements.}
 	\label{fig:cpu-shift1}
 \end{listing}

\begin{figure}[htb]
	\centering
	\includegraphics[width = 0.49\textwidth]{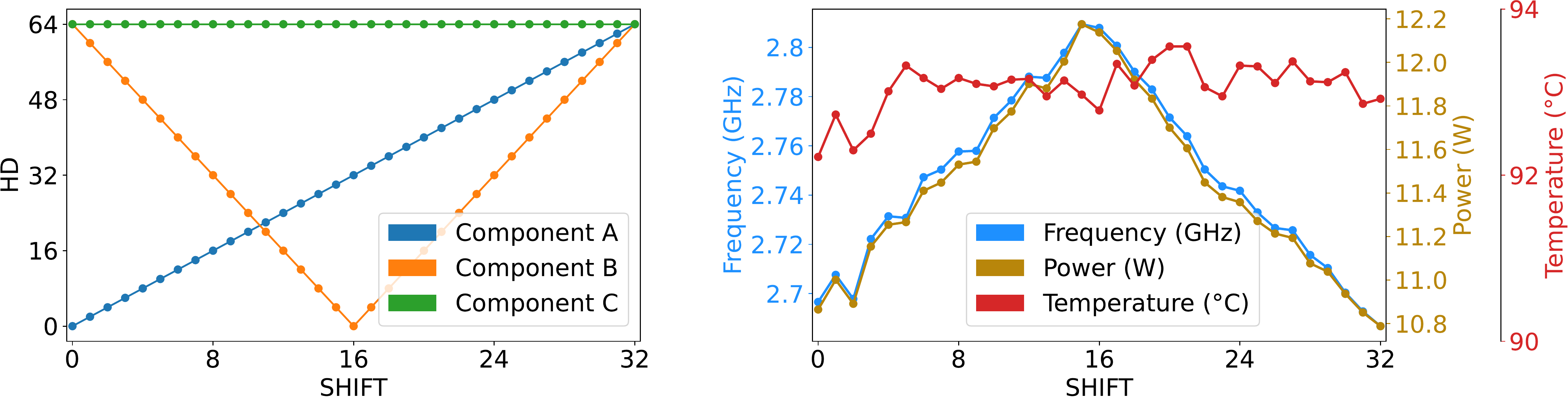}
	\vspace{-1.5em}
	\caption{HD model components vs.\ average measurements for the workload in \cref{fig:cpu-shift1}.}
	\label{fig:m1_shift1}	
\end{figure}

For the evaluation, we use the code in \cref{fig:cpu-shift1}, which uses the \texttt{lsl} and \texttt{lsr} instructions. The \texttt{lsl} instructions shift the value {\footnotesize\texttt{0x00000000FFFFFFFF}} in \texttt{x8}.
Subsequently, the \texttt{lsr} instructions shift the value {\footnotesize\texttt{0xFFFFFFFF00000000}} in \texttt{x11} to the right.
These values are chosen so that  Component~A is $2 \cdot \textsc{shift}$, Component~C is always 64, and  Component~B varies between 0 and 64 depending on the shift, as shown in \cref{fig:m1_shift1} (Left).
We measure the average frequency, power, and temperature for 20 seconds after the M1 CPU reaches steady state.

\cref{fig:m1_shift1} (Right) summarizes our results. We observe that the HD between two consecutive outputs has strong negative correlations with both power (-0.971) and frequency (-0.974).
Next, as the M1-based MacBook Air is passively cooled and thus thermally constrained, we do not observe a correlation between the HD of two consecutive outputs with the CPU's temperature, presumably as the M1 adjusts frequency and power to maintain a fixed thermal budget.

\smallskip
\begin{takeaway}
\textbf{Takeaway:} Higher HD between two consecutive outputs results in lower CPU steady-state frequency and power.
\end{takeaway}

\begin{figure}[htb]
	\centering
	\includegraphics[width = 0.49\textwidth]{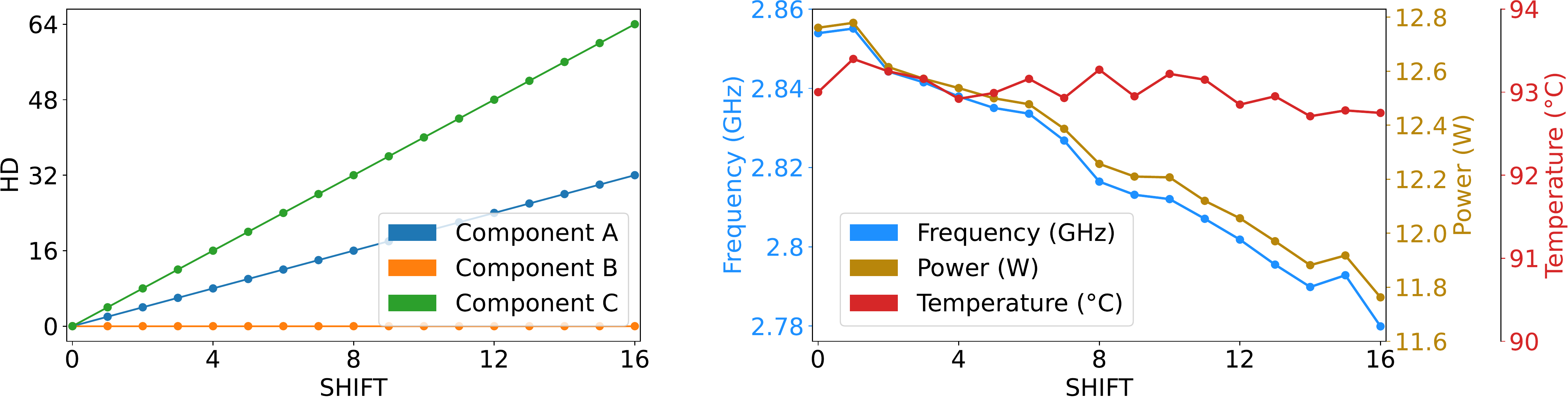}
	\vspace{-1.5em}
	\caption{HD model components vs.\ average measurements for measuring Component~C.}
	\label{fig:m1_shift2}	
\end{figure}

\parhead{Testing Component~C.}
When testing Component~C, we again cannot completely isolate it, so we also vary Component~A.
For this test, we use the code in \cref{fig:cpu-shift1} but modify Lines~1 and~2 to  set 
{\footnotesize\texttt{x8 = 0x0000FFFFFFFF0000 $>>$ {SHIFT}}} and 
{\footnotesize\texttt{x11 = 0x0000FFFFFFFF0000 $<<$ {SHIFT}}}.
This ensures that Component~B, the HD between two consecutive outputs, is constant zero; Component~C is $4 \cdot \textsc{shift}$; and Component~A is $2 \cdot \textsc{shift}$.
These are illustrated in \cref{fig:m1_shift2}.
Finally, our measurement setup is identical to our experiment on Component~B.

We show the results in \cref{fig:m1_shift2} (Right).
Observing \cref{fig:m1_shift2} (Right), we see the HD between two consecutive inputs showing strong negative correlations with both power (-0.969) and frequency (-0.972), similarly to our test of Component~B.
Likewise, the M1 MacBook Air's temperature does not exhibit strong correlation, as the device adjusts its frequency and power in order to maintain its thermal envelope. 

\smallskip
\begin{takeaway}
\textbf{Takeaway:} Higher HD between two consecutive inputs results in lower CPU steady-state frequency and power.
\end{takeaway}

\begin{figure}[htb]
	\centering
	\includegraphics[width = 0.49\textwidth]{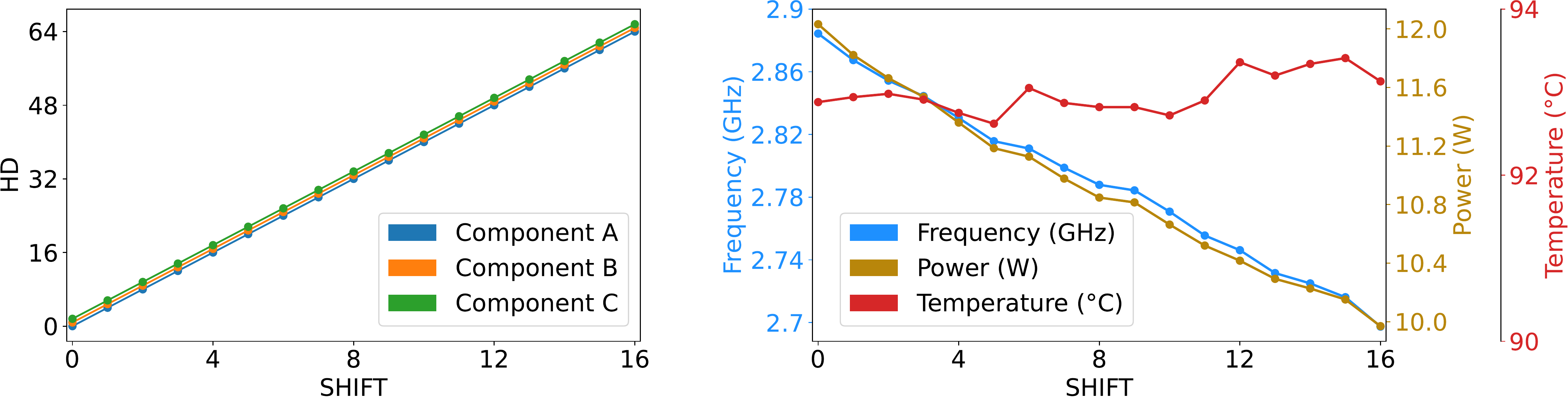}
	\vspace{-1.5em}
	\caption{HD model components vs.\ average measurements for the workload in  \cref{fig:cpu-ror4}.
}
	\label{fig:m1_ror4}
\end{figure}

\parhead{Combining Components.}
We now aim to see how modifying the HD of consecutive inputs and outputs affects the measurements.
For that, we repeat the measurement setup from our tests of Components B and C but use the code in  \cref{fig:cpu-ror4} which applies the \texttt{ror} instruction in-place, allowing us to vary the HD between two consecutive inputs and outputs by controlling the value of \textsc{shift}.
In this workload, Component~A is again $4 \cdot \textsc{shift}$. However, both Component~B and Component~C are $4 \cdot \textsc{shift}$, as shown in \cref{fig:m1_ror4} (Left).

\begin{listing}[htb]
\begin{minted}{c}
x19= x20= x21= x22= x23= x24= 0xFFFF0000FFFF0000
x9 = SHIFT
.Lloop0:
    ror     x19, x19, x9
    ror     x20, x20, x9
    ror     x21, x21, x9
    ror     x22, x22, x9
    ror     x23, x23, x9
    ror     x24, x24, x9
b   .Lloop0
\end{minted}
	\vspace{-0.99em}
	\caption{Workload for evaluating the combined effects of Components~B and~C on the measurements.}
	\label{fig:cpu-ror4}
\end{listing}

\noindent
\cref{fig:m1_ror4} (right) shows our results, where the steady-state frequency and power both have a correlation coefficient of -0.99 to the HD between consecutive inputs and outputs.
We note that in prior experiments, the steady-state CPU frequency ranged 2.7--2.8\,GHz (\cref{fig:m1_shift1}) or 2.78--2.86\,GHz (\cref{fig:m1_shift2}).
However, in this experiment the frequency range of 2.7--2.9\,GHz is twice as big, showing that the effects of Components~B and~C are additive.

\begin{takeaway}
	\textbf{Takeaway:} The effect of HD on frequency throttling between two consecutive inputs and outputs is additive.
\end{takeaway}

\parhead{Modeling Hamming Weight Leakage.} We have also attempted to model the CPU power, frequency and temperature behavior using the Hamming weight (HW) model via the \texttt{and} instruction. 
We did not observe a strong correlation in this model. See \cref{appendix:hw_cpu_diff_data} for more details.

\section{Frequency Side Channels on Integrated and Discrete GPUs}
\label{sec:gpu}
We now investigate the throttling behavior of Apple and Intel integrated GPUs (iGPU), in addition to discrete Nvidia and AMD GPUs.
We design experiments that are analogous to our throttling primitives on CPUs, but instead use GPU kernels running in an infinite loop.
More specifically, kernels are parallelizable routines that are compiled to an intermediate language, and translated by the GPU driver to platform-specific instructions.  
As a result, we demonstrate that GPUs also exhibit instruction-dependent and data-dependent throttling.

\parhead{Experimental Setup.}
We run our experiments on several different GPUs, whose characteristics we summarize in \cref{table:gputested}. 
Akin to the CPU experiments, we require the ability to measure GPU frequency, power consumption, and temperature.
On the Intel iGPU, we collect all three measurements from the \texttt{intel-gpu-tools} utility~\cite{intelgputools}.
For Apple iGPUs, the \texttt{SocPowerBuddy}~\cite{SocPowerBuddy} tool also reports this information, allowing us to follow the same methodology as Apple CPUs.
For discrete GPUs, we use the \texttt{nvidia-smi} tool~\cite{nvidia-smi} for the RTX 3060, and various \texttt{sysfs} files in Linux populated by the \texttt{amdgpu} driver~\cite{amdgpu} for the RX 6600.

All GPUs run with their latest driver versions.
We use macOS Ventura for the Apple iGPUs, and Ubuntu 22.04 LTS for the discrete GPUs and Intel iGPU.

\begin{table}[htb]
	\footnotesize
\begin{tabular}{lrrr}
	\toprule
	\textbf{GPU Model}                       & \textbf{\# EU} 	& \textbf{Freq. (MHz)} & \textbf{\# P-states} \\
	\midrule
	Intel Iris Xe (i7-1280P)					& 96	& 400 - 1450  & -- \\
	\rowcolor{blue!5}
	Apple 4th Gen. (M1)				& 128 	& 396 - 1278  & 6 \\
	Apple 5th Gen.  (M2) 			& 320   & 444 - 1398  & 8 \\
	\rowcolor{blue!5}
	AMD Radeon RX 6600              & 1792  & 500 - 2900  & Not Discrete  \\
	Nvidia GeForce RTX 3060         & 3584  & 405 - 2100  & 227 \\
	\bottomrule                      
\end{tabular}
\vspace{-1em}
	\caption{Summary of the GPUs we tested. The \# EU column denotes the number of execution units. --: we could not determine this information due to it being closed-source.}
	\label{table:gputested}
\end{table}

\subsection{Instruction-Dependent Leakage on iGPUs}
\label{subsec:gpu_diff_instr}
We investigate whether different instructions exhibit different frequency, power or thermal behavior on GPUs.
Accordingly, we implement six GPU kernels in OpenCL~\cite{opencl}, where all kernels operate element-wise on a vector of one million numbers, looping for 20K iterations to amplify the signal.
For integers and floating-point numbers, our kernels perform addition (\texttt{add} and \texttt{fadd}), multiplication (\texttt{mul} and \texttt{fmul}), and division (\texttt{div} and \texttt{fdiv}).
To ensure the OpenCL compiler does not eliminate this loop during optimization, we use the \texttt{-cl-opt-disable}\footnote{While this may add noise from redundant stores, it is negligible for our experiment.} flag to disable them.

\begin{figure}[htb]
	\centering
	\includegraphics[height = 0.122\textwidth]{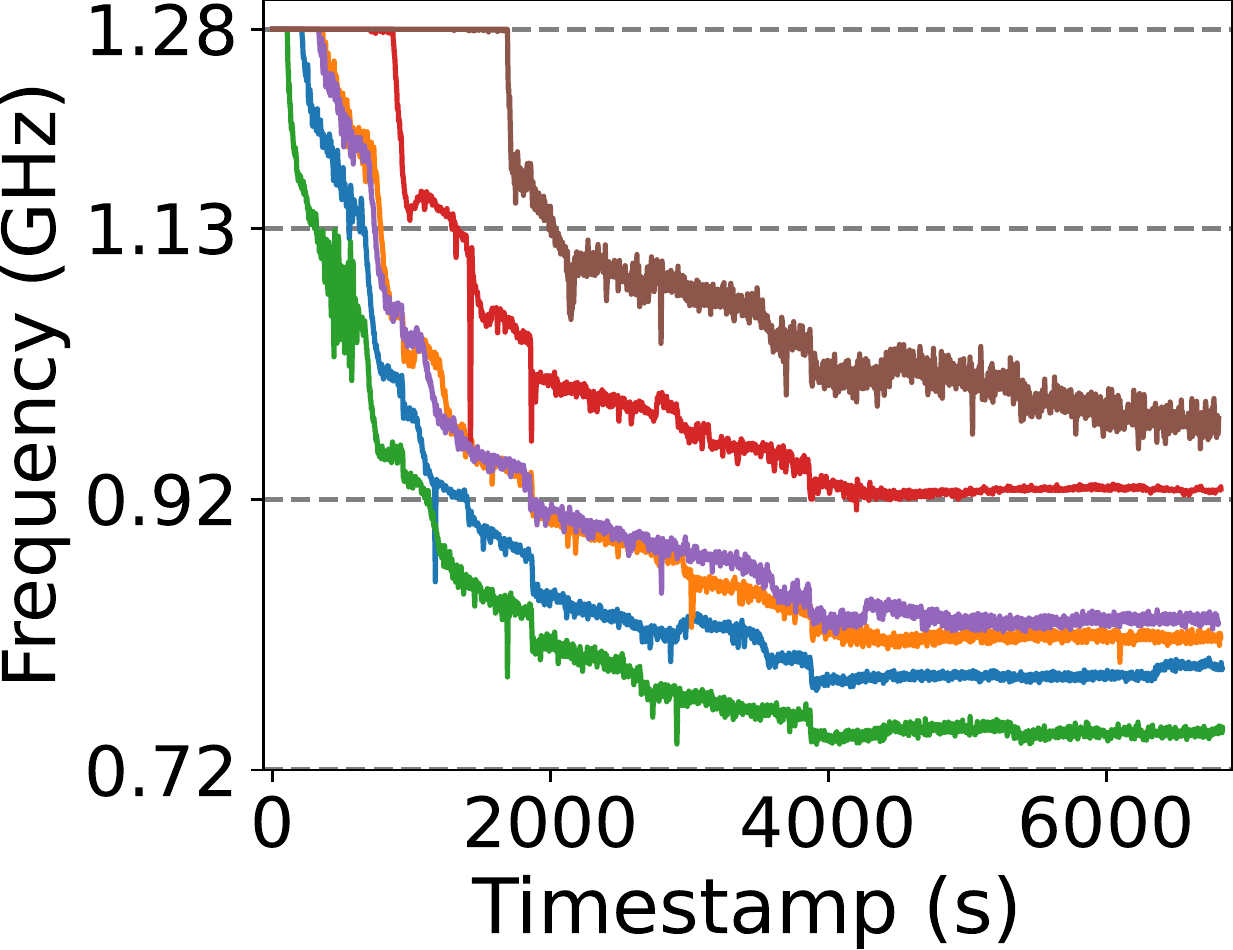}	
	\includegraphics[height = 0.122\textwidth]{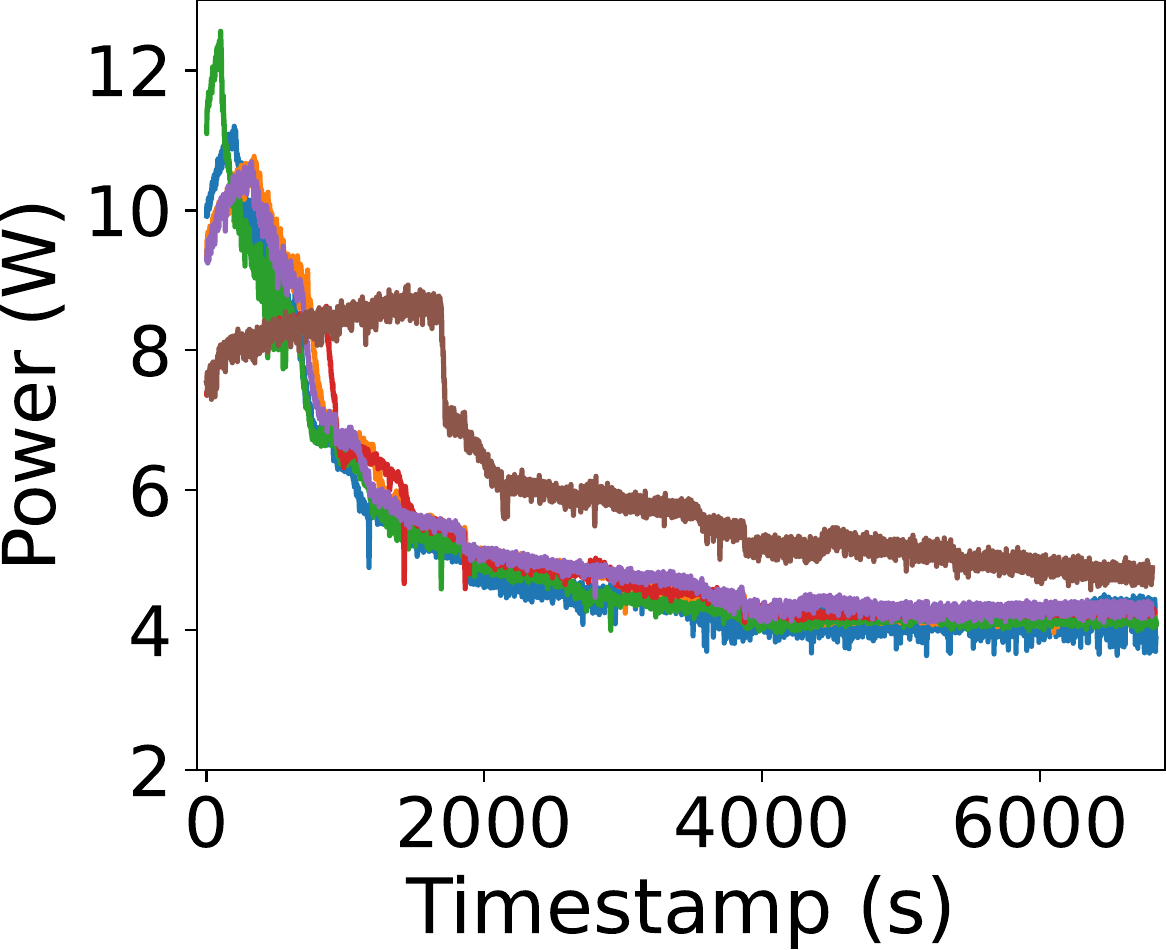}
	\includegraphics[height = 0.122\textwidth]{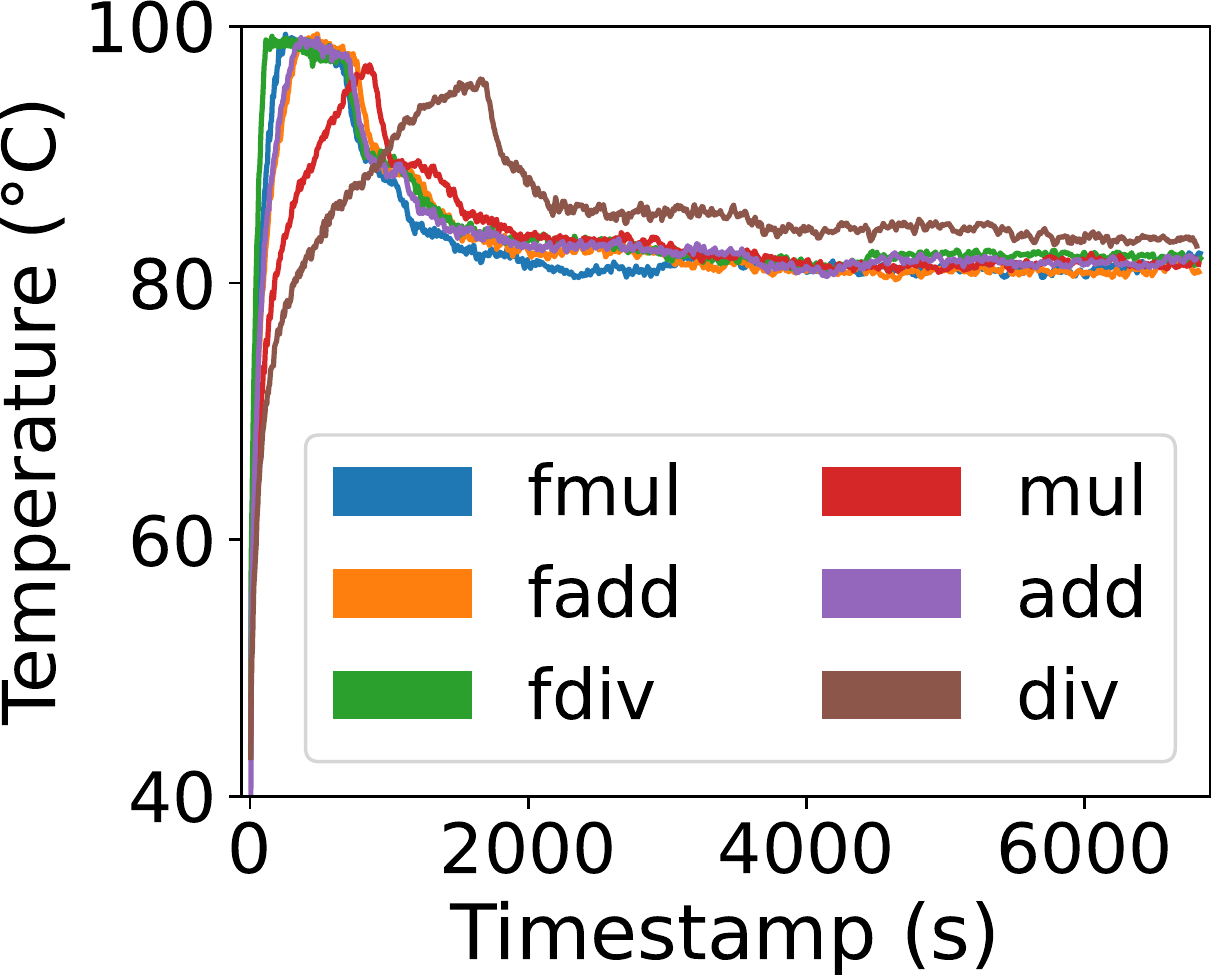}
	\vspace{-0.5em}
	\caption{Traces of frequency (left), power (center), and temperature (right) from the integrated GPU of an Apple M1, measured on the MacBook Air.}
	\label{fig:m1_gpu_diff_instr}	
\end{figure}

\parhead{Distinguishing Instructions on Apple iGPUs.}
\cref{fig:m1_gpu_diff_instr} presents the frequency, power and temperature traces during the execution of our kernels on an Apple M1-based MacBook Air for 7000 seconds.
Before the M1 GPU starts throttling, all the workloads run at the maximum GPU frequency of 1.27 GHz. However, the power and temperature traces of all the workloads, except for \texttt{add} and \texttt{fadd}, are clearly distinguishable. Despite \texttt{div} and \texttt{mul} initially having similar power consumption, the \texttt{mul} workload throttles before \texttt{div}.

After throttling down to a steady state, the frequency values for all the workloads are clearly distinguishable. However, the temperature values for all the workloads at steady state (except for \texttt{div}) are not distinguishable. This indistinguishable nature of steady-state temperature values suggests that the M1 iGPU is thermally constrained. Additionally, the power consumption at steady state can also be used to distinguish some workloads. Specifically, \texttt{add} consumes more power than \texttt{fdiv}, which in turn consumes more power than \texttt{fmul}.

Finally, we observe similar behavior on the iGPU of an M2-based MacBook Air. As both generations of this laptop are thermally constrained, they allow for instruction distinguishability using power and frequency.

\parhead{Ascertaining Thermal Throttling in iGPUs.}
We follow the methodology of \cref{table:M1-cooling} and observe the behavior of the M1 iGPU under different cooling conditions: namely a passively cooled MacBook Air with and without a cooling pad, a MacBook Pro, and a Mac Mini, in increasing order of cooling capacity.  \cref{table:M1-cooling-gpu} summarizes our findings. 

\begin{table}[htb]
	\scriptsize
	\begin{tabular}{lllllllll}
		
		                                        & \multicolumn{2}{c}{ MacBook Air}                     	& \multicolumn{2}{c}{Air+Pad}                   & \multicolumn{2}{c}{MacBook Pro}                            & \multicolumn{2}{c}{Mac Mini}                        \\
		                                        & \multicolumn{1}{c}{fdiv} & \multicolumn{1}{c|}{fmul} 	& \multicolumn{1}{c}{fdiv} & \multicolumn{1}{c|}{fmul}    & \multicolumn{1}{c}{fdiv}     & \multicolumn{1}{c|}{fmul}       & \multicolumn{1}{c}{fdiv} & \multicolumn{1}{c}{fmul} \\ \cline{2-9}
		\multicolumn{1}{l|}{Temp. (\degree C)} 	&     92                & \multicolumn{1}{l|}{92} &   97                  & \multicolumn{1}{l|}{97}   &    89                 & \multicolumn{1}{l|}{86}       &      64               &  59                   \\
		\multicolumn{1}{l|}{Freq. (GHz)} 		&     0.89                 & \multicolumn{1}{l|}{0.96}  &   1.12                   & \multicolumn{1}{l|}{1.20}     &    1.28                     & \multicolumn{1}{l|}{1.28}      &      1.28               &  1.28                   \\
		\multicolumn{1}{l|}{Power (W)} 			&     5.1                  & \multicolumn{1}{l|}{5.4}   &   9.3                   & \multicolumn{1}{l|}{9.6}    &    12.1                      & \multicolumn{1}{l|}{10.7}       &      11.9	           &  10.5                    
	\end{tabular}
	\vspace{-1em}
	\caption{Average GPU temperature, frequency, and power consumption of the \texttt{fdiv} and \texttt{fmul} workloads on M1 iGPU. Air+Pad indicates the MacBook Air with cooling pad.
	}
	\label{table:M1-cooling-gpu}
\end{table}
Akin to our results in \cref{table:M1-cooling}, the Mac Mini and MacBook Pro exhibit sufficient thermal capacity to maintain maximal frequency for both workloads. As these devices are only constrained by their highest frequency, we are able to use temperature and power traces to distinguish \texttt{fdiv} from 
\texttt{fmul}. This is in contrast to both MacBook Air configurations, which are thermally constrained and thus leak via power and frequency.

\begin{figure}[htb]
	\centering
	\includegraphics[height = 0.122\textwidth]{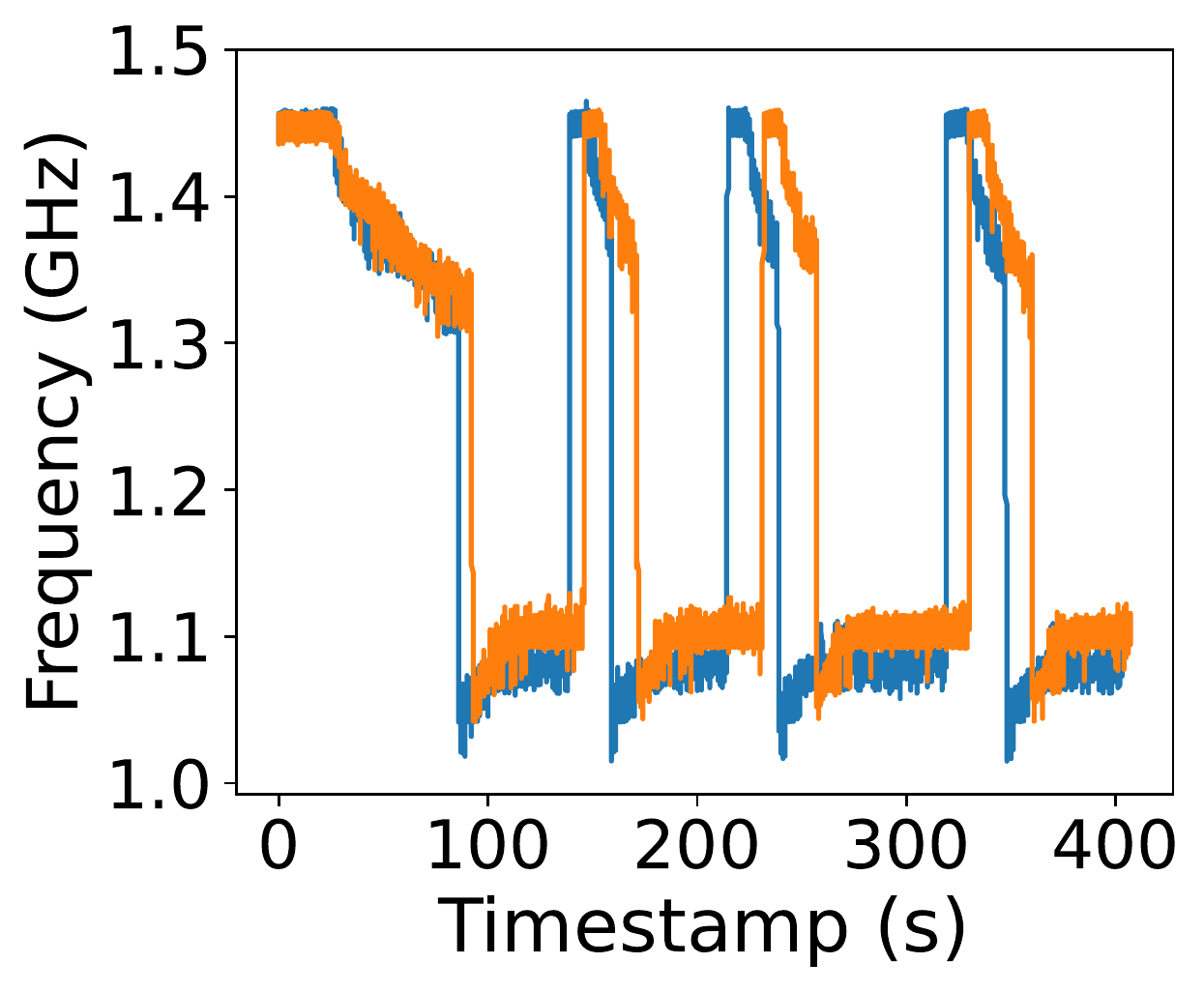}	
	\includegraphics[height = 0.122\textwidth]{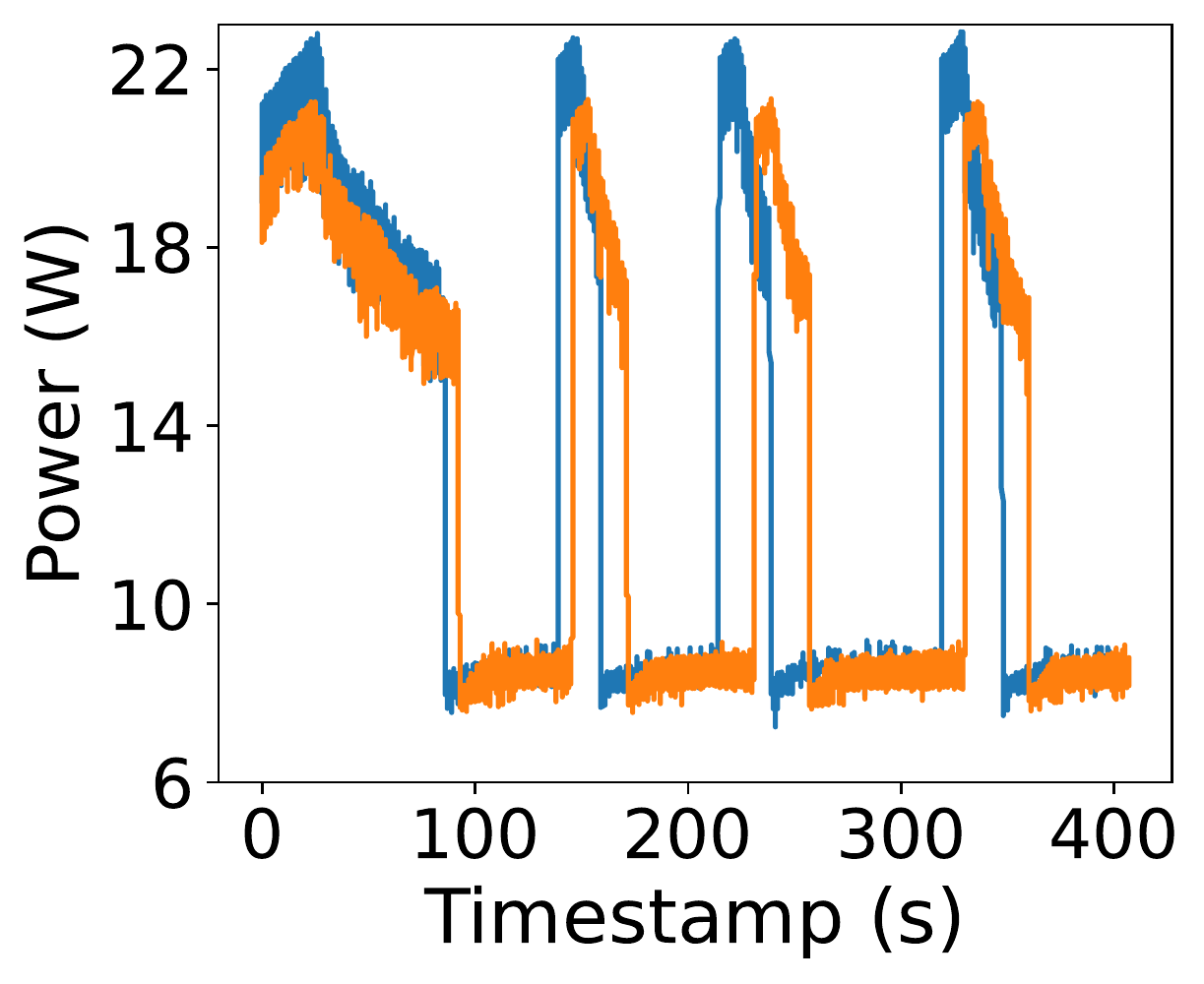}
	\includegraphics[height = 0.122\textwidth]{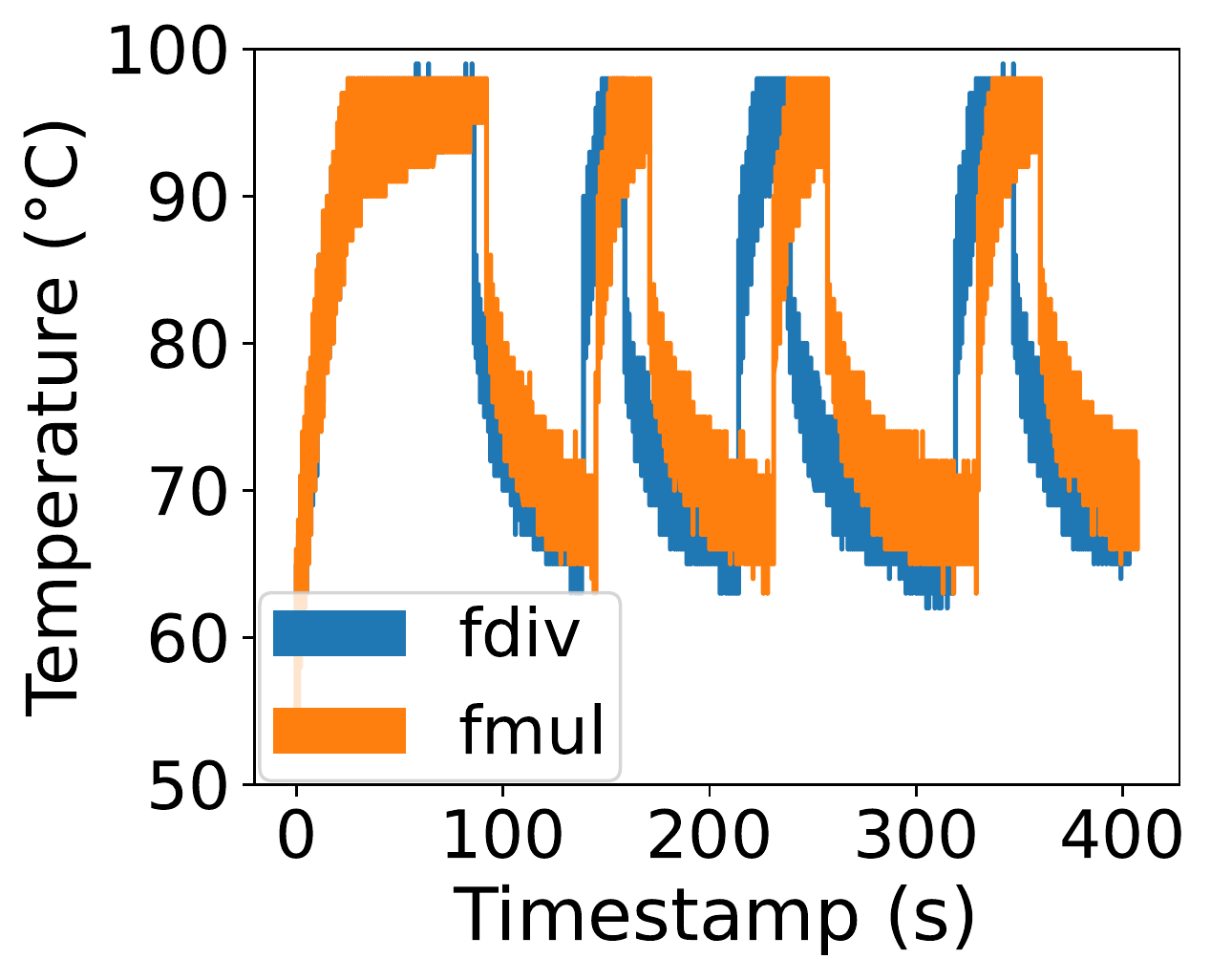}
	\vspace{-0.8em}
	\caption{Traces of frequency (left), power (center), and temperature (right) of \texttt{fdiv} and \texttt{fmul} on an Intel Iris iGPU.}
	\label{fig:intel_gpu_diff_instr}	
\end{figure}

\parhead{Distinguishing Instructions on Intel iGPUs.}
In addition to Apple iGPUs, we also investigated the behavior of an Intel Iris iGPU on our \texttt{fdiv} and \texttt{fmul} kernels in an i7-1280P (Alder Lake) inside an actively cooled Thinkpad X1 Carbon. \cref{fig:intel_gpu_diff_instr} summarizes our findings. As shown, despite active cooling, the  Intel iGPU fails to converge on a steady state frequency, power, and temperature. Instead, the iGPU continuously alternates between two frequencies, presumably in order to meet its sub-100 \degree C thermal budget. Nonetheless, 
\texttt{fdiv} and \texttt{fmul} can still be distinguished in spite of the iGPU's instability,  as the \texttt{fdiv} instruction operates at a slightly lower-end frequency and temperature, while requiring more peak power compared to the \texttt{fmul} instruction. 

\subsection{Distinguishing Instructions on dGPUs}
\begin{figure}[htb]
	\includegraphics[height = 0.119\textwidth]{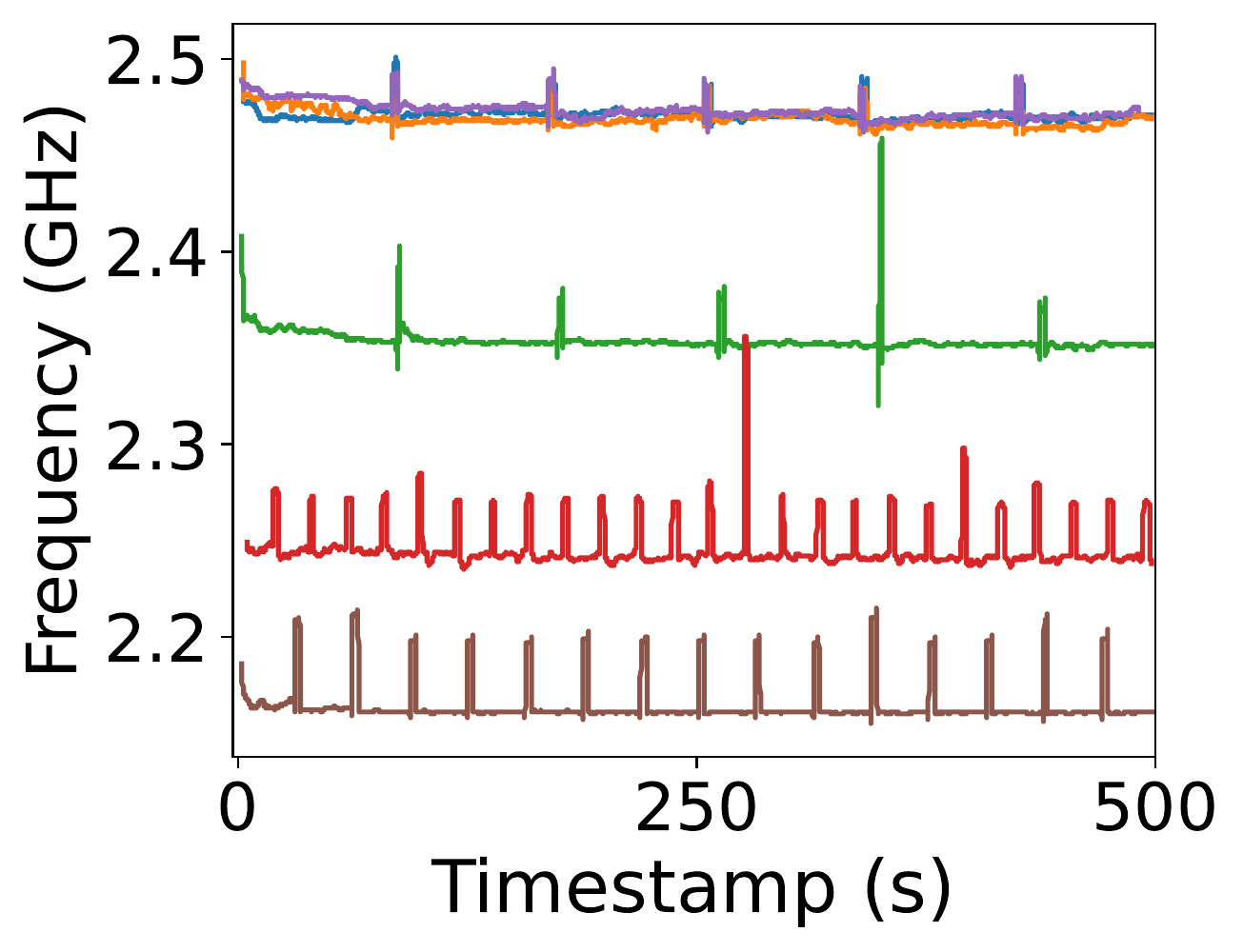}	
	\includegraphics[height = 0.119\textwidth]{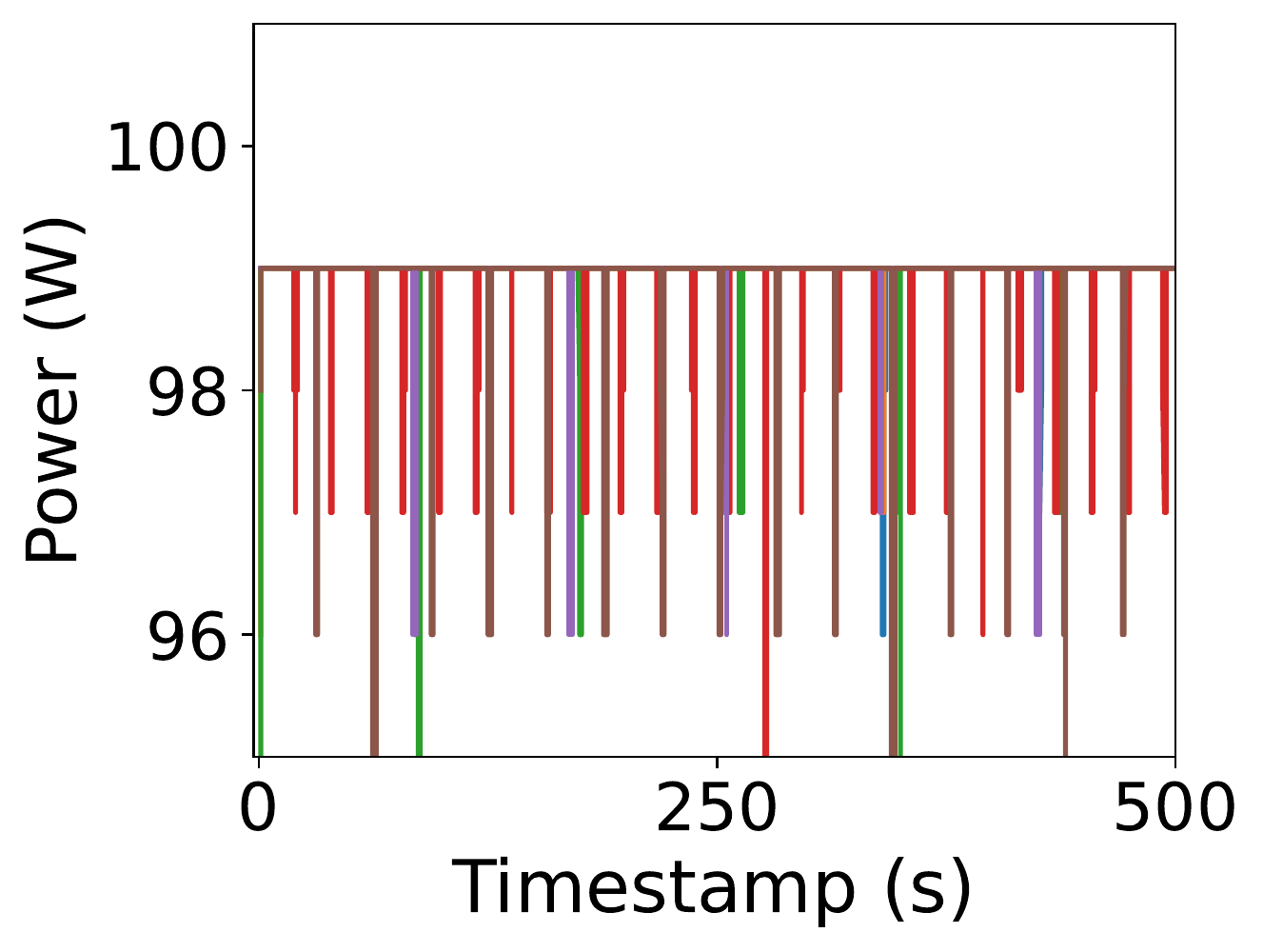}
	\includegraphics[height = 0.119\textwidth]{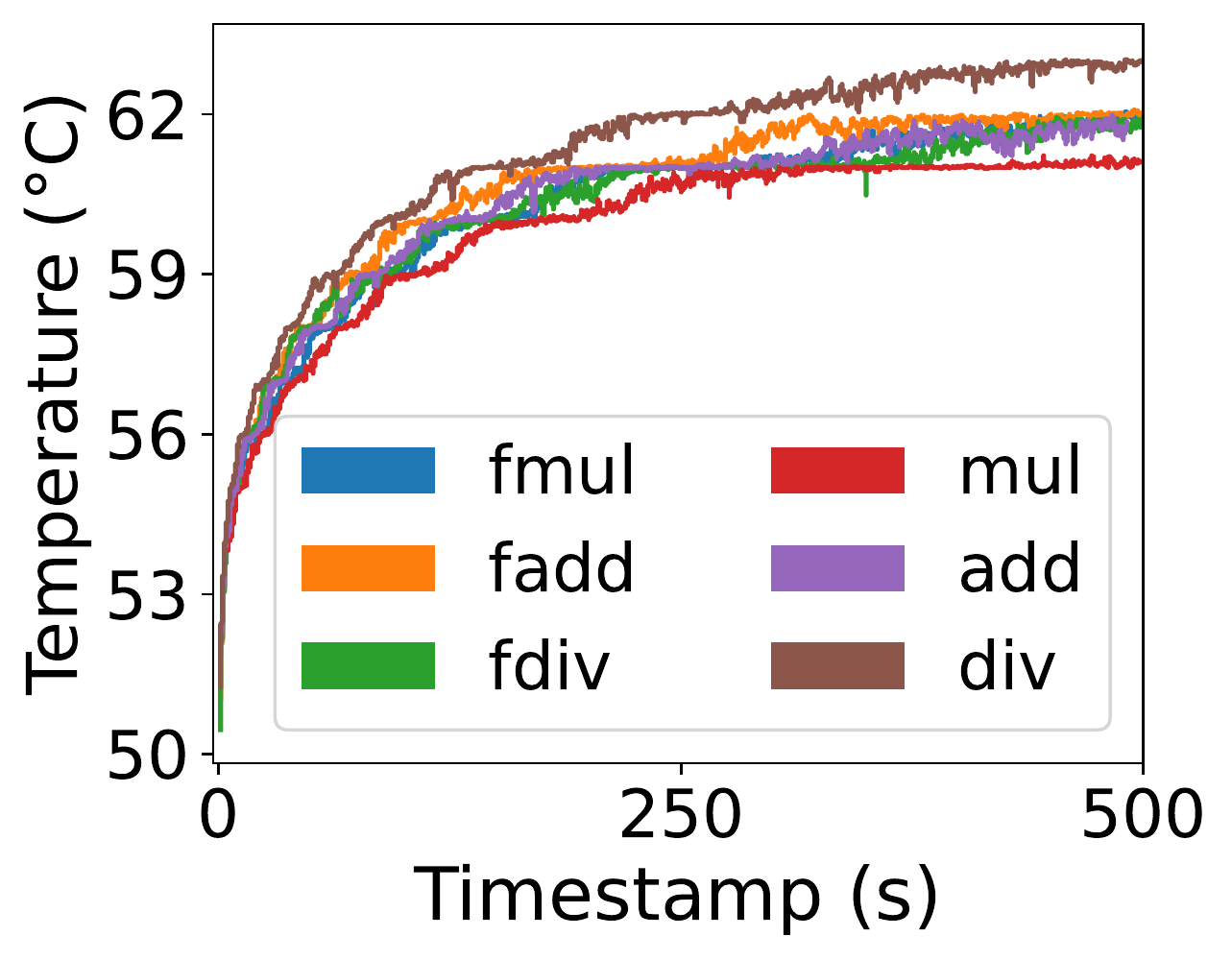}
	\vspace{-1.5em}
	\caption{Traces of frequency (left), power (center), and temperature (right) on an AMD Radeon RX 6600 dGPU.}
	\label{fig:6600_gpu_diff_instr}	
\end{figure}

\begin{figure}[htb]
	\centering
	\includegraphics[height = 0.115\textwidth]{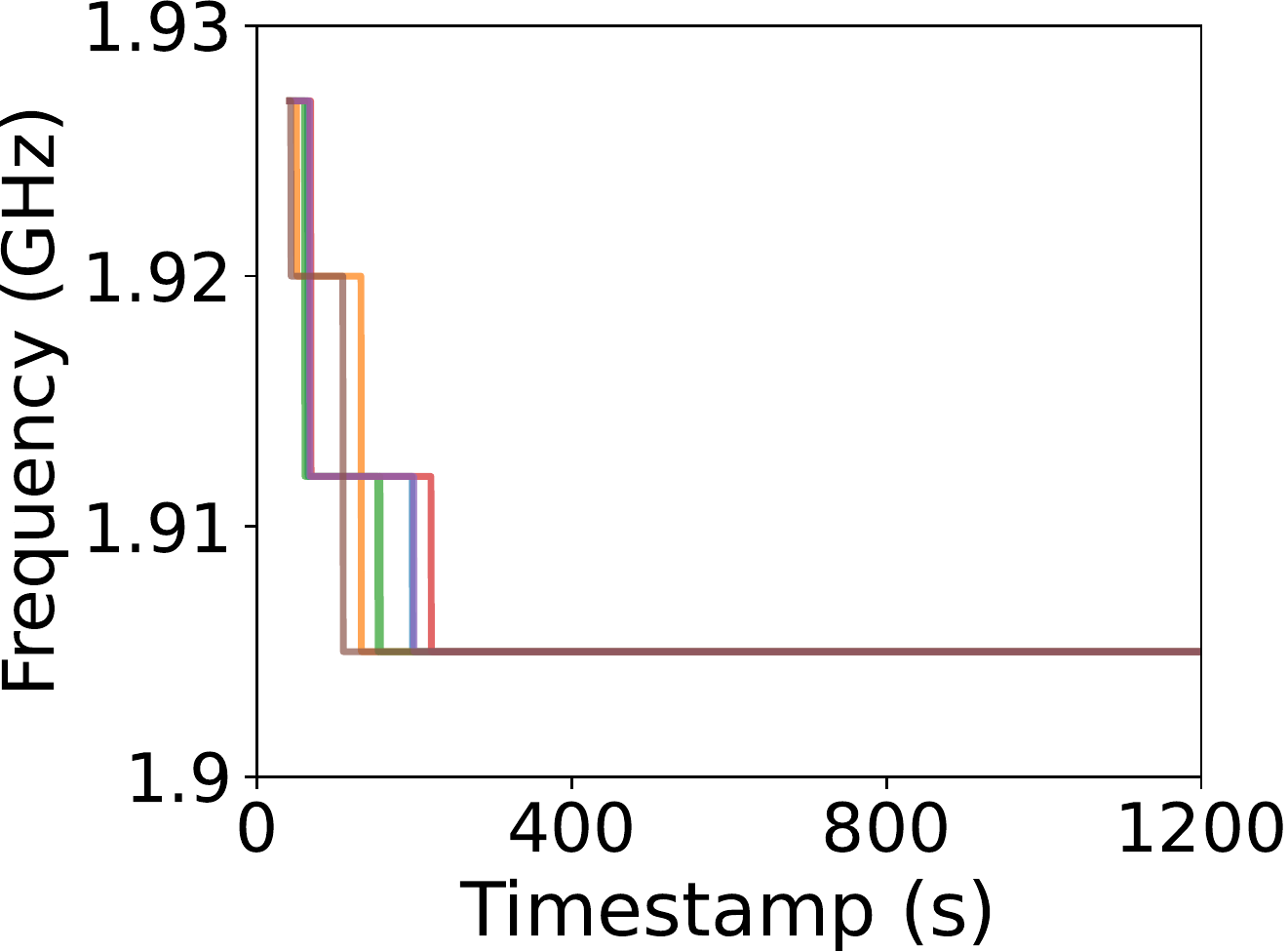}	
	\includegraphics[height = 0.115\textwidth]{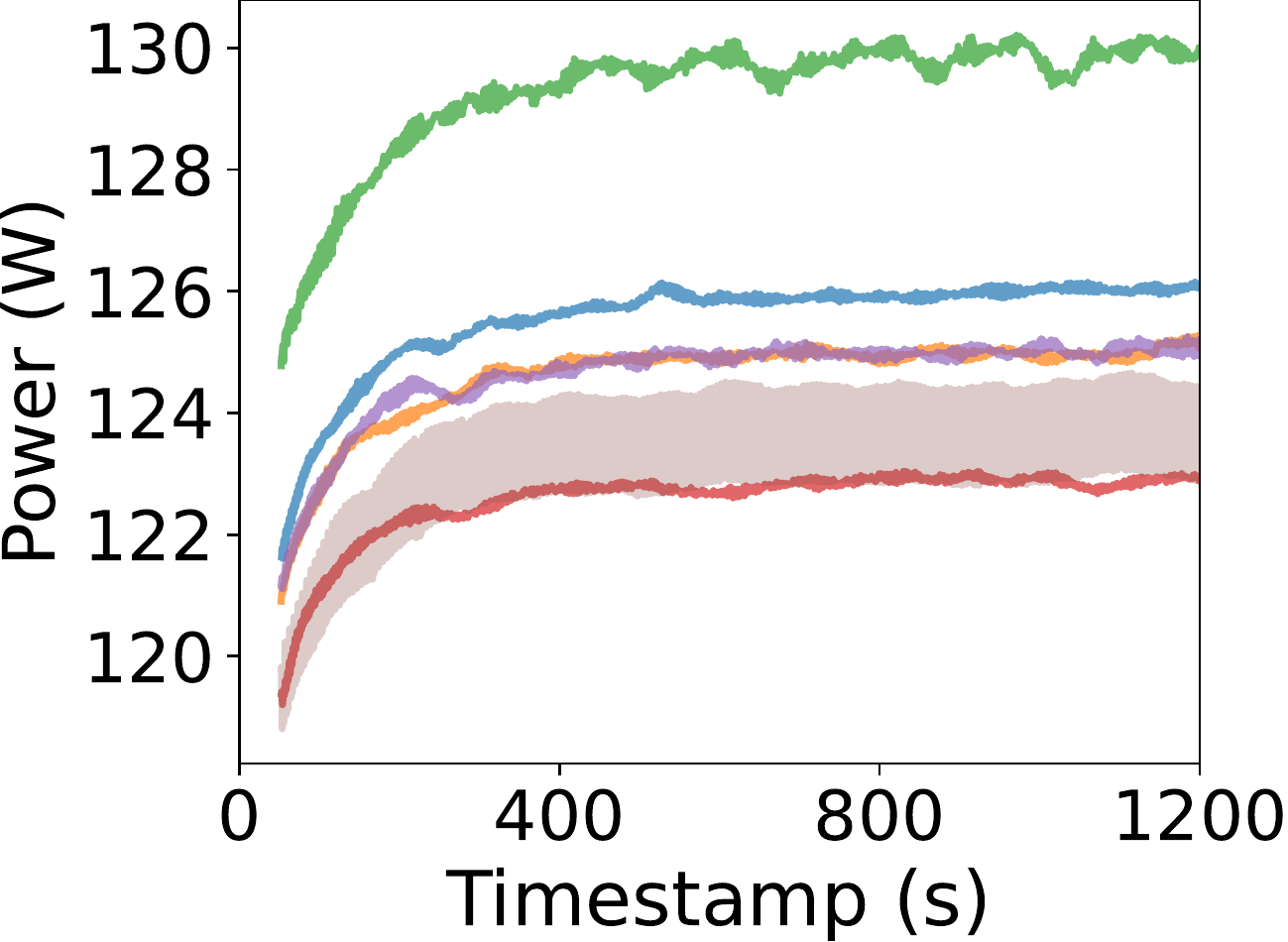}
	\includegraphics[height = 0.115\textwidth]{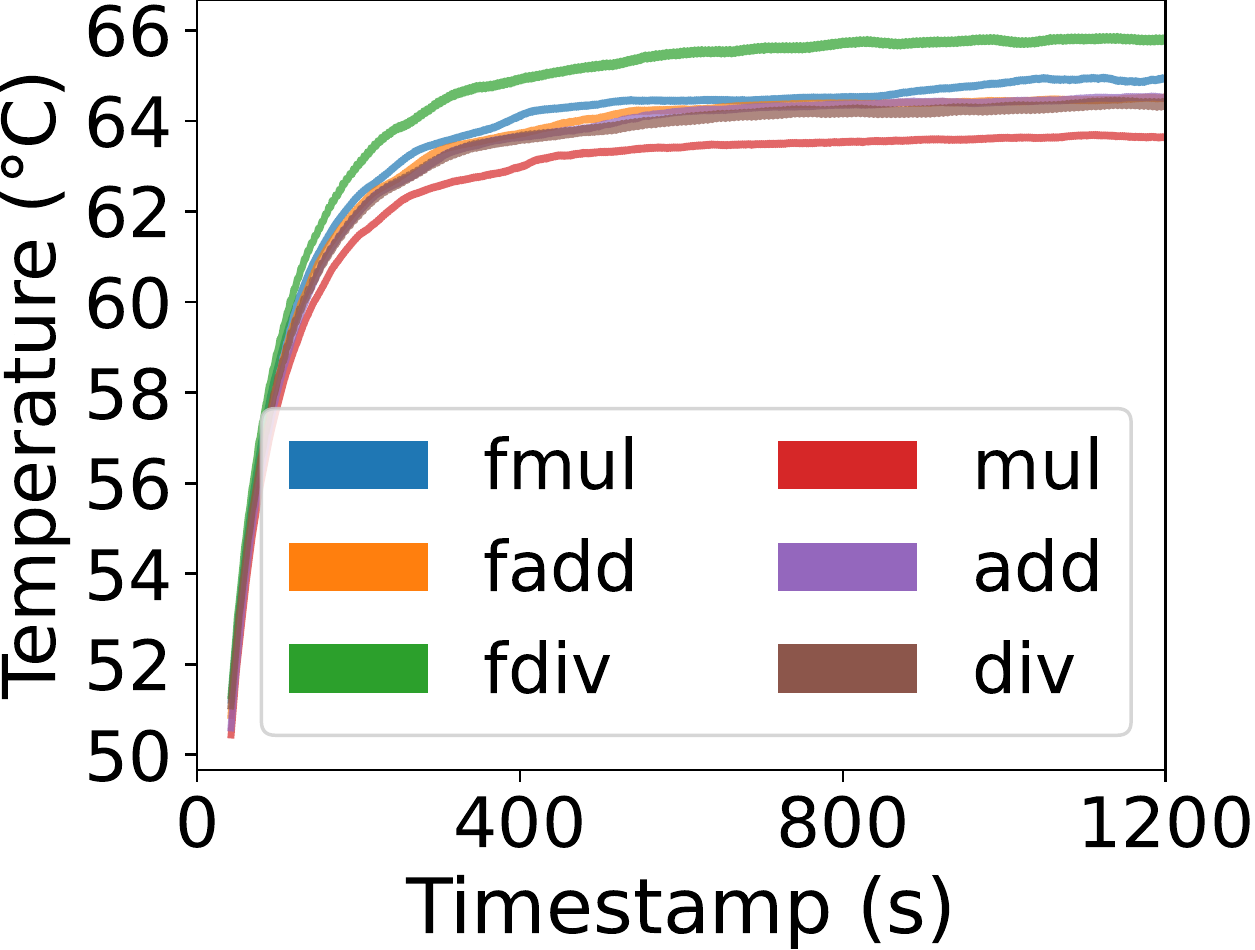}
	\vspace{-1.5em}
	\caption{Traces of frequency (left), power (center), and temperature (right) on an Nvidia GeForce RTX 3060 dGPU.}
	\label{fig:3060_gpu_diff_instr}	
\end{figure}

\cref{fig:6600_gpu_diff_instr,fig:3060_gpu_diff_instr} show the results of our instruction-distinguishing experiment on our Radeon RX 6600 and GeForce RTX 3060 dGPUs with aggressive cooling profiles.

\parhead{Power Constrained: AMD Radeon RX 6600.}
Inspecting the frequency plot of \cref{fig:6600_gpu_diff_instr}, we notice the workloads become clustered into (\texttt{add}, \texttt{fadd}, \texttt{fmul}) at about 2.5 GHz, \texttt{fdiv} at 2.35, \texttt{mul} at 2.25 GHz, and \texttt{div} at 2.16 GHz. Furthermore, we highlight a different throttling behavior compared to Apple iGPUs. Unlike Apple iGPUs that appear to be constrained by their thermal budget, the RX 6600 dGPU is constrained by its 100 W power limit. Finally, we observe we can also use temperature to distinguish between certain workloads, such as \texttt{div}, \texttt{fdiv}, and \texttt{mul} in order of decreasing temperature.

\parhead{Frequency Constrained: Nvidia GeForce RTX 3060.} \cref{fig:3060_gpu_diff_instr} summarizes our experiments on the Nvidia GeForce RTX 3060 dGPU. While our RTX 3060 card has a nominal frequency of 1.32-1.78 GHz, we notice that under full load the dGPU actually overclocks itself to a steady state frequency of 1.905 GHz. The card then exhibits a frequency constraint, adjusting its power and temperature to respect the 1.905 GHz limit.
This, in turn, creates instruction-dependent power and temperature curves. For power, \texttt{fdiv} and \texttt{fmul} are clearly discernible. Moreover, \texttt{mul} and \texttt{div} overlap, but the latter exhibits a greater variation in power consumption. Lastly, \texttt{add} and \texttt{fadd} do not appear to be distinguishable. Remarkably, this clustering of workloads occurs exactly in the same manner on the temperature curve.

\begin{takeaway}
	\textbf{Takeaway:} The frequency, power, and temperature of 
	integrated and discrete GPUs are instruction-dependent.
\end{takeaway}

\subsection{Modeling Hamming Distance Leakage}
\label{subsec:gpu_diff_data_hd}
Having demonstrated that frequency, power and temperature can be used to distinguish instructions on integrated and discrete GPUs, we now show that GPUs exhibit data-dependent leakage in the  Hamming Distance (HD) model. 

\parhead{Constructing a HD-dependent Workload.}
To investigate the effect of HD on the GPU's behavior, we implement a kernel that performs bitwise shift left and shift right operations on each element of a 32-bit unsigned integer vector, as shown in \cref{fig:sh-kernel}.
The kernel starts by shifting the value {\footnotesize\texttt{0x0000FFFF}} to the left by \textsc{shift} bits, followed by shifting the result to the right by \textsc{shift} bits to reinstate the original value. The combined Hamming Distance (HD) between the inputs and outputs of the two instructions is $4 \cdot \textsc{shift}$. We run the kernel on four vectors in a loop for 100K iterations.
Finally, akin to our experimental setup in \cref{subsec:hd_cpu_diff_data}, we measure average frequency, power, and temperature for 20 seconds once the GPU is at steady state, and report correlation coefficients.

\begin{listing}[htb]
\begin{minted}{js}
// left = right = SHIFT
// val0= val1= val2= val3= 0x0000FFFF
uint reps = 100000;
while (reps--) {
  val0 = val0 << left; val0 = val0 >> right;
  val1 = val1 << left; val1 = val1 >> right;
  val2 = val2 << left; val2 = val2 >> right;
  val3 = val3 << left; val3 = val3 >> right;
}
\end{minted}
	\vspace{-0.99em}
	\caption{Our bit-shifting workload to analyze the effect of HD on GPU frequency and power consumption.}
	\label{fig:sh-kernel}
\end{listing}

\begin{figure}[htb]
	\includegraphics[height=0.11\textwidth, width = 0.32\linewidth]{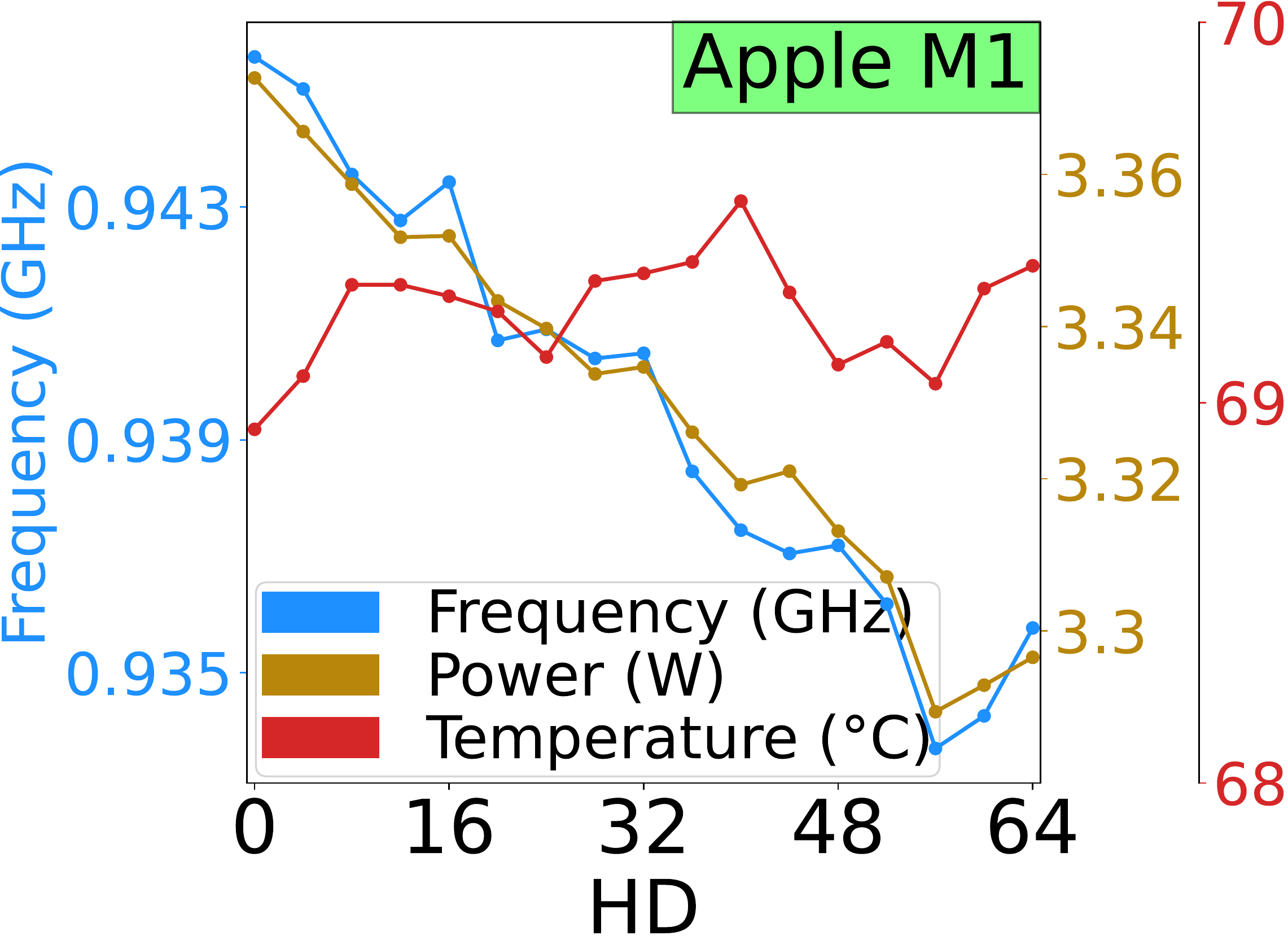}
	\includegraphics[height=0.11\textwidth]{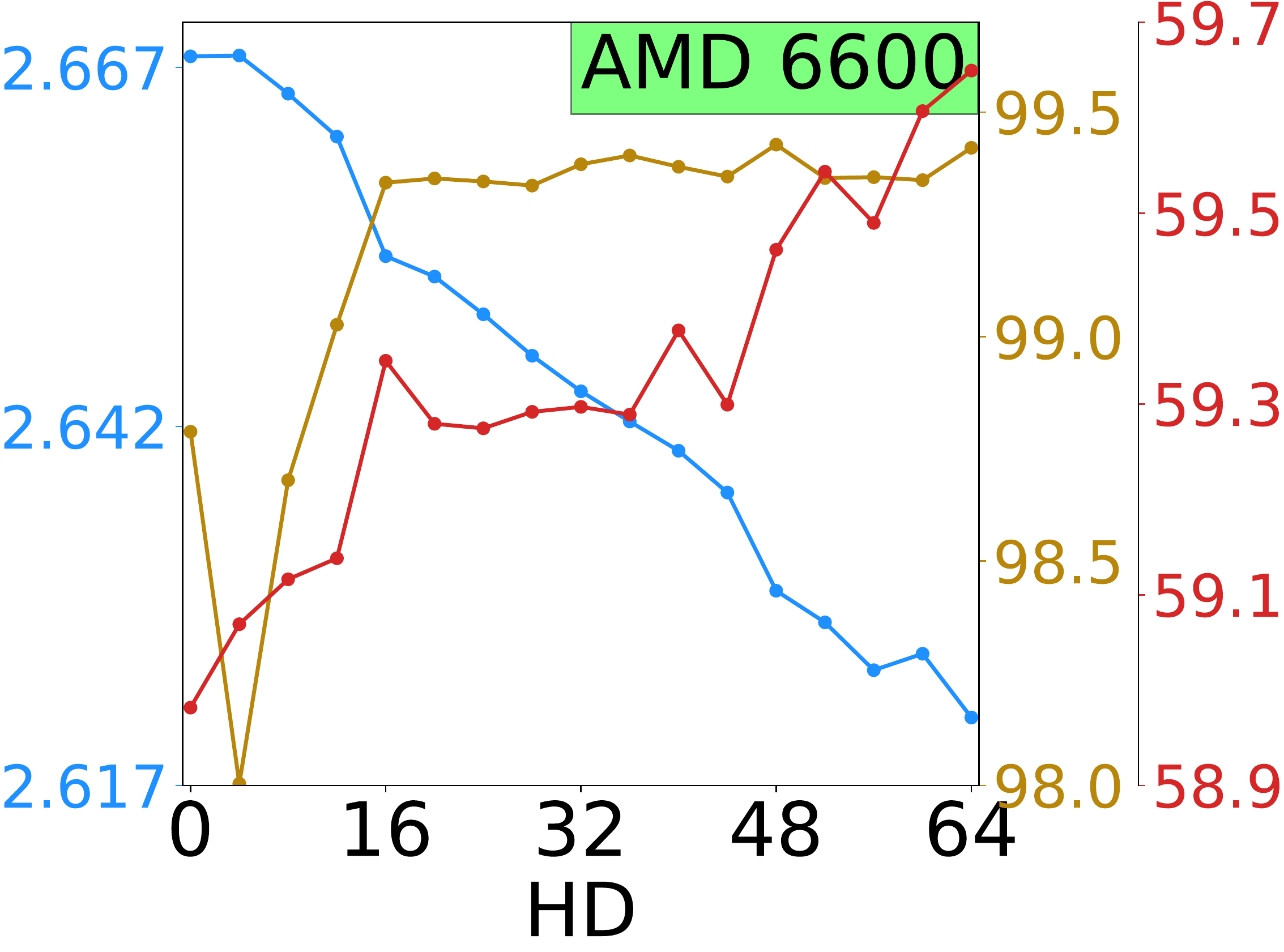}
	\includegraphics[height=0.11\textwidth]{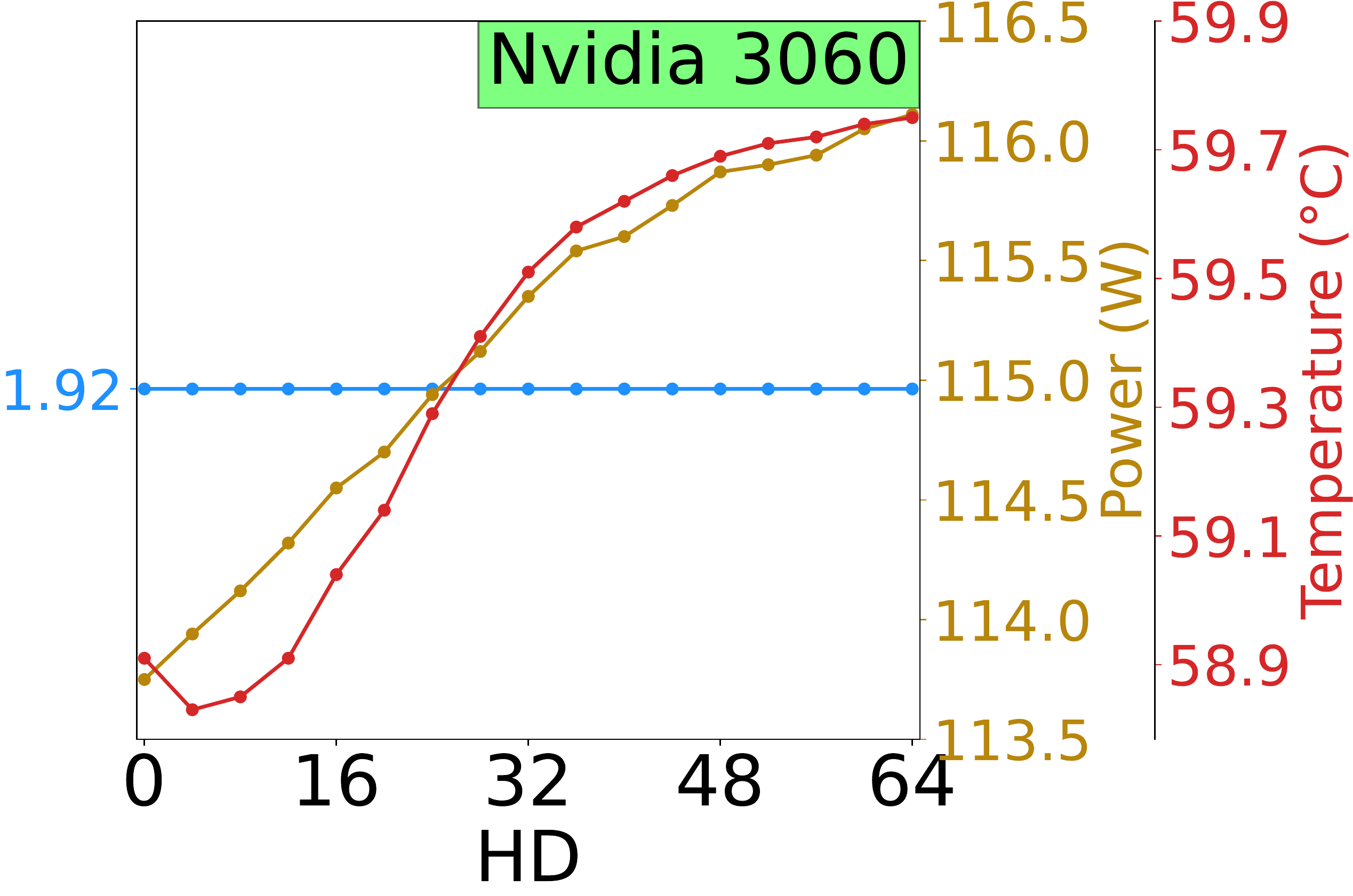}
	\vspace{-1.5em}
	\caption{Traces of frequency, power consumption and temperature on the M1-based MacBook Air (Left), AMD Radeon RX 6600 (Center), and Nvidia GeForce RTX 3060 (Right) resulting from our HD-dependent workload in \cref{fig:sh-kernel}. }
	\label{fig:gpu_hd}	
\end{figure}

\parhead{HD-Dependent Frequency on Apple iGPUs.}
\cref{fig:gpu_hd} (Left) shows our results on the M1 MacBook Air.
Here, we note an inverse correlation for the HD between the inputs and outputs of our shift instructions and the iGPU's power (-0.971) and frequency (-0.951). With prior work~\cite{chandrakasan1995minimizing} showing that heat dissipation in a CMOS circuit scales directly with HD, we conjecture that our thermally constrained MacBook Air is forced to reduce its power and frequency as the HD increases to maintain its thermal budget. Finally, we observe similar behavior on the iGPU of an M2-based MacBook Air. 

\parhead{HD-Dependent Frequency Throttling on Discrete GPUs.}
\cref{fig:gpu_hd} (Center) presents the result of the same experiment repeated on our AMD Radeon RX 6600 dGPU. Here, we observe a direct correlation between the HD and steady-state power (0.663) and temperature (0.798), coupled with an inverse correlation between the HD and frequency (-0.994). Finally, \cref{fig:gpu_hd} (Right) presents our results on an Nvidia GeForce RTX 3060 dGPU. Here, we see that the GPU is frequency constrained, maintaining the same 1.92 GHz frequency across all HDs. However, as typical with frequency constrained devices, we also note a direct correlation between the workload's HD and the device's temperature (0.914) and power consumption (0.964). 

\smallskip
\begin{takeaway}
	\textbf{Takeaway:} The frequency, power, and temperature of discrete and integrated GPUs are dependent on HD.
\end{takeaway}

\subsection{Modeling Data-dependent Throttling via Hamming Weight}
\label{subsec:gpu_diff_data_hw}
Augmenting different data on the same kernel operation being distinguishable by HD, we now demonstrate data-dependent GPU behavior in the Hamming Weight (HW) model. 

\parhead{Constructing a HW-dependent Workload.}
To model the dependence of the GPU's frequency, temperature, and power consumption on the HW of instruction operands, we implement a kernel performing element-wise \texttt{and} operations on a vector, shown in \cref{fig:and-kernel}.
To maximize the signal, we run the \texttt{and} operations in a loop of 10K iterations.
Finally, we increment the vector's elements across 33 different HW from 0 (no bits set) to 32 (all bits set).  
We disable OpenCL's compiler optimizations similarly to \cref{subsec:gpu_diff_instr}, and repeat the experimental setup in \cref{subsec:hd_cpu_diff_data} and \cref{subsec:gpu_diff_data_hd}.

\begin{listing}[htb]
\begin{minted}{js}
//left= right= 0x0, 0x1, 0x3... 0xFFFFFFFF
uint reps = 10000;
while (reps--) {
  value = left & right; value = left & right;
  value = left & right; value = left & right;
  value = left & right; value = left & right;
  value = left & right; value = left & right;
}
\end{minted}
	\vspace{-1.49em}
	\caption{Our bitwise-and workload to analyze the affect of HW on GPU frequency and power consumption. 
	}
	\label{fig:and-kernel}
\end{listing}

\begin{figure}[htb]
	\centering
	\includegraphics[height=0.113\textwidth]{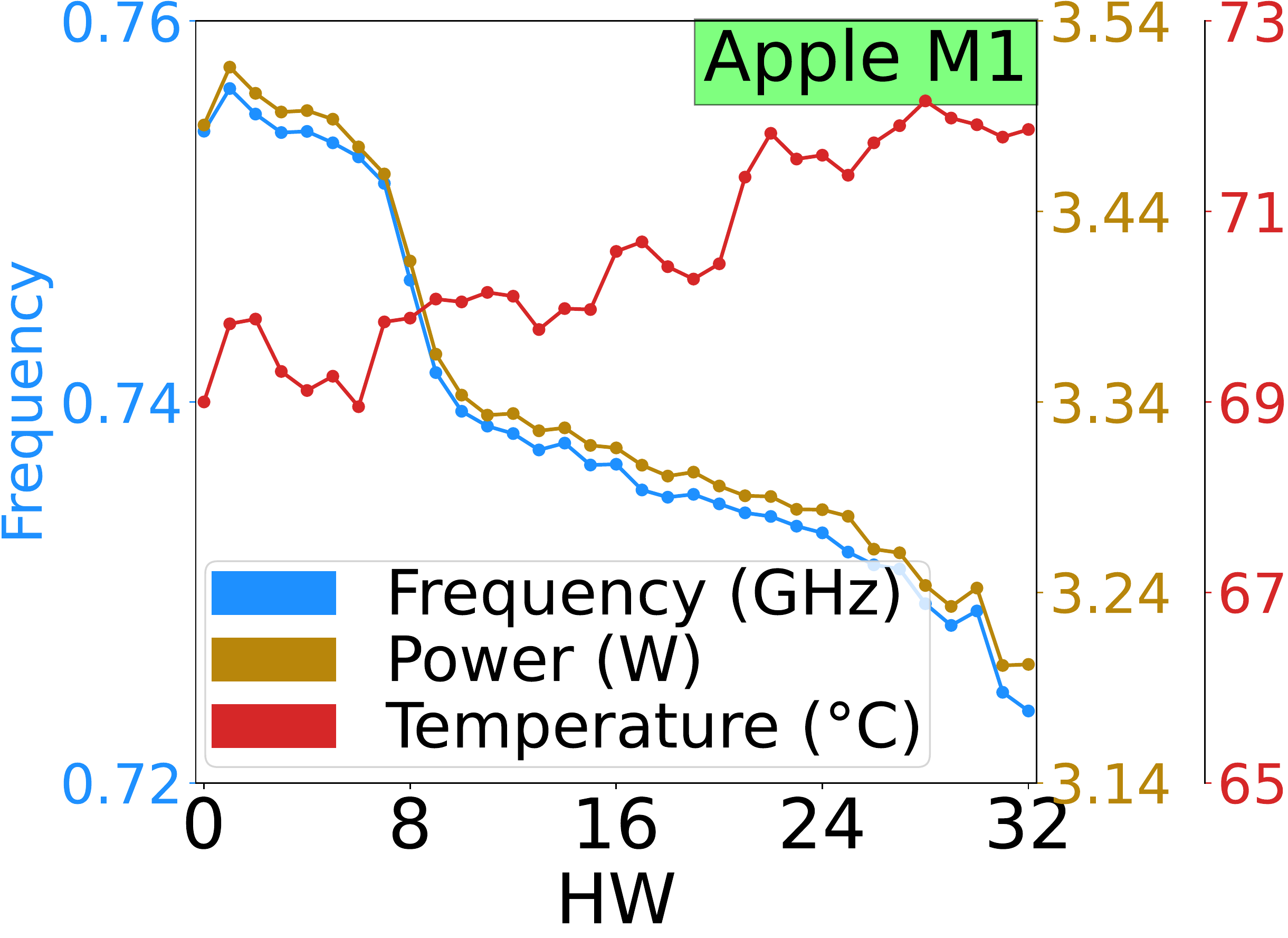}
	\includegraphics[height=0.113\textwidth]{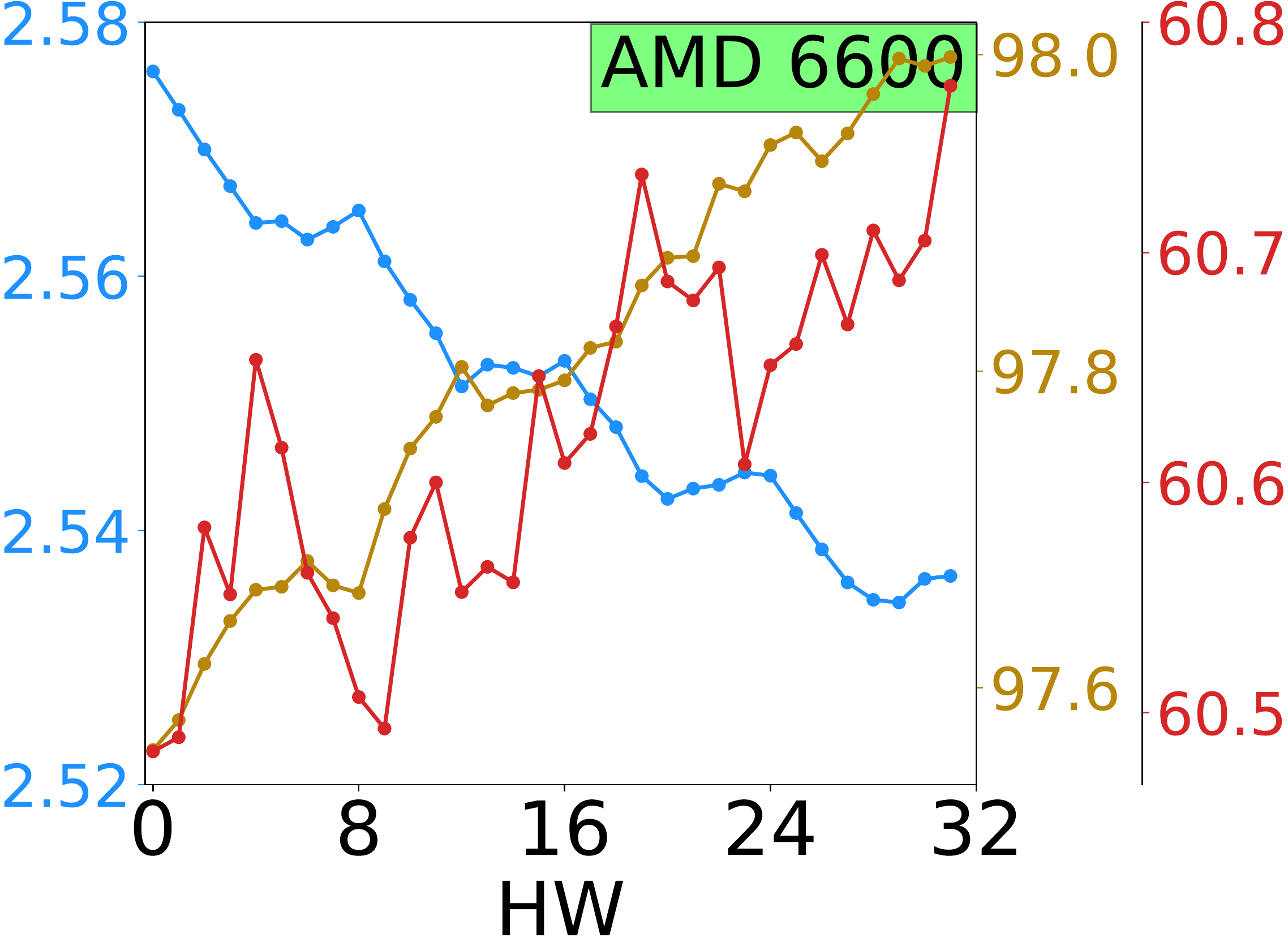}
	\includegraphics[height=0.113\textwidth]{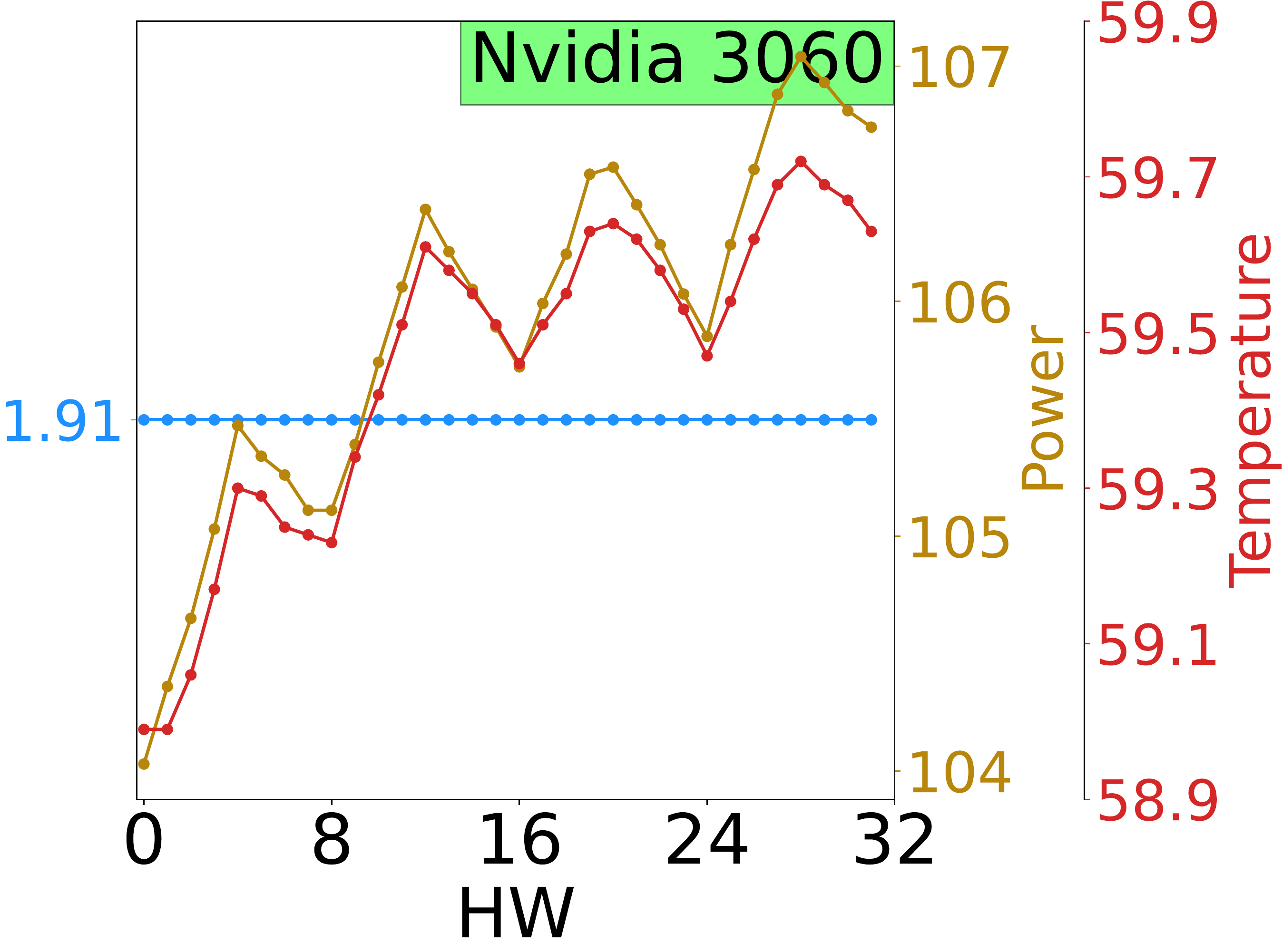}
	\vspace{-0.5em}
	\caption{Traces of frequency, power consumption and, temperature on the Apple M1 iGPU (Left), AMD Radeon RX 6600 (Center), and Nvidia GeForce RTX 3060 (Right) resulting from our HW-dependent workload in \cref{fig:and-kernel}.}
	\label{fig:gpu_hw}	
\end{figure}

\parhead{HW-dependent Frequency Throttling on Apple iGPUs.}
We present the results from the M1 MacBook Air in \cref{fig:gpu_hw} (Left).
We observe that as the HW increases, the steady-state frequency and power consumption both decrease consistently, with correlation coefficients of (-0.961) and (-0.963) respectively.
Given that our M1 iGPU is thermally constrained, we attribute this phenomenon to the higher HW workloads generating more heat, and therefore undergoing more throttling.
Our measurements on the M2 MacBook Air show similar trends, with its iGPU also throttling due to thermal limits.

\parhead{HW-dependent Frequency Throttling on Discrete GPUs.}
We repeat the HW experiment on both discrete GPUs, and show the results for the AMD Radeon RX 6600 in \cref{fig:gpu_hw} (Center).
Here, we see a negative correlation (-0.982) between HW and steady-state GPU frequency. As this device is power constrained to a power budget of about 100W, we conjecture that the dGPU throttles down based on HW in order to maintain this power budget. Next, focusing within the dGPU's power consumption, we see a range of 97.5 to 98.1 W, which is directly correlated (0.985) to the workload's HW. Finally, as the same circuit given more power generally tends to run hotter, we see likewise an average difference of 0.3 \degree C directly proportional to HW (0.740). 

\cref{fig:gpu_hw} (Right) presents our results on an Nvidia GeForce RTX 3060 dGPU. As in prior experiments, we observe that this device is frequency constrained, maintaining a frequency of 1.91 GHz regardless of HW. Finally, we also observe a direct correlation between the workload's HW and the device's power consumption (0.878) and temperature (0.716). 

\smallskip
\begin{takeaway}
	\textbf{Takeaway:} The frequency, power, and temperature of discrete and integrated GPUs are dependent on HW.
\end{takeaway}

\section{Attacking DVFS on Intensive Workloads} \label{sec:attacks-intensive}
Having demonstrated instruction- and data-dependent behavior of CPUs and GPUs during intensive workloads in terms of power consumption, frequency, and temperature, in this section we proceed to demonstrate browser-based pixel stealing attacks using such workloads.   
More specifically, we aim to constrain devices on power or temperature such that they will exhibit differences in frequency based on the color of targeted pixels. This in turn leads to timing differences that are observable by \js-based attackers, allowing them to deduce the pixels' color. 

\subsection{Observing Data-dependent Frequency and Timing From the Web Browser}\label{sec:svg-filters}
The basis for our attacks stems from applying SVG filters on pixels the attacker cannot read. Because the Hamming weights of white and black pixels are 24 and 0 in the RGB space respectively, we expect computation on white pixels to result in lower frequencies and longer runtime compared to black pixels. We note that at the time of writing, Chrome renders SVG filters using the GPU while Safari uses the CPU.

\begin{listing}[htb]
\begin{minted}{js}
let last_render_ts = 0.0;
const tick = now => {
  render_time = now - last_render_ts;
  last_render_ts = now;
  // Applying filter f0 to element:
  // <filter id = "f0">
  //  <feGaussianBlur in = "SourceGraphic"/>
  // </filter>
  element.style.filter = "url(#f0)";
  window.requestAnimationFrame(tick);
};
window.requestAnimationFrame(tick);
\end{minted}
	\vspace{-1em}
	\caption{Our function that repeatedly applies SVG filters and measures the elapsed time for each render.}
	\label{fig:raf-timing}
\end{listing}

\parhead{Inducing Data-dependent Behavior with SVG Filters.}
We create a stack of several \texttt{feColorMatrix} and \texttt{feGaussianBlur} filters.
More specifically, \texttt{feColorMatrix} multiplies a transformation matrix of floating-point numbers to the RGB values of each pixel, changing its color.
Moreover, \texttt{feGaussianBlur} applies a two-dimensional Gaussian function per pixel, requiring several floating point exponentiations.
This filter stack has the effect of performing computations on the Hamming weight of the target pixel repeatedly to an extent that constrains the system's power or temperature budget.
As a result, the stack causes the system to throttle its frequency differently depending on pixel color.

\parhead{Observing Data-dependent Behavior via Time.}
Having induced data-dependent execution frequency on the machine's GPU or CPU, we now observe this behavior by timing the filter's execution. 
For this, we use \js's \texttt{requestAnimationFrame} function (rAF) as shown in line 12 of \cref{fig:raf-timing}.
rAF invokes a user-provided callback function \texttt{tick} (lines 2-11), which receives a timestamp \texttt{now} (line 2) of when it was called.
First, we measure the elapsed time (line 3) between filter applications as a proxy variable for frequency.
Next, we render the filter in line 9 by setting the CSS \texttt{filter} style.
We recursively call rAF on line 10, repeating this procedure.

\parhead{Leakage Source.}
Next, we verify that the source of the timing difference is data-dependent frequency throttling.
To that aim, we compare the difference in rendering times between black and white pixels in two settings: the device's default configuration and when the device frequency is constant. \cref{fig:m1_rca} summarizes our findings for GPUs, and we show our experimental setup and CPU results in \cref{appendix:filter-rca}.  

\begin{figure}[htb]
	\centering
	\includegraphics[height = 0.12\textwidth]{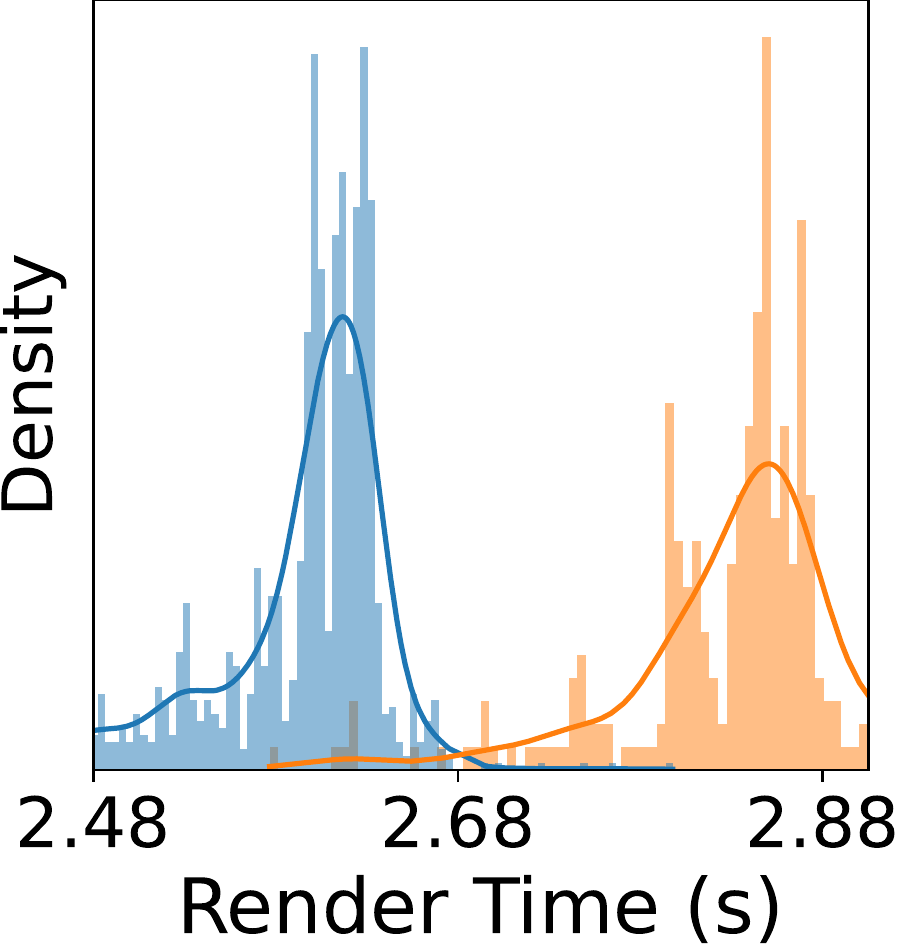}
	\includegraphics[height = 0.12\textwidth]{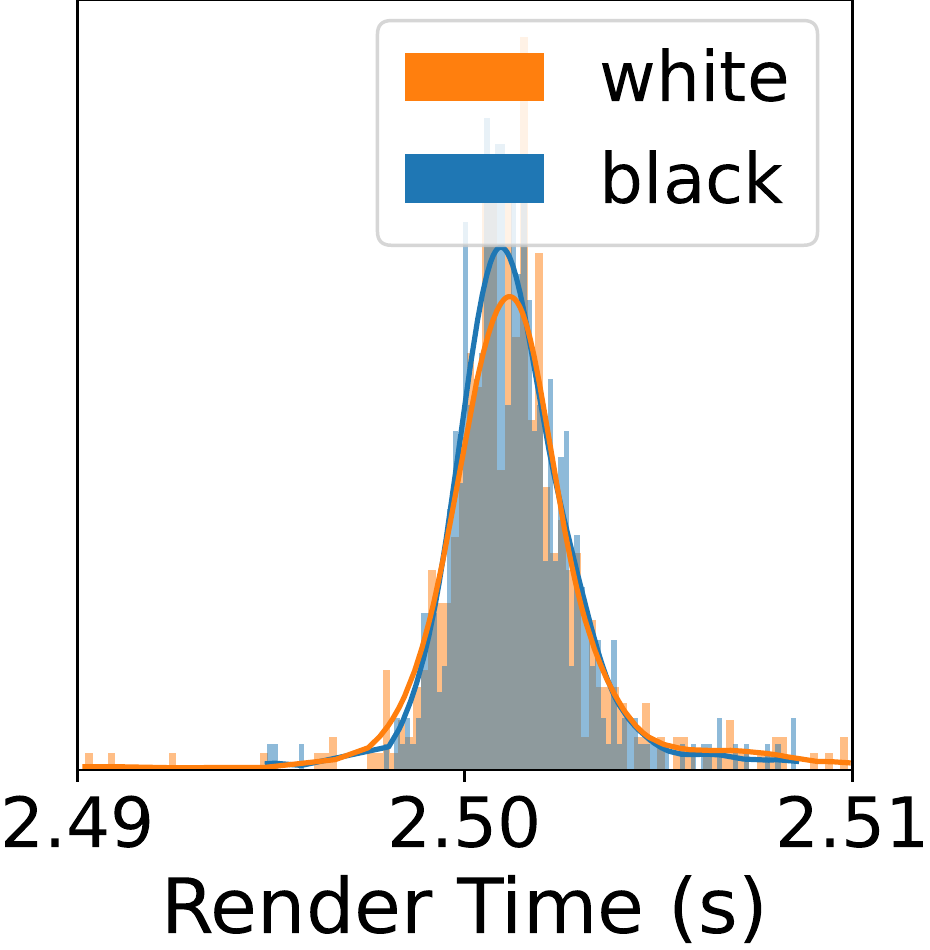}	
	\hfill	
	\includegraphics[height = 0.12\textwidth]{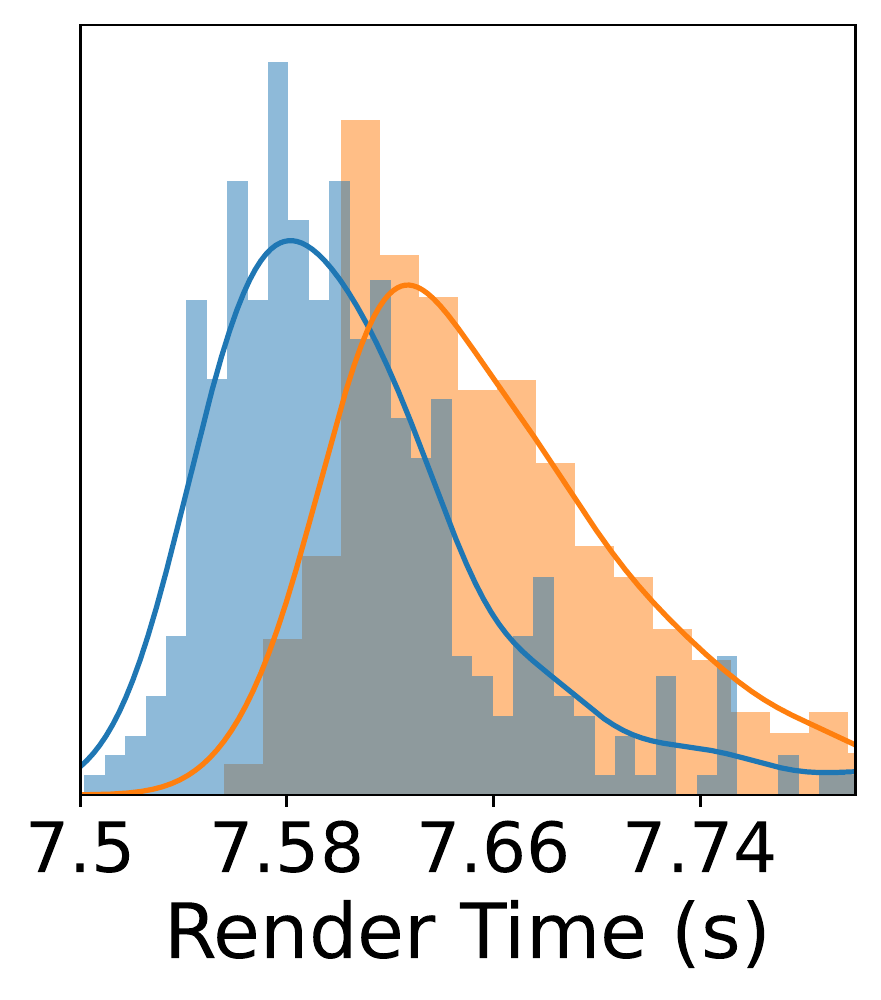}
	\includegraphics[height = 0.12\textwidth]{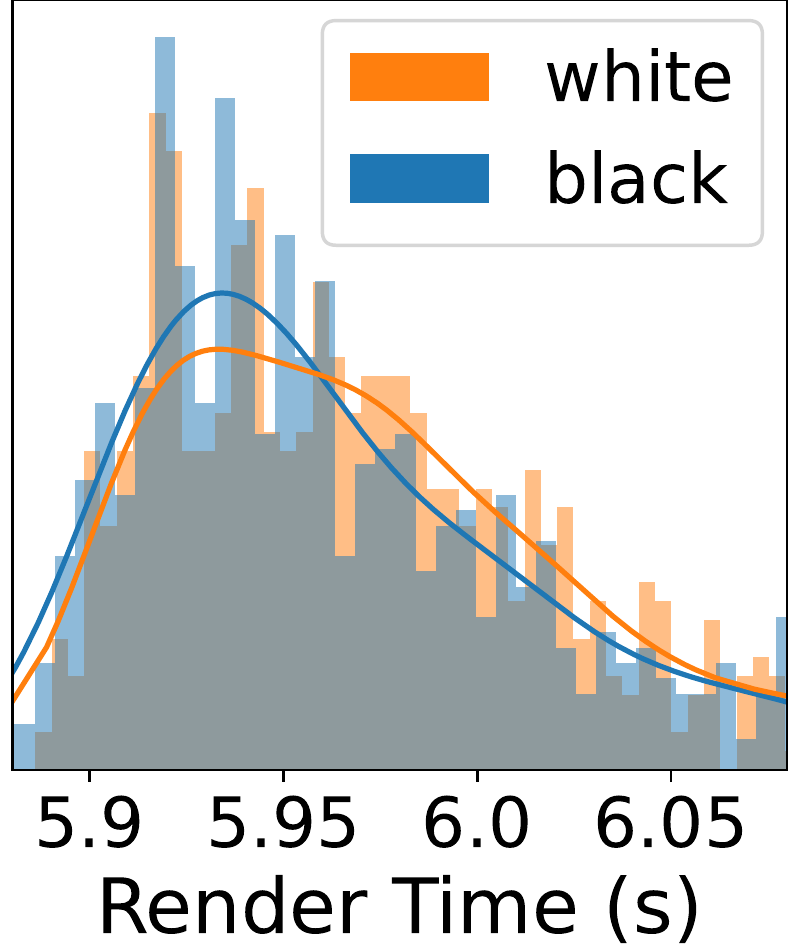}
	\vspace{-0.5em}
	\caption{Filter rendering times on M1 (left two plots) and RX 6600 (right two plots) GPUs for white and black pixels on Chrome. In each pair: left plot is from the default configuration, while the right plot is when the frequency is constant.}
	\label{fig:m1_rca}	
\end{figure}

More specifically, the left two plots show the Apple M1's iGPU and the right two plots show the RX 6600, both running Chrome.
In each pair of plots, the default configuration is shown to the left, and the constant-frequency setting is on the right.
Here, we observe the timing difference disappears when the frequency is made constant, and therefore conclude the difference originates from frequency.

\subsection{Stealing Pixels in Chrome}
Using our filter stack and \texttt{requestAnimationFrame} callback function for filter application and timing, we now steal pixels from an unaffiliated target site from Chrome.
We put the target site as an \texttt{iframe} element in our attacker page.
We assume the target visits our page, and that the \texttt{iframe} renders sensitive information about them.
We cannot inspect the \texttt{iframe}'s contents from \js, but can compute an SVG filter on top of it, laying ground to our attack.

\begin{figure}[htb]
	\centering
	\includegraphics[width=0.8\linewidth]{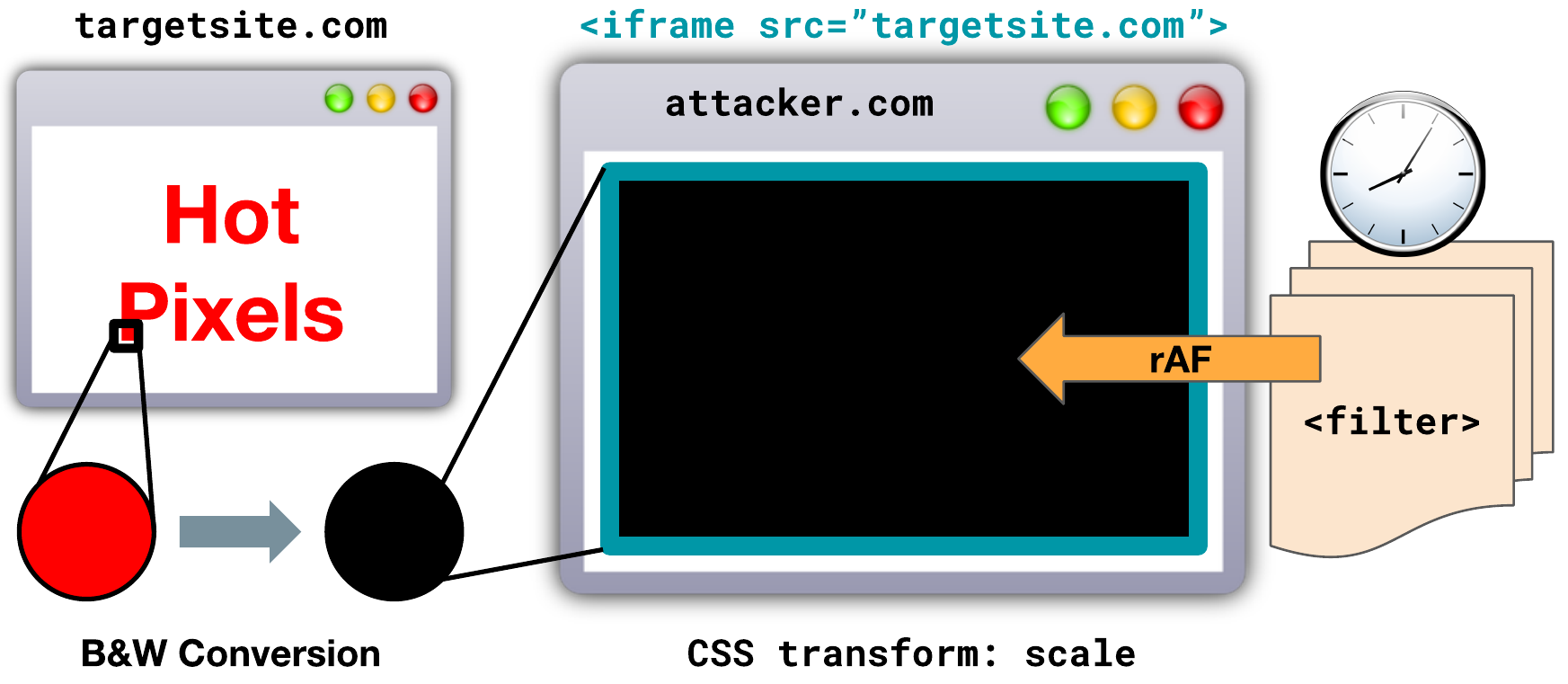}	
	\vspace{-0.5em}
	\caption{An overview of our technique for stealing pixels.}
	\label{fig:filters}	
\end{figure}

\parhead{Experimental Setup.}
Our attacks begin with a calibration phase, wherein we apply and time our filter stack on known pixels to set a ground-truth timing threshold.
Subsequently, we start the stealing phase, 
where we apply and time our filter stack on a target pixel from an \texttt{iframe} element whose contents we cannot access for pixel stealing (see 
\cref{fig:filters}). 

More specifically, we pick a pixel from the target page and use CSS to convert the pixel value to black or white, maximizing the difference in HW.
We then use the \texttt{scale} CSS transform to propagate the target pixel across the entire browser window. 
Next, we use \cref{fig:raf-timing} to repeatedly apply the filter stack, measuring about 200-400 filter renders to overcome \js's 
coarse timer resolution. 
Finally, we classify the color of the target pixel using the timing threshold we obtained during calibration.

\begin{figure}[htb]
	\frame{\includegraphics[height = 0.055\textwidth]{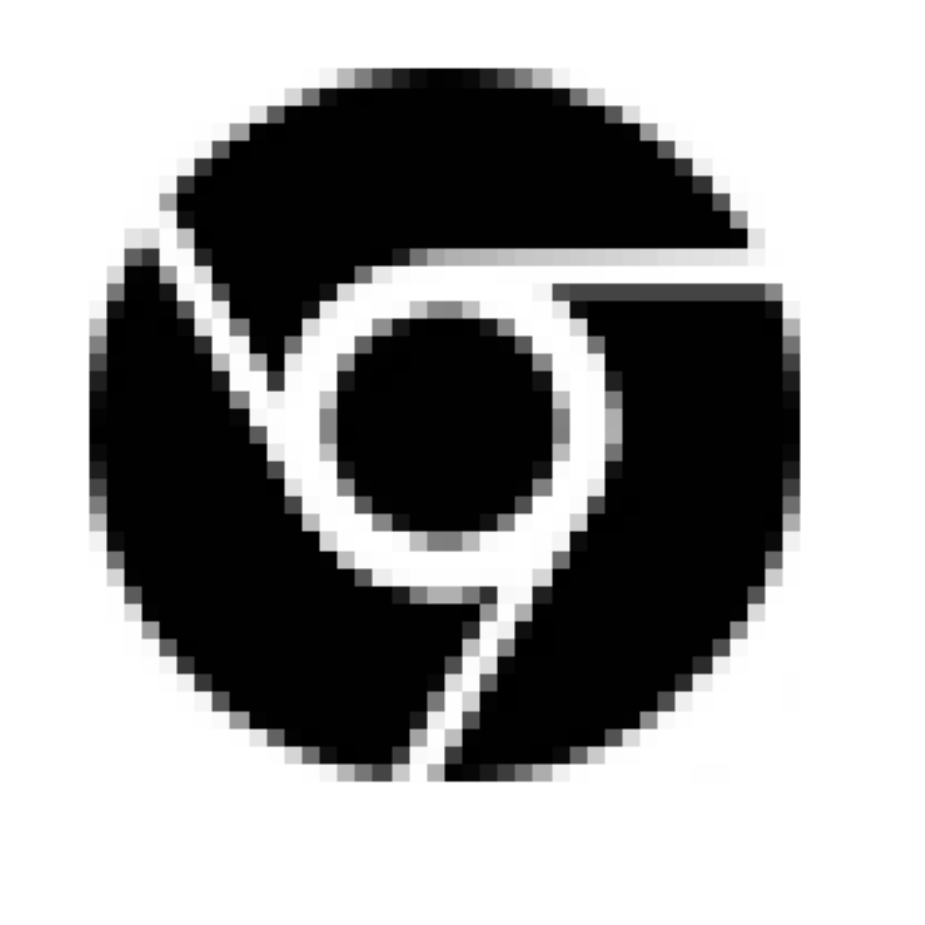}}
	\includegraphics[height = 0.055\textwidth]{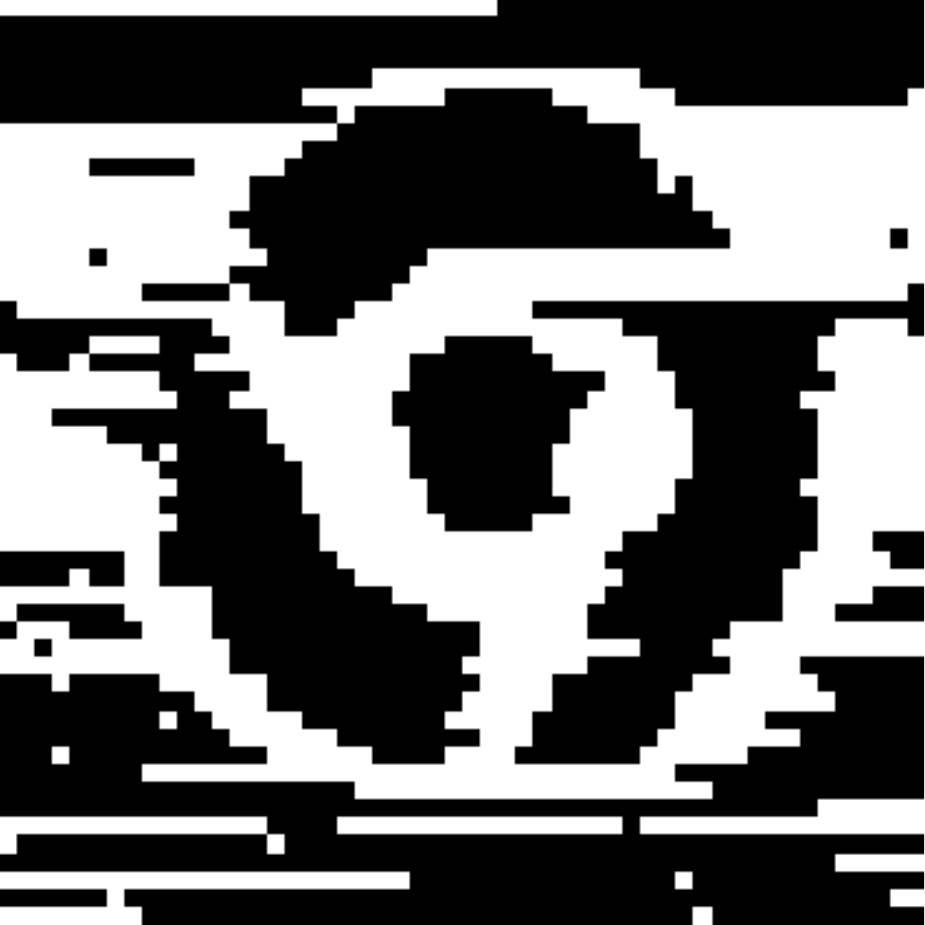}
	\includegraphics[height = 0.055\textwidth]{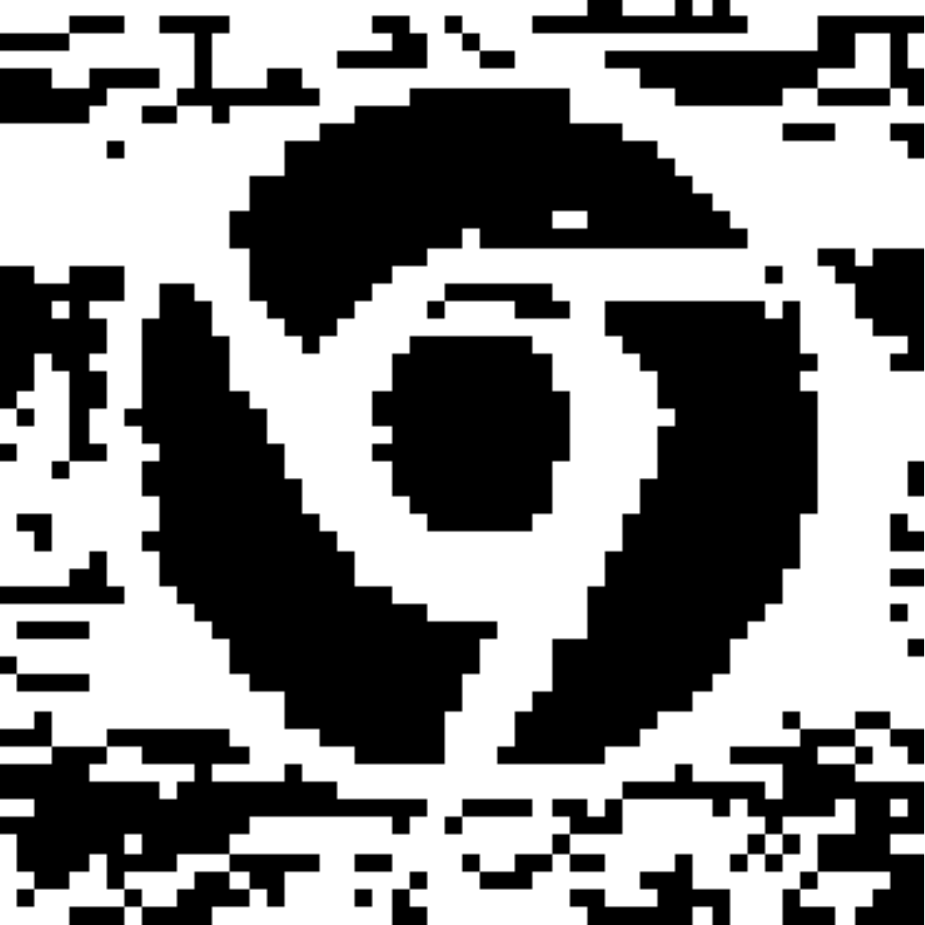}
	\includegraphics[height = 0.055\textwidth]{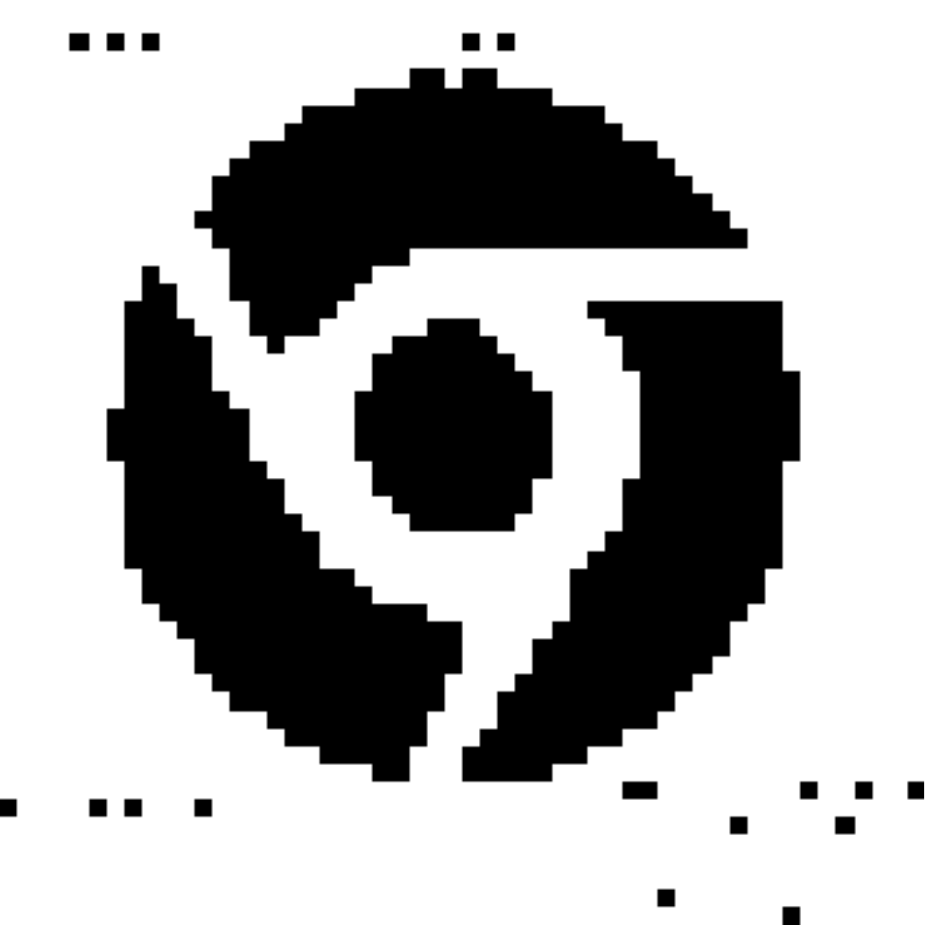}
	\includegraphics[height = 0.055\textwidth]{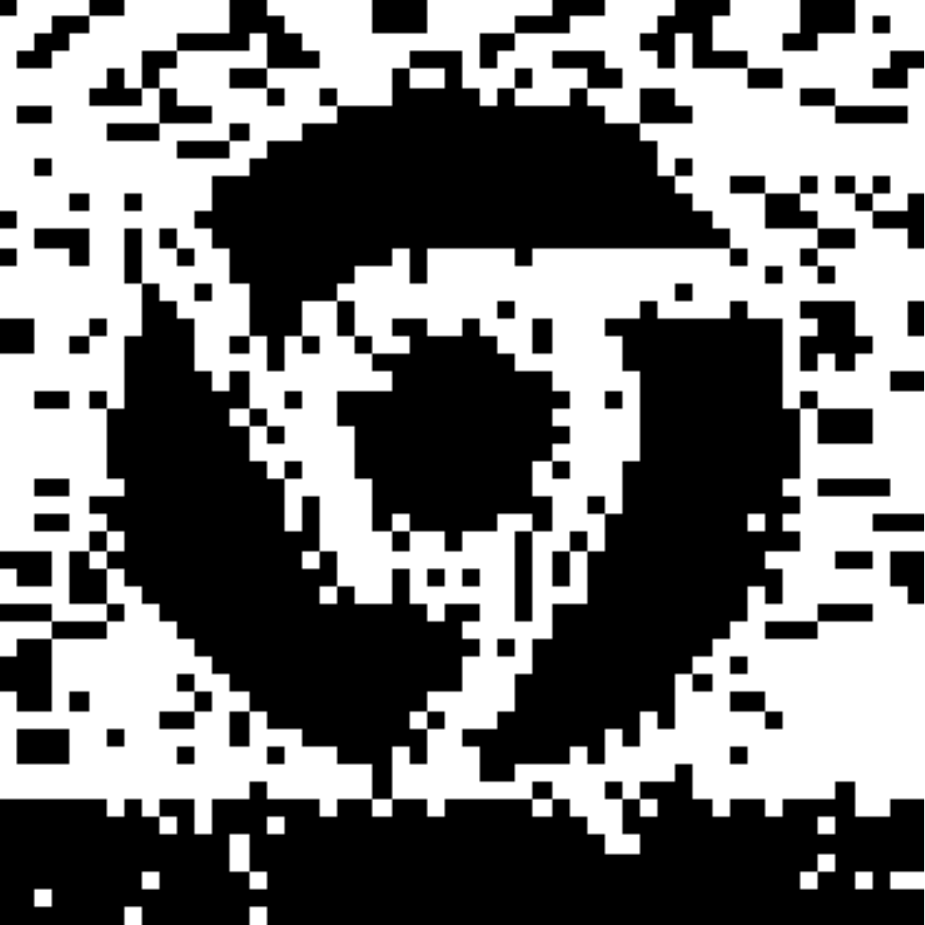}
	\includegraphics[height = 0.055\textwidth]{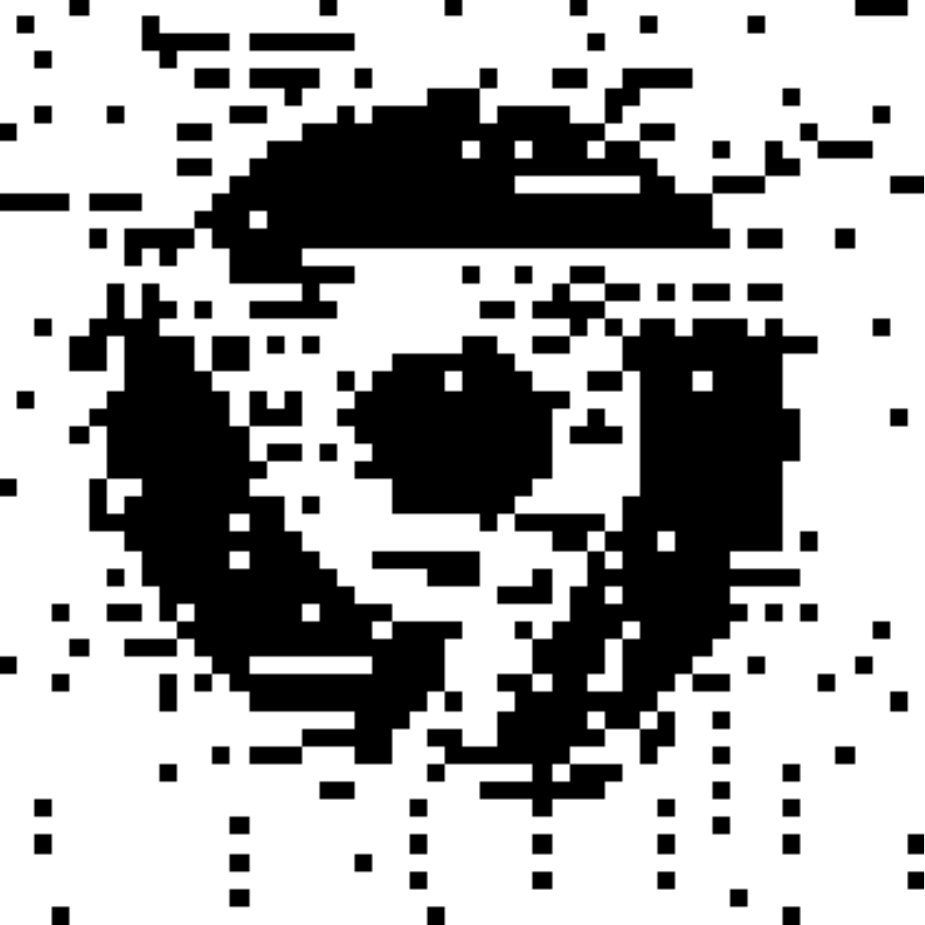}
	\includegraphics[height = 0.055\textwidth]{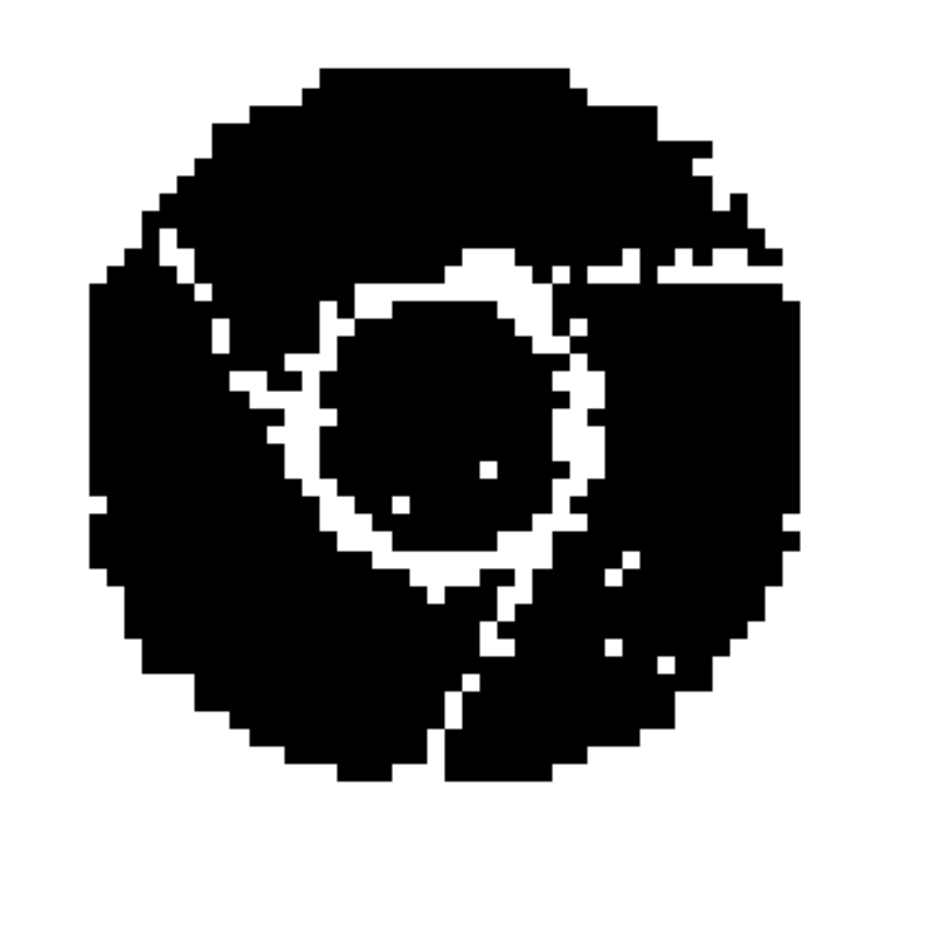}
	\includegraphics[height = 0.055\textwidth]{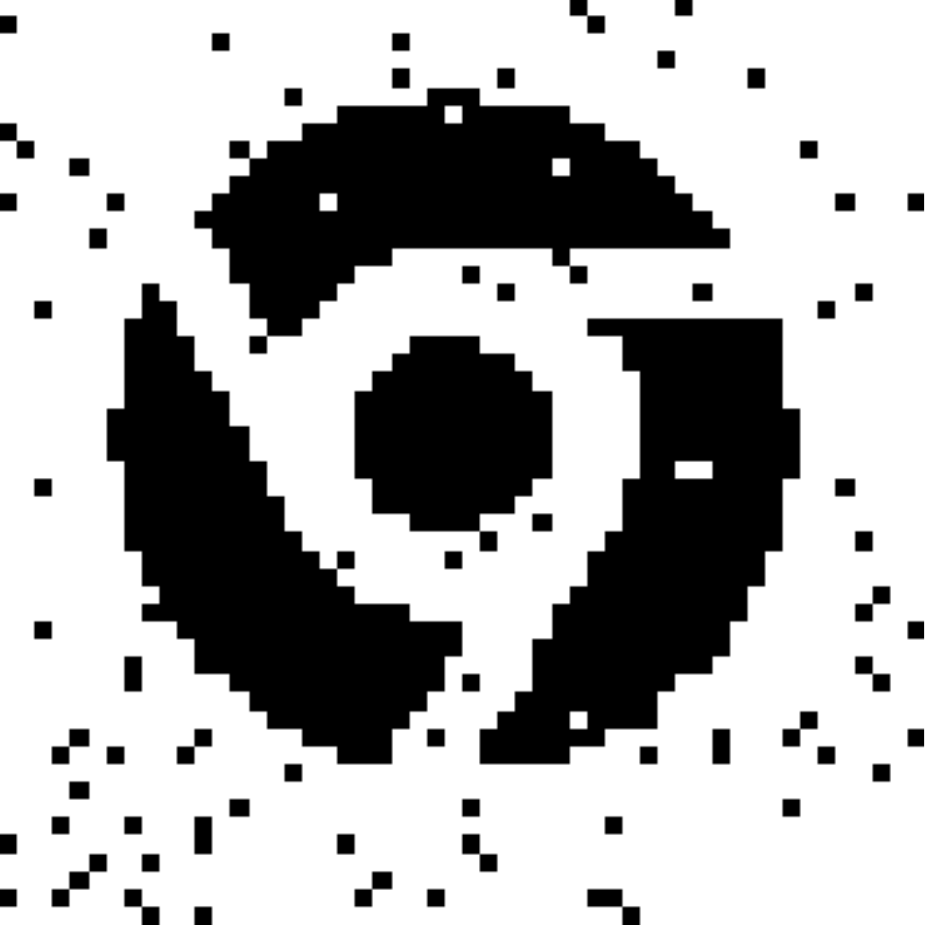}
	\footnotesize
	\centering
	\begin{tabular}{lrr}
		\vspace{-0.8em}
		& &
		\\
		\textbf{Device}         &\textbf{Time/Pixel (s)} & \textbf{Accuracy (\%)} \\
		\midrule
		Apple MacBook Air (M1)  & 22.4 & 60.0\\
		\rowcolor{blue!5}
		Apple MacBook Air (M2)  & 20.8 & 67.0 \\
		Google Pixel 6 Pro & 9.6 & 81.8\\
		\rowcolor{blue!5}
		OnePlus 10 Pro & 18.9 & 70.0 \\
		Nvidia GeForce RTX 3060 & 8.7 & 75.4 \\
		\rowcolor{blue!5}
		AMD Radeon RX 6600 & 8.1 & 94.0 \\
		Intel Iris Xe (i7-1280P) & 22.6 & 77.0 \\
		\bottomrule
	\end{tabular}
	\vspace{-1em}
	\caption{Chrome pixel stealing results (from left to right): Original Image, M1 MacBook Air, M2 MacBook Air, Pixel 6 Pro, OnePlus 10 Pro, Nvidia RTX 3060, AMD RX 6600, Intel Iris Xe. The table summarizes recovery rates and accuracies.}
	\label{fig:logos}	
\end{figure}

\parhead{Empirical Results.}
\cref{fig:logos} (top) presents the results of our end-to-end pixel stealing attack, recovering a picture of the Chrome logo via the GPUs of M1 and M2 MacBook Airs, RTX 3060, RX 6600, OnePlus 10 Pro, Google Pixel 6 Pro, and Intel Iris.
Furthermore, we report the pixel recovery rate and average accuracy of all test devices in \cref{fig:logos} (bottom).
As can be seen, an accuracy of 70\% to 90\% is practically sufficient to recover the image, particularly more so for edges in the image (where the pixel color transitions) than textures.
That is, while almost all of the pixels corresponding to an edge in the original image must be misclassified for that edge to `disappear' in our recovered image, only a couple noisy pixels can make it difficult to determine whether a texture in the original image is smooth or grainy.

\begin{figure}[htb]
	\centering
	\includegraphics[width=\linewidth]{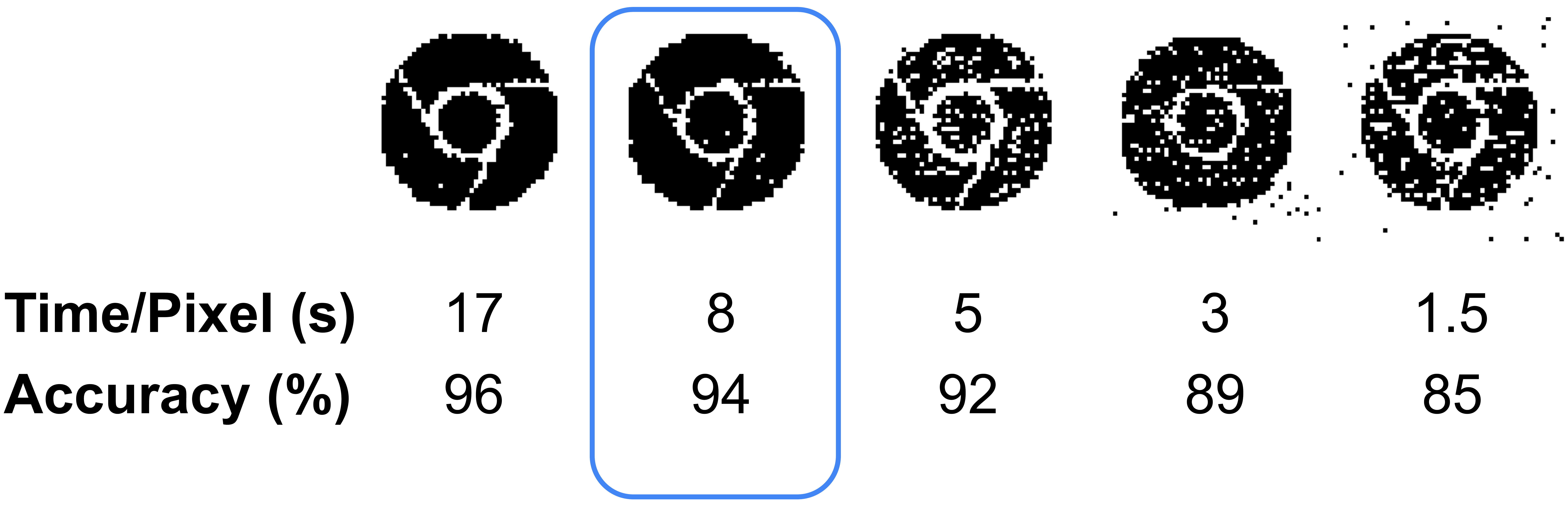}
	\vspace{-2.2em}
	\caption{Results for pixel stealing on AMD RX 6600, with varying amounts of sampling period per pixel. The baseline period of 8 seconds is highlighted.}
	\label{fig:logos_tradeoff}
\end{figure}

\parhead{Comparing Accuracy with Throughput.}
Next, we measure the tradeoff between accuracy and leak rate on the RX 6600 in \cref{fig:logos_tradeoff}. The second-from-left Chrome logo was reconstructed by measuring for 8 seconds per pixel (s/pixel), similarly to the 8.1 s/pixel on the RX 6600 leading to the second-from-right Chrome logo in \cref{fig:logos}. We consider this logo's accuracy and throughput as our baseline. In the leftmost logo of \cref{fig:logos_tradeoff}, we increase the sampling period to more than double, at 17 s/pixel. Conversely, the right three logos of \cref{fig:logos_tradeoff} are reconstructed from decreased sampling periods, namely 5, 3, and 1.6 s/pixel.

The benefits of sampling for 17 seconds generally do not seem to outweigh the cost in throughput, with a 2\% increase in accuracy over the baseline for less than half the leak rate. Meanwhile, the right three logos show improved throughput at notable costs in accuracy, with 2\%, 5\%, and 9\% drops in accuracy over the baseline when sampling for 5, 3, and 1.6 seconds respectively. Visually, the edges remain partially intact and discernible akin to our comments on \cref{fig:logos}, but we observe a sharp decrease in the accuracy of textures for all sampling periods lower than our baseline. 

\subsection{Sniffing History on Safari}
As a countermeasure against pixel stealing, Safari does not send cookies for \texttt{iframe} elements unless they are from the same origin as the attacker's parent page. This fundamentally eliminates pixel stealing attacks, as the \texttt{iframe} will not contain any non-public user-specific information. 

\parhead{History Sniffing.} However, an attacker can still recover the target's history by placing links to sensitive pages on the attacker's own site. As links are often displayed in different colors after the user has visited them, 
querying the value of the \texttt{visited} CSS selector once trivially revealed if a user had accessed a specific hyperlink~\cite{visited-selector}. With modern browsers always reporting the \texttt{visited} CSS selector as `not accessed', we now mount history sniffing attacks using SVG filters. 

\parhead{Attack Setup.} Accordingly, we use the \texttt{visited} selector to set the color of hyperlinks to black for unvisited and white for visited links, and subsequently apply our filter stack to apply stress on the CPU.\footnote{Recall that Safari uses CPU rendering for SVG filters.}
In contrast to pixel stealing, since the hyperlink's pixels are all identical, it suffices to perform the attack on just a single pixel per hyperlink.
We perform our attack on 50 of the top Alexa websites, where we manually visit half of them at random and leave the other half unvisited.

We report their pixel recovery rates and accuracies in \cref{table:history_sniffing}.
We note the CPU, in comparison to the GPU, takes much longer to arrive at a frequency difference that we can observe via timing.
We conjecture that this is due to signal-to-noise ratio, as the 
CPU executes significantly more unrelated processes than the GPU.
Nonetheless, once the difference does become clear-cut, we observe higher accuracies overall, including near-perfect accuracy on the iPhone devices.

\begin{table}[htb]
	\footnotesize
	\centering
	\begin{tabular}{lrrrr}
		\toprule
		\textbf{Device}         &\textbf{RR (s)} & \textbf{Acc. (\%)} & \textbf{FPR (\%)} & \textbf{FNR (\%)}  \\
		\midrule
		MacBook Air (M1) & 270 & 88.8 & 8.9 & 15.1 \\
		\rowcolor{blue!5}
		MacBook Air (M2)  & 187 & 94.8 & 5.9 & 3.3 \\
		iPhone 12 & 211 & 99.0 & 1.2 & 0.0 \\
		\rowcolor{blue!5}
		iPhone 13 & 183 & 99.3 & 0.0 & 2.5 \\
		\bottomrule
	\end{tabular}
	\vspace{-1em}
	\caption{Recovery rate (RR), accuracy, false positive rate (FPR), and false negative rate (FNR) of our history-sniffing attack on Safari across our test devices.}
	\label{table:history_sniffing}
\end{table}

\section{Attacking DVFS on Light Workloads} \label{sec:attacks-light}
Moving away from attacks that recover pixels via heavy workloads, in this section we consider lighter attacks that fingerprint websites based on
frequency adjustments performed by the DVFS mechanism. More specifically, 
we show that individual websites cause bursts of computation on the GPU at different times and of varying intensity. Measuring these bursts results in a pattern that is unique to a website, allowing unprivileged code to profile the target's web activity. 

\parhead{Experimental Setup.}
Following the methodology of \cite{cook2022there}, we filter sites from the Alexa Top 500 whose content is offensive, login-protected (e.g., where the main page is only a login window), or near-identical to another site because it only differs in locale.
After this filtering step, we visit a list of the top 100 remaining websites, which we show in \cref{appendix:fingerprinted-sites}.
We use the Selenium~\cite{selenium} library to automate loading each website with Chrome 108.
We let each website load for 15 seconds, and collect 150 traces of the power consumption and frequency of the iGPU integrated in a M1-based Mac Mini machine. To collect the measurements we use \texttt{SocPowerBuddy}~\cite{SocPowerBuddy}, which can access both sensor data without the need of any elevated privileges.

\parhead{Trace Collection.}
In the preparation phase, we repeat trace collection on the Mac Mini for 10 folds.
We use 9 folds as the training data, and exclude one fold to use as the validation data.
8 hours later, we begin the attack phase by collecting traces for test data on an M1 MacBook Air.
We separate the devices for training and test data collection and leave a time gap to demonstrate that the model's learned decision boundary can generalize robustly to different form factors and thermal budgets for the M1, as well as frequently updating webpages (e.g., social media).
Finally, we use the model of \cite{shusterman2021prime+}, consisting of a small CNN and LSTM neural network.

\begin{figure}[htb]
	\centering
	\includegraphics[width=0.45\textwidth]{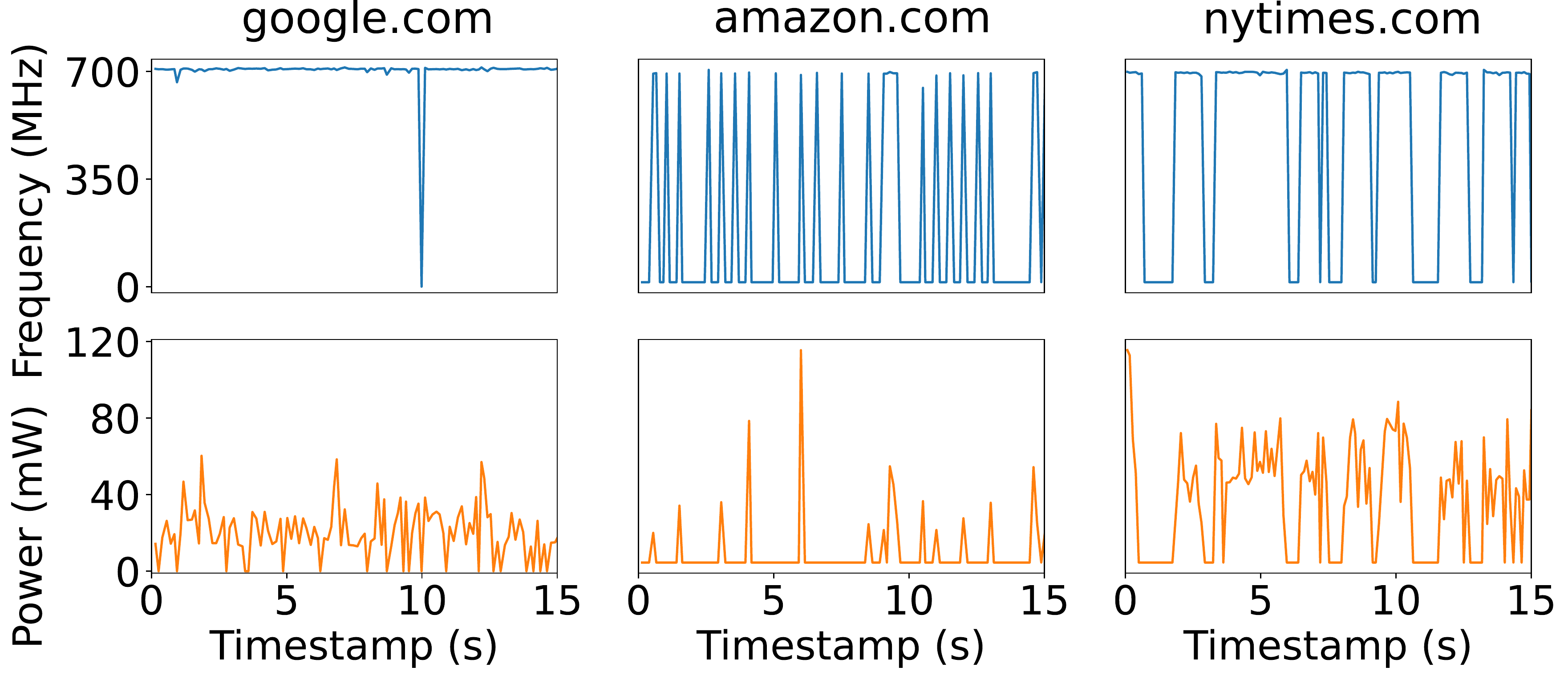}
	\vspace{-0.5em}
	\caption{Apple M1 GPU traces on Mac Mini. The top plot shows GPU frequency, while the bottom plot shows GPU power consumption. \label{fig:fingerprinting}}
\end{figure}

\parhead{Observing Website Distinguishability.}
First, we show a motivating example of how websites can be fingerprinted via GPU frequency and power.
\cref{fig:fingerprinting} shows traces for Google (left), Amazon (center), and the New York Times (right), with GPU frequency on top and power on the bottom.
We observe that while GPU frequency mostly fluctuates at either 0 or 700 MHz, the frequency spikes over time cause the websites to be distinguishable.
GPU power varies likewise over time, and even more over range at anywhere in between 0 and 120 mW.

\parhead{Website Classification Results.}
We show our model's top-1, top-2, and top-5 classification accuracies for the validation and test data in \cref{table:model-acc}.
On the validation data, the classifier guesses the correct website in its top-5 guesses with an accuracy of 0.76.
On the test data, the accuracy decreases to 0.49,
owing to the changes in website content and the different power and thermal capacity of the
passively cooled MacBook Air compared to the actively cooled Mini.  
Nonetheless, we show the classifier maintains an order-of-magnitude improvement over the baseline accuracy, despite changes in machines and cooling methods.

\begin{table}[htb]
	\footnotesize
	\centering
	\begin{tabular}{llrrrr}
		\toprule
		\textbf{Device} &\textbf{Data} &\textbf{Top 1} & \textbf{Top 2} & \textbf{Top 5} & \textbf{Baseline}	\\
		\midrule
		Mac Mini & Validation & 0.51 & 0.62 & 0.76 & 0.01 \\
		\rowcolor{blue!5}
		MacBook Air & Testing & 0.27 & 0.37 & 0.49 & 0.01 \\
		\bottomrule
	\end{tabular}
	\vspace{-0.9em}
	\caption{Classification accuracies of our model. The baseline accuracy is the probability that a random guess is correct.}
	\label{table:model-acc}
\end{table}

\section{Limitations and Mitigations}\label{sec:limit+mit}

\parhead{Limitations of Thermally Constrained Devices.}
Our pixel stealing and history sniffing attacks require the target machine to reach steady state, which is achieved almost instantly for power-constrained devices. However, for thermally constrained devices, it takes a considerable amount of time to attain steady state, depending on how quickly the machine can reach thermal equilibrium.
Furthermore, as the variations in frequency and power consumption are small, our experiments require longer sampling durations to observe discernible timing differences. This limits our leakage rate to about 0.1 bits per second. Thus, while our work serves as a leakage rate lower bound, we leave the task of developing faster DVFS attacks to future work.

\parhead{Hardware-based Mitigations.}
We discuss hardware mitigations primarily for secret-dependent frequency behavior, 
as unlike frequency that can be measured via the passing of time, access to temperature and power consumption sensors can be blocked via API changes. 
First, observing from \cref{sec:cpu_diff_instr,sec:cpu-data-dependent,subsec:gpu_diff_instr} that the Apple M1 SoC does not exhibit instruction- or data-dependent frequency when it is actively cooled,
we foresee that active cooling for thermally constrained devices will mitigate our attacks.

Moving away from cooling, we note that our attacks stem from data-dependent behavior of DVFS algorithms under stress. Thus, a practical compromise is to run the system well below its power or thermal budgets, such that it can accommodate for the difference in power and heat for different instructions or data without having to throttle the frequency.

\parhead{Software-based Mitigations.}
One mitigation for pixel-stealing attacks is to isolate cookies from cross-origin iframes, enforcing all content displayed in iframes not to contain secrets. Such mitigation is already deployed in Safari~\cite{safari-itp}, and is currently under consideration by Chrome developers. More systematically, although it requires a specification change to the HTML standard, prohibiting SVG filters from being applied to iframes or hyperlinks would mitigate both pixel stealing and history sniffing attacks.
Finally, our website fingerprinting attack can be mitigated by OS vendors removing unprivileged access to sensor readings.

\section{Conclusion}
In this paper, we discover that operational constraints cause information about instructions or data to leak via differences in frequency, power consumption, or temperature, depending on the design of each device.
We show this phenomenon is pervasive, demonstrating leakage from software-accessible sensor readings on CPUs and both integrated and discrete GPUs, across several vendors and form factors.
Finally, we demonstrate the privacy risk when the sensor readings are accessible or inferrable with pixel stealing, history sniffing, and website fingerprinting attacks.

As DVFS is widely used in heavyweight chips and also is a crucial component for them to balance performance with energy efficiency, it is possible that the currently known affected devices and attacks are the tip of the iceberg. Since disabling DVFS entails severe practical drawbacks, DVFS-based attacks may persist for the years to come. As such, we leave the task of understanding the true power of DVFS-based leakage, beyond cryptography and SVG filters, to future work.

\ifAnon\else
\section*{Acknowledgments}
This research was partially supported by 
the Air Force Office of Scientific Research (AFOSR) under award number FA9550-20-1-0425, 
  an ARC Discovery Early Career Researcher Award DE200101577; %
  an ARC Discovery Project number DP210102670; %
the Defense Advanced Research Projects Agency (DARPA) under Award number HR00112390029, 
  the Deutsche Forschungsgemeinschaft (DFG, German Research Foundation) under Germany's Excellence Strategy - EXC 2092 CASA - 390781972; %
the National Science Foundation under grant CNS-1954712, 
as well as gifts from Cisco and Qualcomm. 

The views and conclusions contained in this document are those of the authors and should not be interpreted as representing the U.S. Government.
\fi

\bibliographystyle{plainnat}
{
	\setlength{\bibsep}{3pt plus 2pt minus 2pt}
	\def\UrlBreaks{\do\/\do-}
	\bibliography{gpubleed}
}

\appendix

\section{Modeling Hamming Weight Leakage}
\label{appendix:hw_cpu_diff_data}
To understand the effect of Hamming weight (HW) on power consumption, temperature and frequency, we conduct an experiment where we execute the \texttt{and} instruction in a loop on data with varying HW.
We run this instruction with both operands set to the same value to eliminate variations in HD.
The Arm assembly for this workload is shown in \cref{fig:cpu-and}.

\begin{listing}[htb]
\begin{minted}{c}
x8 = x9 = INPUT
.Lloop0:
    and     x19, x8, x9
    and     x20, x8, x9
    and     x21, x8, x9
    and     x22, x8, x9
    and     x23, x8, x9
    and     x24, x8, x9
b   .Lloop0
\end{minted}
	\vspace{-1.49em}
	\caption{Our workload to analyze the effect of HW on the frequency and power consumption of the Apple M1 CPU.}
	\label{fig:cpu-and}
\end{listing}

\begin{figure}[htb]
	\centering
	\includegraphics[width = 0.45\textwidth]{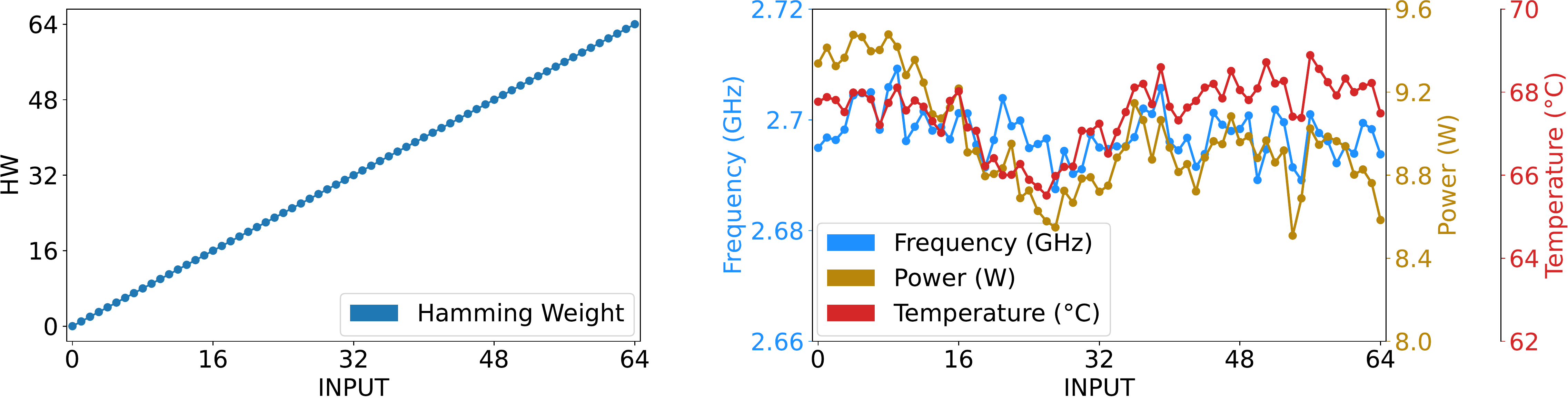}
	\vspace{-0.5em}
	\caption{ HW for each input (left), frequency, power consumption, and temperature (right) resulting from \cref{fig:cpu-and} after averaging the values between 400 and 800 seconds.}
	\label{fig:m1_hw}	
\end{figure}

We run the workload in an infinite loop, with a total of 65 different inputs ranging from $HW = 0$ to $HW = 64$ (setting bits from least significant to most significant). We terminate the infinite loop by signaling once the traces have been collected.
We show the results of this experiment in \cref{fig:m1_hw}.
We did not observe any correlation between power consumption and frequency to the HW of the operand.
Hence, we are unable to conclusively model the frequency and power consumption as a function of operand HW.

This behavior can be explained using our takeaways from \cref{subsec:hd_cpu_diff_data}: we conjecture that the difference in power consumption of the ALU is insufficient to cause a discernible difference in steady state frequency that depends on the value of the input and output operands.
That is, the difference in frequency throttling can only be observed from the bit flips in the internal input and output buffers. 

\section{Ascertaining Frequency as the Primary Source of Timing Differences} \label{appendix:filter-rca}
Here, we describe our experimental setup and additional results for ensuring that frequency causes the timing difference between black and white pixels.

\parhead{Experimental Setup.}
We define the baseline difference as the timing difference between rendering filters on black pixels and white pixels when each device is in its default configuration.
Next, we compare the baseline with the timing difference when we clamp the frequency such that it is constant, and define this as the experimental difference.

We perform our analysis on the Nvidia RTX 3060, AMD RX 6600, and Apple M1.
On the RTX 3060 and RX 6600, we clamp the frequency to several P-states below the peak P-state to avoid power- or temperature-induced behavior.
We achieve this with the \texttt{nvidia-smi} utility~\cite{nvidia-smi} on the RTX 3060, and the \texttt{radeon-profile} utility~\cite{radeon-profile} for the RX 6600.

In contrast, Apple does not provide an interface to clamp the frequency of the M1 CPU or GPU.
Observing from our previous experiments that the M1 Mac Mini has sufficient power and thermal budgets to always maintain its highest CPU and GPU P-states, we use the Mac Mini to measure the experimental difference.
We then use the M1 MacBook Air to measure the baseline, knowing that it exhibits data-dependent behavior on frequency.

\begin{figure}[htb]
	\centering
	\includegraphics[height = 0.11\textwidth]{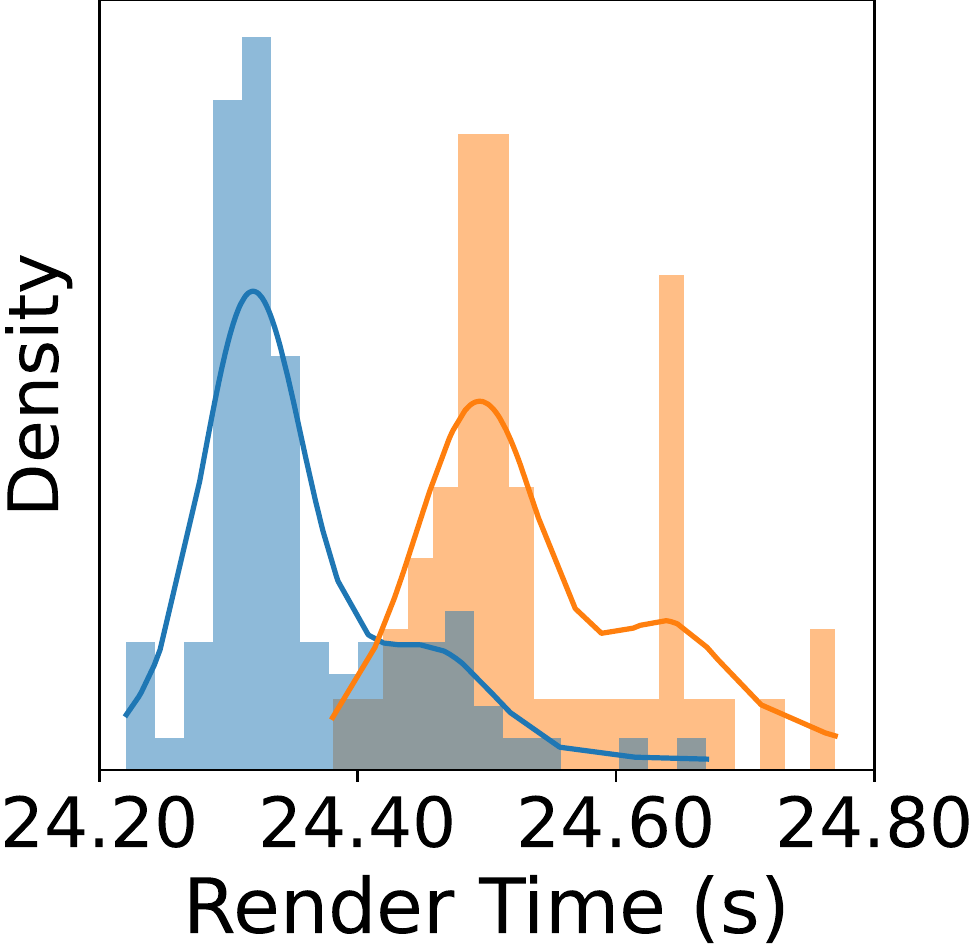}
	\includegraphics[height = 0.11\textwidth]{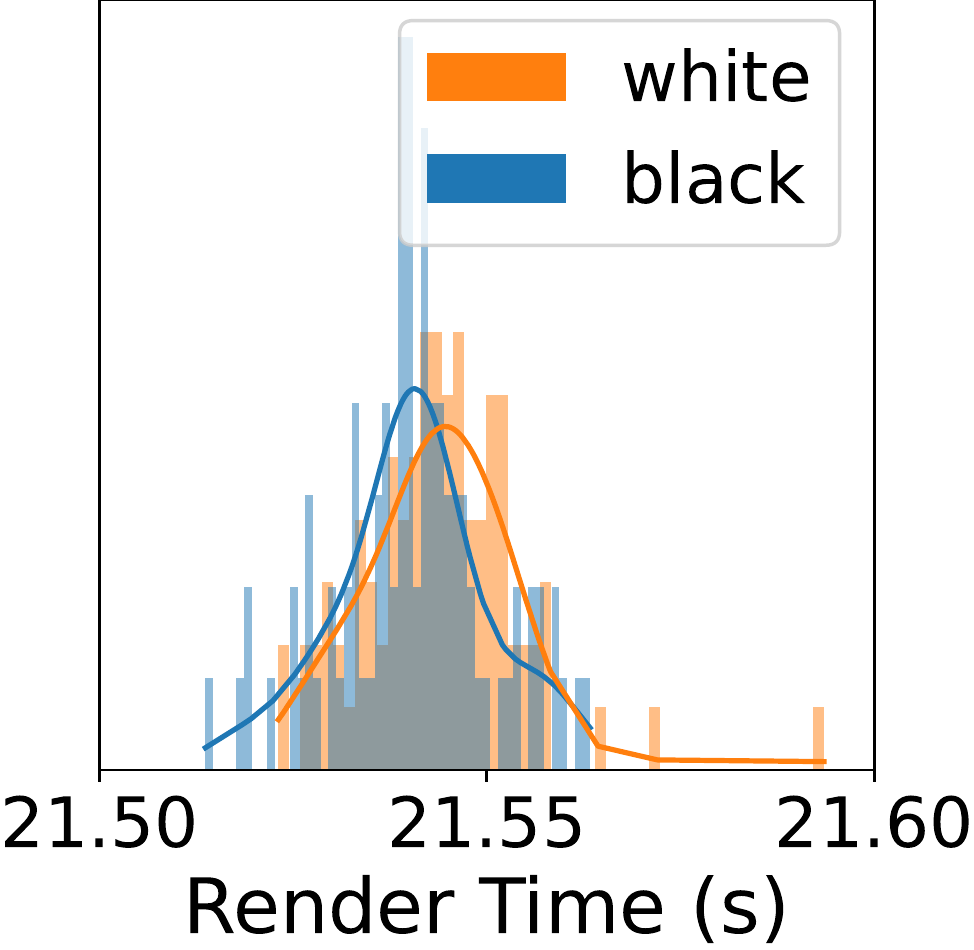}
	\vspace{-0.5em}	
	\caption{Filter rendering times on M1 CPU for white and black pixels for History Sniffing on Safari. The left plot is from the default configuration, while the right plot is when the frequency is constant.}
	\label{fig:safari_rca}	
\end{figure}

We measure the rendering time for a fixed number of iterations of \texttt{requestAnimationFrame}.
We time 100 iterations on the M1 GPU, 250 iterations on the M1 CPU for Safari, and 250 iterations on the RX 6600.
Noticing the coarse timer resolution in both Chrome and Safari, we patch the browsers to supply high-resolution timestamps of at least microsecond granularity, without rounding.
Finally, we show our results for History Sniffing on Safari in \cref{fig:safari_rca}.

\newpage
\section{Website Fingerprinting via GPU Frequency and Power}
\label{appendix:fingerprinted-sites}
We collected traces from the following websites:

\begin{table}[htb]
	\footnotesize
	\centering
	\begin{tabular}{lll}
		google.com	&	youtube.com	&	google.com/maps	\\
		cnn.com	&	amazon.com	&	twitter.com	\\
		wikipedia.org	&	weather.com	&	duckduckgo.com	\\
		walmart.com	&	ebay.com	&	espn.com	\\
		office.com	&	github.com	&	stackoverflow.com	\\
		binance.com	&	bbc.com	&	imdb.com	\\
		w3schools.com	&	geeksforgeeks.org	&	airbnb.com	\\
		vimeo.com	&	soundcloud.com	&	imgur.com	\\
		usps.com	&	tiktok.com	&	target.com	\\
		taobao.com	&	spotify.com	&	sohu.com	\\
		paypal.com	&	nytimes.com	&	msn.com		\\
		apple.com	&	aliexpress.com	&	naver.com	\\
		zoom.us	&	yelp.com	&	tripadvisor.com		\\
		rottentomatoes.com	&	steampowered.com	&	adobe.com	\\
		booking.com	&	hotels.com	&	globo.com	\\
		foxnews.com	&	chase.com	&	wellsfargo.com	\\
		example.com	&	bilibili.com	&	zhihu.com	\\
		reddit.com	&	yahoo.com	&	bing.com	\\
		twitch.tv	&	udemy.com	&	etsy.com	\\
		zillow.com	&	qq.com	&	pinterest.com	\\
		linkedin.com	&	baidu.com	&	bestbuy.com	\\
		speedtest.net	&	expedia.com	&	bankofamerica.com	\\
		openai.com	&	medium.com	&	canva.com	\\
		weibo.com	&	dzen.ru	&	sina.com.cn	\\
		douban.com	&	mail.ru	&	iqiyi.com	\\
		tmall.com	&	avito.ru	&	wordpress.com	\\
		flipkart.com	&	nih.gov	&	mediafire.com	\\
		rakuten.co.jp	&	tumblr.com	&	youku.com	\\
		sogou.com	&	linktr.ee	&	intuit.com	\\
		researchgate.net	&	aws.amazon.com	&	1688.com	\\
		hao123.com	&	nicovideo.jp	&	epicgames.com	\\
		alipay.com	&	xiaohongshu.com	&	shutterstock.com	\\
		uol.com.br	&	dailymail.co.uk	&	samsung.com	\\
		apartments.com	&	&	\\
	\end{tabular}
\end{table}

\end{document}